\documentclass{emulateapj}

\newcommand\eq{\begin{equation}}
\newcommand\eeq{\end{equation}}
\newcommand\eqn{\begin{eqnarray}}
\newcommand\eeqn{\end{eqnarray}}

\begin{document}

\shorttitle{The dwarf galaxy population in nearby groups}
\shortauthors{Carrasco, Mendes de Oliveira \& Infante}

\title{The Dwarf Galaxy Population in Nearby Groups. The data\altaffilmark{1}}
\altaffiltext{1}{Based upon data collected on the 1.3m Warsaw telescope and 2.5m
Du Pont telescope at Las Campanas Observatory, Chile}

\author{Eleazar R. Carrasco\altaffilmark{2}}
\affil{Gemini Observatory, Southern Operations Center, AURA, Casilla 603, 
La Serena, Chile; rcarrasco@gemini.edu}

\altaffiltext{2}{Visiting Astronomer, Las Campanas Observatory. Las \ 
Campanas Observatory is operated by the Carnegie Institution of
Washington.}

\author{Claudia Mendes de Oliveira}
\affil{Instituto de Astronomia, Geof\'{\i}sica e Ci\^encias
Atmosf\'ericas,  Departamento de Astronomia, Universidade de S\~ao Paulo,
Rua do Mat\~ao 1226,  Cidade Universit\'aria, 05508-900, SP, Brazil;
oliveira@astro.iag.usp.br}

\and 

\author{Leopoldo Infante}
\affil{Departamento de Astronom\'{\i}a y Astrof\'{\i}sica, Pontificia
Universidad  Cat\'olica de Chile, Vicu\~na Mackenna 4860, Casilla 306,
Santiago 22, Chile;  linfante@astro.puc.cl}

\slugcomment{Revised version. Submitted 2006 May 22}

\begin{abstract} 
We used $V$ and $I$ CCD  photometry to search for low-surface
brightness dwarf galaxies in the central ($<0.5$ h$^{-1}$ Mpc) region
of the loose groups NGC 6868 (Telescopium) and NGC 5846, the compact
group HCG 42 and the poor cluster IC 4765. We used the parameters given
by the exponential profile fit (central surface brightness, scale
length and limiting diameter) to identify 80 low surface-brightness
dwarf candidates with magnitudes  $17 < V < 22$ mag ($-16.7 > M_{V} >
-11.4$ at the assumed distances of the groups) and with colors
$V-I<1.5$ mag at the limiting isophote of   25.8 V mag/arcsec$^2$. The
selected galaxies have central surface  brightnesses $\mu_{0}>22.5$
V  mag/arcsec$^{2}$, scale lengths $h>1.5$\arcsec,  and diameters
larger than 1.2 h$^{-1}$  kpc. Twenty of the eighty galaxies  are
extended low surface-brightness galaxies that were detected only on
smoothed  images, after masking all high surface brightness objects.
The  completeness fraction in  the detection of the low surface-brightness  
dwarf galaxies is $\sim80$\% for $V\la20$  and
$22.5<\mu_{0}<24.5$  V mag/arcsec$^{2}$, and below 50\% at fainter
magnitudes and central surface  brightnesses ($\mu_{0}>24.5$ V
mag/arcsec$^{2}$ and $V>20$ mag). In this last bin, the completeness
increases to $\sim 80$\% when we search for galaxies in smoothed
images instead.

The detected low surface-brightness dwarf galaxies are highly
concentrated towards the center of the four groups in the inner 250
h$^{-1}$ kpc. At larger radii, the projected number density is similar
to the value found in the control fields. The best fit power-law slope
of the surface density  distribution is, on average, $\beta \sim -1.5$ 
($R < 250$ h$^{-1}$ kpc), in  agreement with the values found for
satellites dwarfs around isolated E/S0 galaxies and in X-ray groups.
The distribution of the low surface-brightness dwarf galaxies in the 
$M_{V} - \mu_{0}$ plane does not show a clear correlation, suggesting
that  the correlation noted by other studies could be produced by
selection  effects. The low surface-brightness dwarf galaxies follow a
well defined  color-magnitude relation, extending for more than ten
magnitudes (from bright ellipticals to faint dwarfs). A similar well
defined color-magnitude relation from giants to dwarfs is known to be
valid for galaxy clusters but it is the first time that it is
demonstrated in the sparse environments of groups.

The spectroscopic follow-up shows that only 78 of the 412 galaxies with
measured velocities in the fields of these groups are group members. We
were able to obtain the radial velocity for 11 of the 80 low-surface
brightness  dwarf galaxies. Five of these galaxies are group members
while six are in the background of the groups. We infer, then, that more
than 90\% of the 80 low surface-brightness dwarf galaxies in our
sample must be bonafide group members. In addition, new
structures along the group's lines-of-sights were discovered. These new
structures are groups and poor  clusters that extend to $\sim 0.3$ in
redshift space.
\end{abstract}


\section{Introduction}

Groups are the most common environment of galaxies in the Universe. Most
galaxies are associated to these poor systems and, due to their low
velocity dispersions, these systems are good laboratories to study the
different physical processes that regulate galaxy formation and evolution.

Nearby galaxy groups are characterized by their small numbers of
bright galaxies (typically 5--7 galaxies brighter than M$^{*}$). Usually
redshifts are known only for the brightest members, and no redshift
information is available for the galaxies at fainter magnitudes. 

The faint-end of the luminosity distribution is the regime where  the
dwarf galaxies dominate. Dwarf galaxies are exceptional laboratories 
for the study of many physical processes, such as structure formation,
galaxy  evolution, star formation, and dark matter distribution.
However, the low luminosity and surface brightness of these objects have
limited most of the previous studies done so far. Because of their
low-luminosities, spectroscopic observations of these objects are
extremely difficult and time consuming.  Except for the Local Group and
other nearby structures \citep[e.g. M81, Sculptor, LeoI, ComaI:][1999]{jer98,jer00,tre02,bre98,flint01}  
the physical characterization of dwarf galaxies and its implication to
galaxy formation and evolution is poorly constrained. In the last few
years, the advent of a new generation of telescopes and new large-format
CCD detectors opened a unique possibility to search and detect the
low surface-brightness dwarf galaxy population of nearby systems. It became
possible to obtain deep photometry for statistically significant 
samples of these galaxies in large areas of nearby groups and clusters
\citep[e.g.][]{car01,tre02,tre05}.

This is the second paper of a series dedicated to survey nearby
groups with the aim of identifying the population of low surface
brightness dwarf galaxies (LSBD) down to  M$_V =-13$ mag. The main
goal is to identify the LSBD and to characterize and study the
physical properties and the Galaxy Luminosity Function of the groups
(the latter will be presented in  a later paper).   The program
started with the observation of the Dorado group at  $cz\sim1200$
km s$^{-1}$ \citep{car01} and continued with the observation of  other four
nearby groups with $1000<cz<4500$ km s$^{-1}$, described here. In this
paper we present  the results of the photometric and spectroscopic
study of the galaxy population   in NGC 6868, NGC 5846, HCG 42 and
IC 4765.

The paper is organized as follows.  In Section 2 we describe the sample.
The observations, the data reduction and the estimation of the
photometric errors and completeness of the spectroscopic sample is
presented in section 3. In section 4 we explain the criteria
used to select the LSBD sample, the determination of the completeness
fractions in the detection of Local Group-like dwarf  galaxies at the
distance of the groups,  and the photometric analysis  of the LSBD
galaxies detected. Section 5 is  devoted to analyse the spectroscopic
survey, while in section 6 we present the main results. Finally, our
conclusions are presented in Section 7.

For all calculations, the following cosmological parameters were
assumed: $\Omega_{M}=0.2$, $\Omega_{\Lambda}=0.8$ with a Hubble
constant of H$_{0}=75$ km s$^{-1}$ Mpc$^{-1}$. Here $h$ is a 
dimensionless constant defined as the Hubble constant divided
by 75 km s$^{-1}$ Mpc$^{-1}$, i.e. $h=H/75$.

\section{The sample}

The four groups studied were selected from the catalog of
\citet{gar93}. The groups vary in richness and density, with galaxy
populations which differ widely in morphological fractions. Some
of their characteristics are described below.

The Hickson Compact Group 42 \citep[HCG 42;][]{hic82} is dominated
by the giant elliptical galaxy NGC3091 (E3) and by the galaxies HGC 42b
(SB0), HCG 42c (E2) and HCG 42d (E2).  The group has a galaxy density of 
$\sim25$ gal/deg$^2$ inside a radius of 0.3  h$^{-1}$ Mpc an 
brighter than $M_{V}\sim-18$ mag \citep{zab98} with
a high number of faint galaxies superimposed onto the region. In
fact, \citet{roo94} found evidence that this system could be
associated to the poor group NGC 3091 \citep[or LGG186;][]{gar93},
suggesting that HCG 42 is an intermediate structure: a compact group
inside a more extended system \citep{deC94}. X-ray maps from the
ROSAT satellite analyzed by \citet{ebe94} showed that this group has
an extended X-ray emission. Further analysis by \citet{pon96} and
\citet{mul98}  showed that HCG 42 has two components in the gas 
emission, one centered around NGC 3091 galaxy, with
L$_{X}\sim10^{41}$ h$^{-2}$ erg s$^{-1}$  and  a second, more
extended, that could be associated to the global potential well of
the  group \citep{mul98,mul03}. The assumed distance to
HCG 42, shown in Table 1 (col 1), is based on the
distance to the galaxy NGC 3091, obtained by the D$_{n} - \sigma$
method \citep{fab89}.

The NGC 5846 group \citep[GH150;LGG929][]{gel83,gar93} is dominated
by the giant elliptical galaxy of the same name (E0) and by the
galaxies NGC 5846A (cE2), NGC 5850 (SB(r)b), NGC 5845 (E1) and NGC
5839 (SAB(r)). For galaxies $M_{V}\sim-18$ mag, the group has a
density of $\sim11$ gal/deg$^2$ inside a radius of 0.3  h$^{-1}$ Mpc
\citep{zab98}. This group is part of an extended cloud of galaxies
detected in the Virgo-Libra direction. Also, it  shows an extended
X-ray emission, with two clear components, one, with a luminosity
L$_{X}\sim10^{41}$ h$^{-2}$ erg s$^{-1}$ centered on NGC 5846 and
a second, more extended, with L$_{X}\sim10^{42}$ h$^{-2}$ erg s$^{-1}$
\citep{mul98,mul03} associated to the intragroup medium. The assumed
distance to this group (see Table 1) is based on the average
distance to the five brightest galaxies, determined through the
surface brightness fluctuation method \citep{ton00}.

IC 4765 \citep[Sersic 129-01, PavoII, DC1842-63, Abell S805, LGG422;][]{ser74,dre80,abe89,gar93} 
is defined as a poor cluster of galaxies (Bautz-Morgan I, richness
0). However, it could also be treated as a rich group.  It is
located close to the galactic plane ($b\sim-23$ deg) and it is
dominated by the D galaxy IC 4765 and by a barred spiral galaxy IC 4749
(SB(rs)bc and 11 arcmin south of IC 4765). Following \citet{qui75},
this group is formed by two structures in interaction: one poor,
spiral rich, centered around IC 4749 and another rich sub-system
centered around IC 4765.  The group shows X-ray emission
L$_{X}=3.6\times10^{43}$ h$^{-2}$ erg s$^{-1}$ \citep{jon99} and a moderate
{\em cooling flow} \citep{whi97}. There is no information in the
literature about the distance to this cluster. The value in Table \ref{tab1}
has been calculated assuming that the redshift is a good indicator
of the distance.

The last studied group, NGC 6868 \citep[Telescopium, LGG430;][]{gar93},
extends for more than 2 degrees on the sky and shows an elongated
form in the East-West direction. It is dominated by the giant
elliptical galaxy NGC 6868 (E2) and by the galaxies ESO233-G035
(S0), NGC6870 (SAB(rs)ab) and NGC 6861D (SA(s)). The group shows a
low X-ray emission with L$_{X}\sim10^{41}$ h$^{-2}$ erg s$^{-1}$ centered
on NGC 6868 galaxy \citep{beu99}. \citet{gar93} identified 11
galaxies brighter than $M_{V}=-18$ as group members, while
\citet{ram96}  identified five other members. The
assumed distance to this group  is based on the average distance to
the five brightest group galaxies determined through  the Surface
brightness fluctuation method \citep{ton00}.

The group parameters shown in Table 1 are (1) group name;
(2) and (3) the equatorial coordinates; (4) the distance; (5)
velocity dispersion; (6) bolometric X-ray luminosity; (7) and (8)
the interstellar extinctions  in the direction of the groups. 
Extinction values have been obtained from the maps of 
Schlegel et al. (1998) using the relations $A_{V}=3.1 \times E(B-V)$ 
and A$_{I}=1.5\times E(B-V)$ from \citet{car89}.  

\section{Observation and Data Reduction}

\subsection{Imaging}

The groups were observed in February and March 1998, with the  1.3m
Warsaw telescope (Udalsky et al. 1997) located in Las Campanas 
Observatory, Chile, using the SIT1 $2048 \times 2048$ CCD detector. 
With a pixel size of 24$\mu$m and a scale of 0.414\arcsec/pix, the
detector  covers an area of $14.13\times14.13$ arcmin$^{2}$ on the
sky. 

The groups were imaged through the standard Johnson V and Cousins I
filters. All observations were done under photometric conditions,
except  for one of the fields in  NGC 6868. For this field, a
calibration image  observed under photometric conditions and in a
different night, was  obtained in order to derived the correct
zero-point. The seeing conditions  were relatively good and varied
between 1.2\arcsec and 1.5\arcsec in  both filters. The derived
seeing values per filter for each of the observed  fields are
tabulated in Table \ref{tab2} (see below).

We have observed each group with several different pointings, covering
the central area, following a mosaic pattern. The total areas observed 
in the central region of the group were: for HCG 42, $28 \times 28$ 
arcmin$^2$ (5 fields); for NGC 5846, $42 \times 14$ arcmin$^2$ (3 fields);
for IC 4765, $14 \times 28$  arcmin$^2$ (2 fields); and for NGC 6868, 
$28 \times 14$ arcmin$^2$  (2 fields). These values correspond to a 
physical area of 
$\sim 0.44 \times 0.44$ h$^{-2}$  Mpc$^2$, $0.29 \times 0.10$ h$^{-2}$ 
Mpc$^2$, $\sim 0.24 \times 0.48$  h$^{-2}$ Mpc$^2$ and $\sim 0.26
\times 0.13$  h$^{-2}$ Mpc$^2$ at the distance of HCG 42, NGC 5846, IC
4665 and NGC 6868  respectively. For three of the groups (all but NGC
6868), observations of  control fields in both filters were obtained in
order to estimate the   background and foreground galaxy counts (9
fields). We pointed the   telescope 2-3 deg from the center of the
groups to acquire these control  fields. 

\begin{figure}[!htb]
\figurenum{1}
\centering
\includegraphics[totalheight=4.5cm]{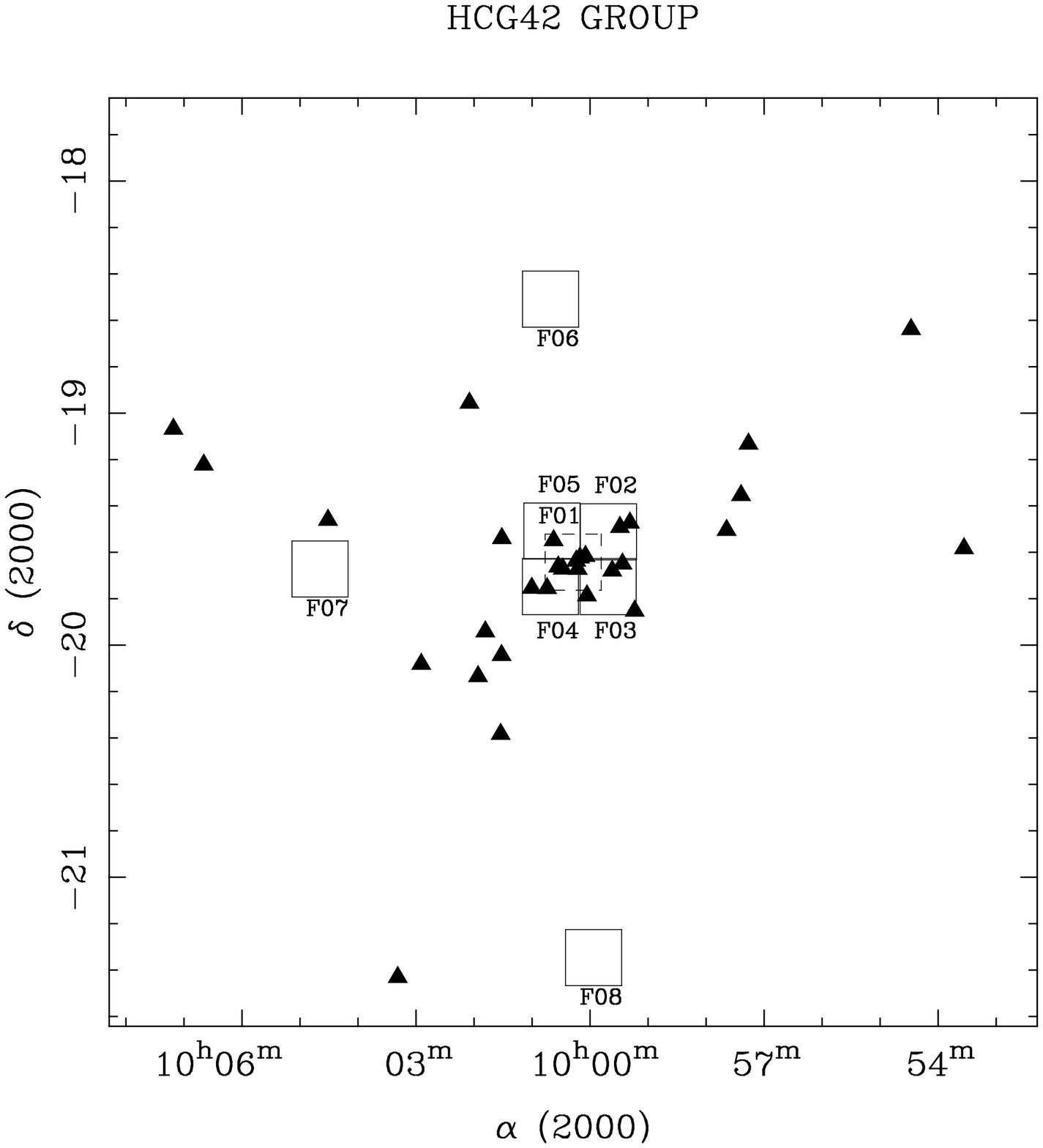}%
\includegraphics[totalheight=4.5cm]{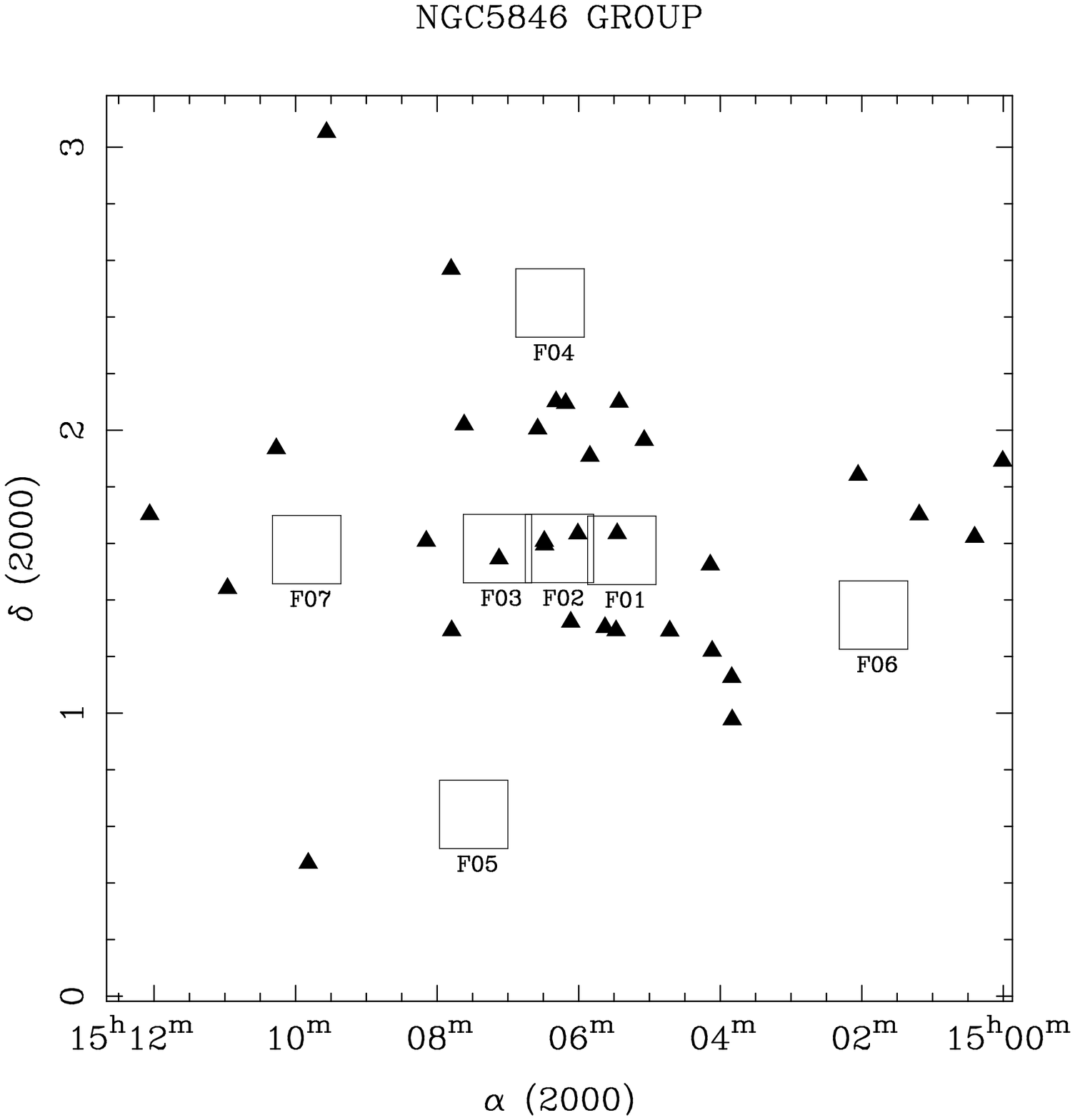}

\includegraphics[totalheight=4.5cm]{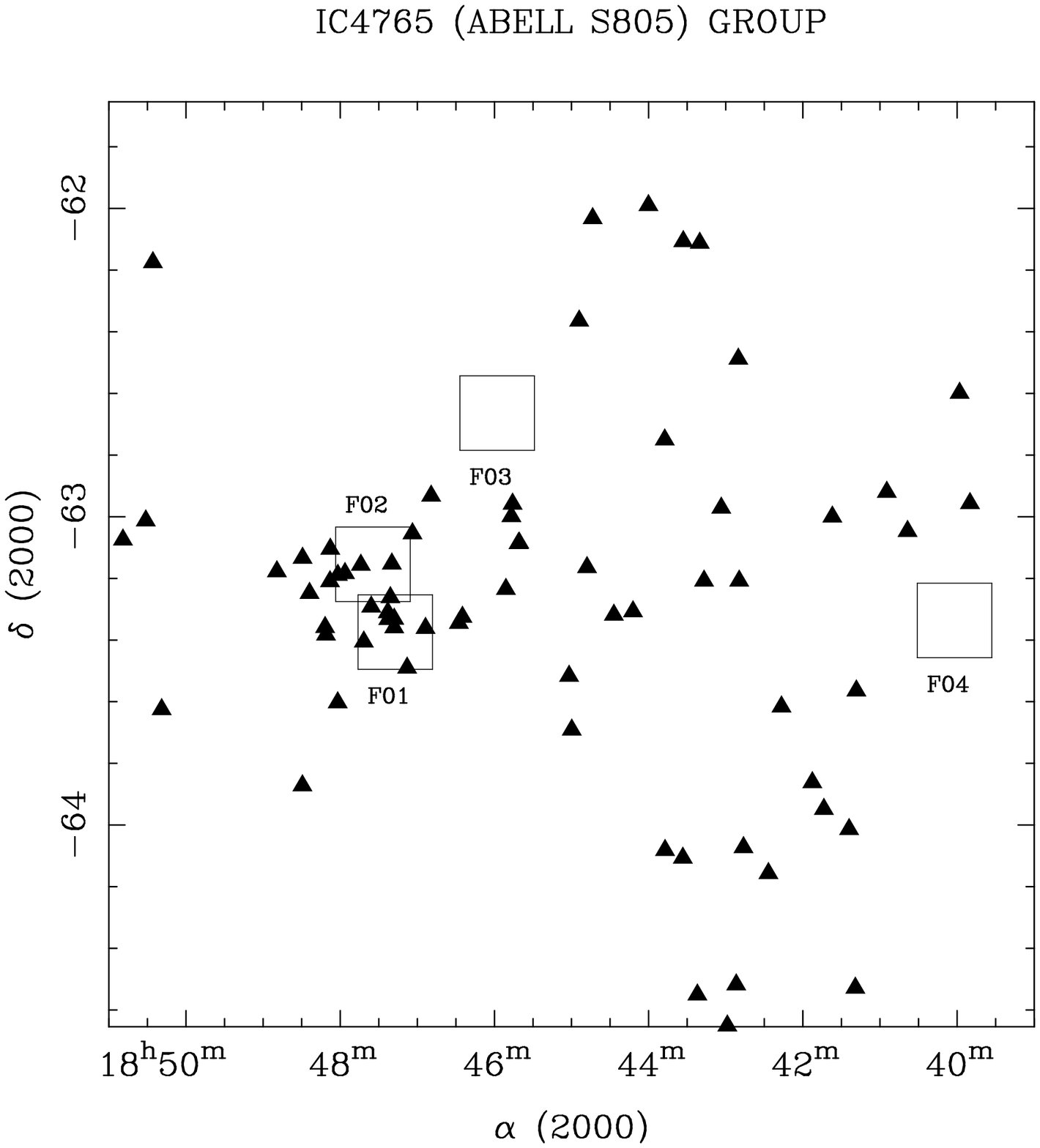}%
\includegraphics[totalheight=4.5cm]{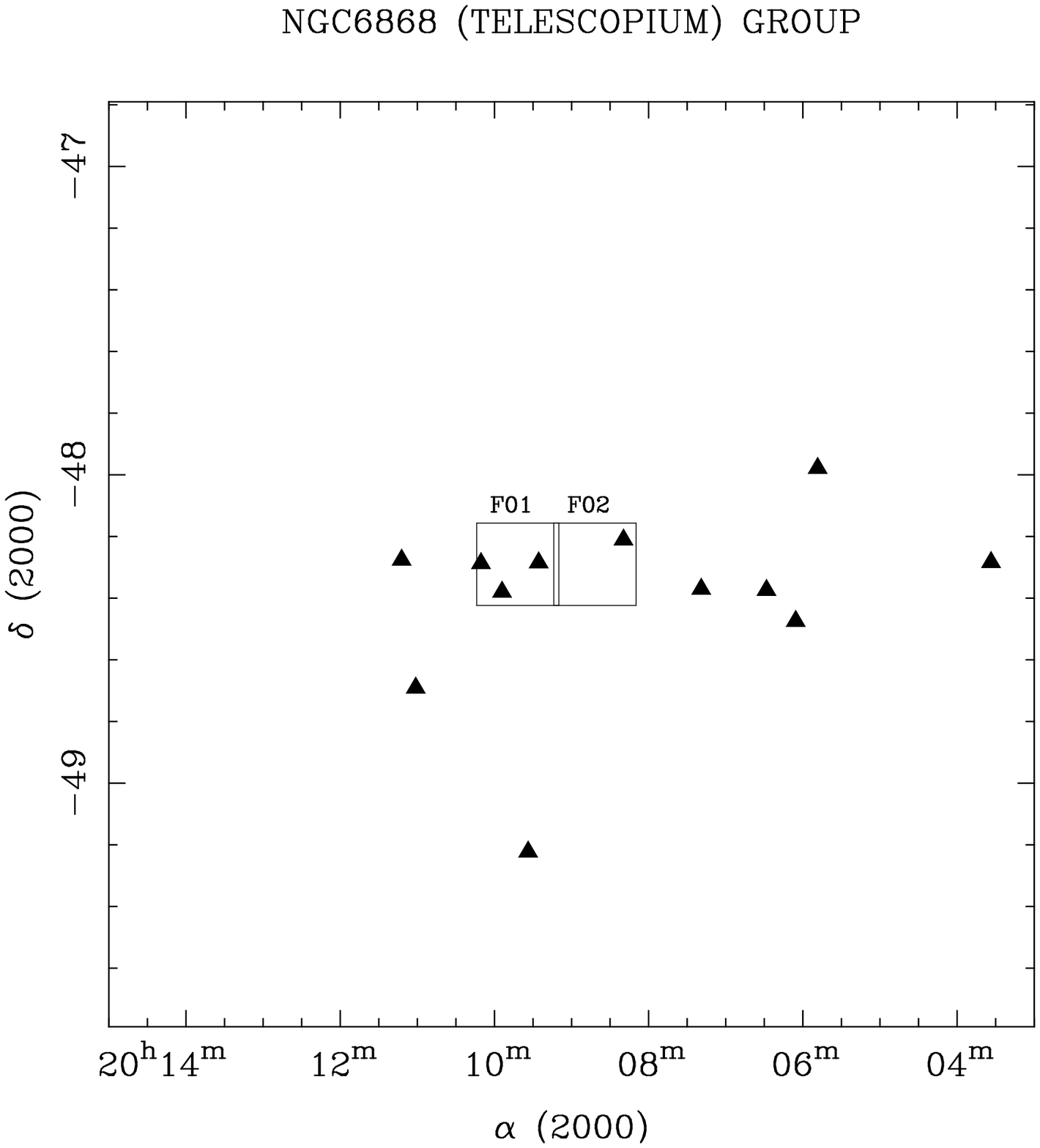}
\caption{Projected distribution  of
all observed fields (squares) in the  area of the groups HCG 42 (upper
left),  NGC 5846 (upper, right), IC 4765  (lower, left) and NGC 6868
(lower,  right). The triangles indicate all  previously catalogued memeber
galaxies with known redshifts in the area of  each group. The size of each
plot is  $\sim 3\times3$ deg$^{2}$, except for  HCG 42 where the size is
$\sim 4\times4$ deg$^{2}$. \label{fig1}}
\end{figure}

All frames were observed with a small overlap between them 
($\sim20-30$ arcsec) in order  to check the photometry  and estimate
the  photometric errors. The details of the observations are contained
in  Table 2. The table shows the number of fields observed in each
group (column  2), the field positions (columns  3 and 4), the filters
used and the total exposure time per filter (columns  5, 6 and 7), the
derived seeing per filter  (columns 8 and 9) and the  total area covered
in each group. The projected  distribution of all observed  fields is
shown in Fig. \ref{fig1}. We also  included in the figure all member
galaxies of the groups with previously  known redshifts (solid 
triangles).  We have used the SExtractor software \citep{ber96} to
detecte the objects and to obtain the photometric parameters. All co-added 
images were previously processed by removing
all bright galaxies and saturated stars. Saturated stars were masked
and replaced by the median value of the nearby sky, while the bright
galaxies were extracted from the frames by fitting their isophotes
using the STSDAS.ISOPHOTES.ELLIPSE and BMODEL programs.  When there
were remaining residuals from the galaxy subtraction process, specially
in cases where elliptical isophotes were a poor fit to the profile of
the galaxy, we simply masked the region and replaced the pixels by the
median nearby sky value.

We have processed the CCD images at the telescope, with the
IRAF\footnote[1]{IRAF is distributed by NOAO, which  is operated by
the Association of Universities for Research in Astronomy Inc.,under
contract with the National Science Foundation.} package, using the
data pipeline scripts written specifically to reduce the data
product obtained with the Warsaw telescope.  The data have been
bias/overscan-subtracted and trimmed. Twilight flats in V and I 
were observed at the beginning and at the end of each night. Flatfield
images were  combined in order to make a master flat which was then used
to flat field the data. The images were registered to a common
position and combined by obtaining  their median value at each
pixel. The final combined images show, on average, an {\em rms}
pixel-to-pixel variation of 1.2\% and 0.8\% in the V- and I-band
respectively.

Calibrations of the magnitudes to the standard system were derived
using observations of standard stars from \citet{lan92}. The tasks
APPHOT and PHOTCAL were used to determine the instrumental
magnitudes, to determine the aperture corrections and to derive the
transformation between the instrumental and standard systems.  The
calculated coefficients associated to the color terms were small and
very stable during the  nights. The zero points were calculated
assuming a color index of $(V-I)=1$, a typical value for low surface
brightness dwarf galaxies in nearby groups and clusters like Virgo
\citep[see for example ][]{imp97,car01}.

\placetable{tab2}

\subsubsection{Detection and photometry}

We have used the V-band images for object detection and to extract the
main photometric parameters. The photometry in the I-band  images were
performed using the parameter ASSOC, i.e. the main photometric
parameters were extracted only for those objects detected in the
V-band images.  The resulting catalogs were then matched to generate a
master catalog per group. All objects with a fixed surface brightness
threshold of 25.8  mag/arcsec$^2$ ($\sim$ 1.1$\sigma$ above the sky
level) and with a minimum area of 10 pix$^2$ in V-band images were
found  and their main photometric parameters obtained. Note that the
1.1 $\sigma$ above  the  sky level corresponds to a limiting surface
brightness of $\sim$  25.8$-$26.0 V mag/arcsec$^2$ in the fields. We
used a threshold of 25.8  mag/arcsec$^2$ in order to have a common
surface brightness cut for  object detection in all observed fields.

We decided to adopt the magnitude given by the parameters
MAG$\_$AUTO as the total magnitude of the objects. This magnitude
gives a more precise estimation of ``total magnitude'' for galaxies.
The MAG$\_$AUTO is based on the Kron's ``first moment'' algorithm
\citep{kro80}. For each object, elliptical apertures are chosen and
the position angle and the ellipticity are defined by the second-order
moments of the light distribution. The maximum aperture is then
defined  as one that has double the  area of the isophotal elliptical
aperture.  Inside this aperture, the integrated  light is used to
construct a  growth curve where the first moment is computed.
The sensitivity lavel in magnitude is very similar in all groups.
The completeness of our sample for point sources fall below 
90\% at magnitudes of $V\sim23$ mag and $I\sim 22.5$ mag.
 
The colors were derived by measuring the fluxes  inside a fixed 
circular aperture. The aperture was chosen  to be slightly larger than 
the average seeing measured in the images. We used a value for the 
aperture of 3 arcsec in diameter in both filters.

The star/galaxy separation has been done using the ``stellarity'' flag. 
All objects with ``stellarity'' index $\le0.35$ have been
selected and flagged as galaxies.  This chosen value is based on our
previous work done in the Dorado Group \citep{car01} and supported
by Monte Carlo simulations (see section 4.3). Galaxies which mimic
Local  Group low surface-brightness dwarf galaxies at the distance
of the groups are  classified in all cases with ``stellarity'' flags
lower than 0.35. To check the  classification we used  two different
approches presented in \citet{car01}; using eye control
and by plotting pairs of parameters \citep[see section 2.4 in ][]{car01}.
In both cases we found that  $\ga 95$\% of the object
were classified identically as galaxies down to $V=21.5$  mag 
and $I=21$ mag. These results are in agreement with those obtained  
from Monte Carlo simulations presented in section 4.3. 

The final catalogs contain all galaxies brighter that $V=21.5$ mag. 
We used this magnitude cutoff due to the large uncertanties in the
classification at fainter magnitudes. This imposed cutoff corresponds
to absolute magnitudes $M_{V}=-12.2$, $-$10.9, $-$12.9 and $-$11.6 at
the distances of HCG 42, NGC 5846, IC 4765 and NGC 6868 respectively.

\subsubsection{Photometric errors}

\begin{figure}[!htb]
\figurenum{2}
\centering
\includegraphics[totalheight=4.5cm]{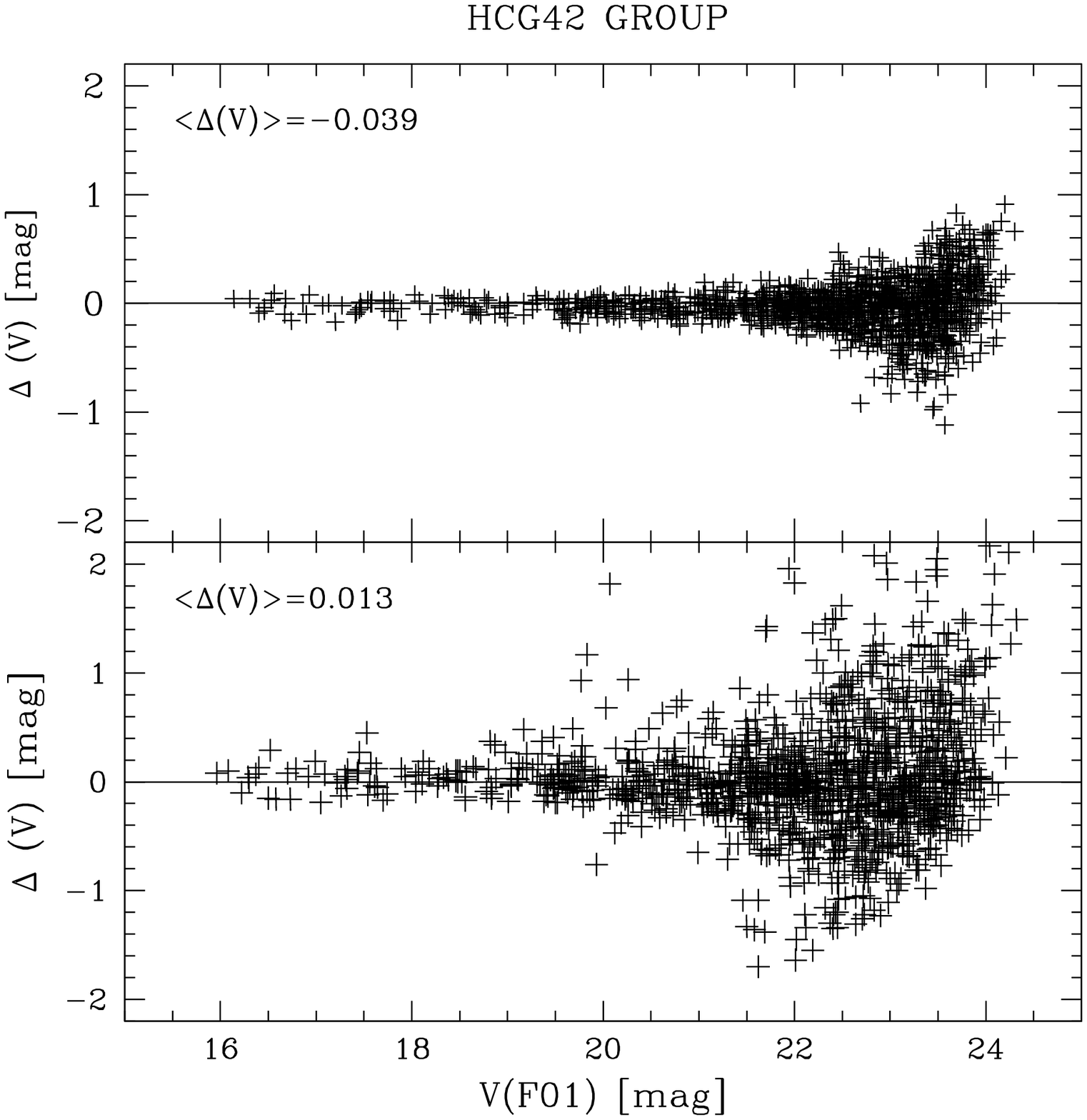}%
\includegraphics[totalheight=4.5cm]{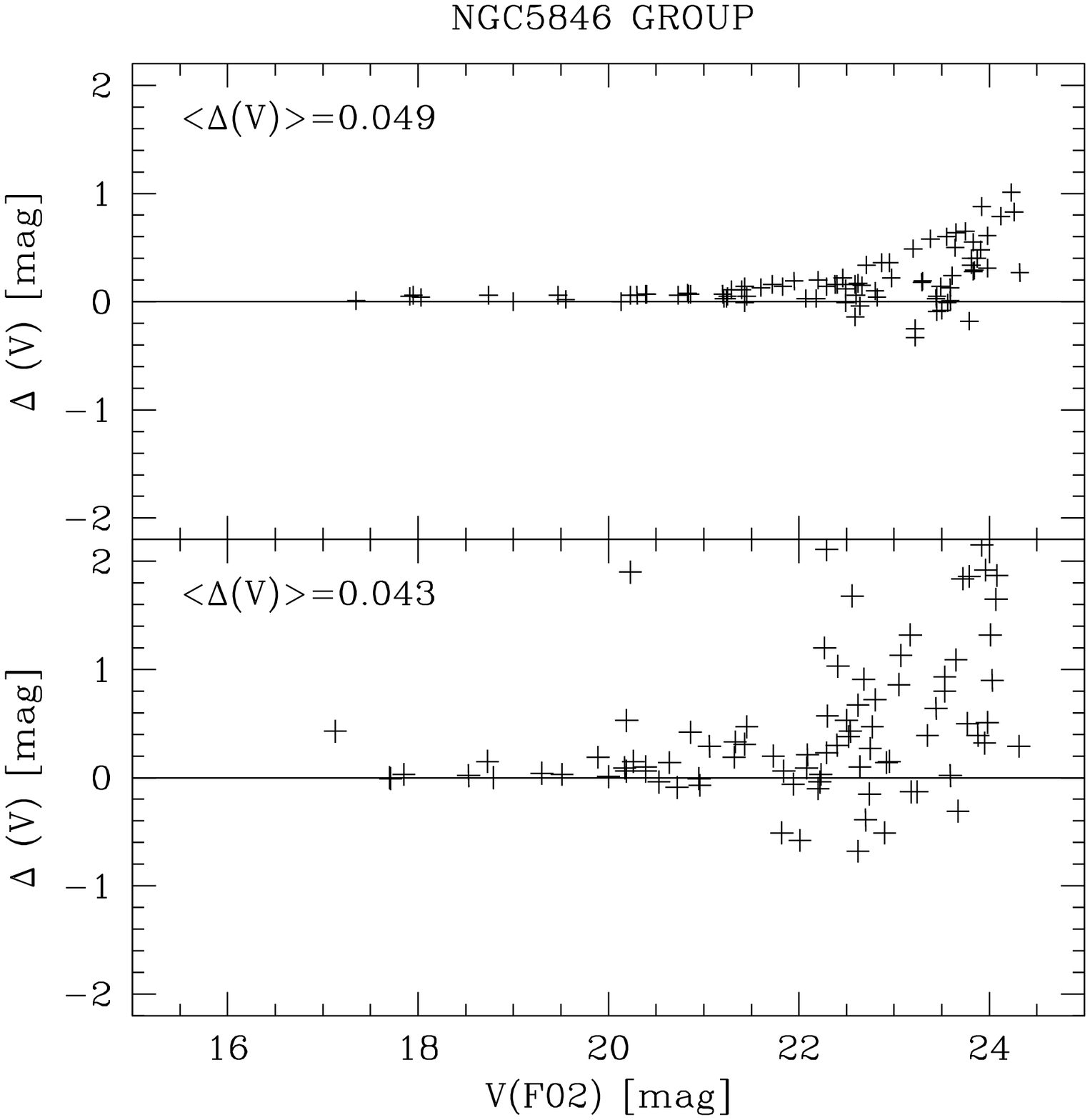} 

\includegraphics[totalheight=4.5cm]{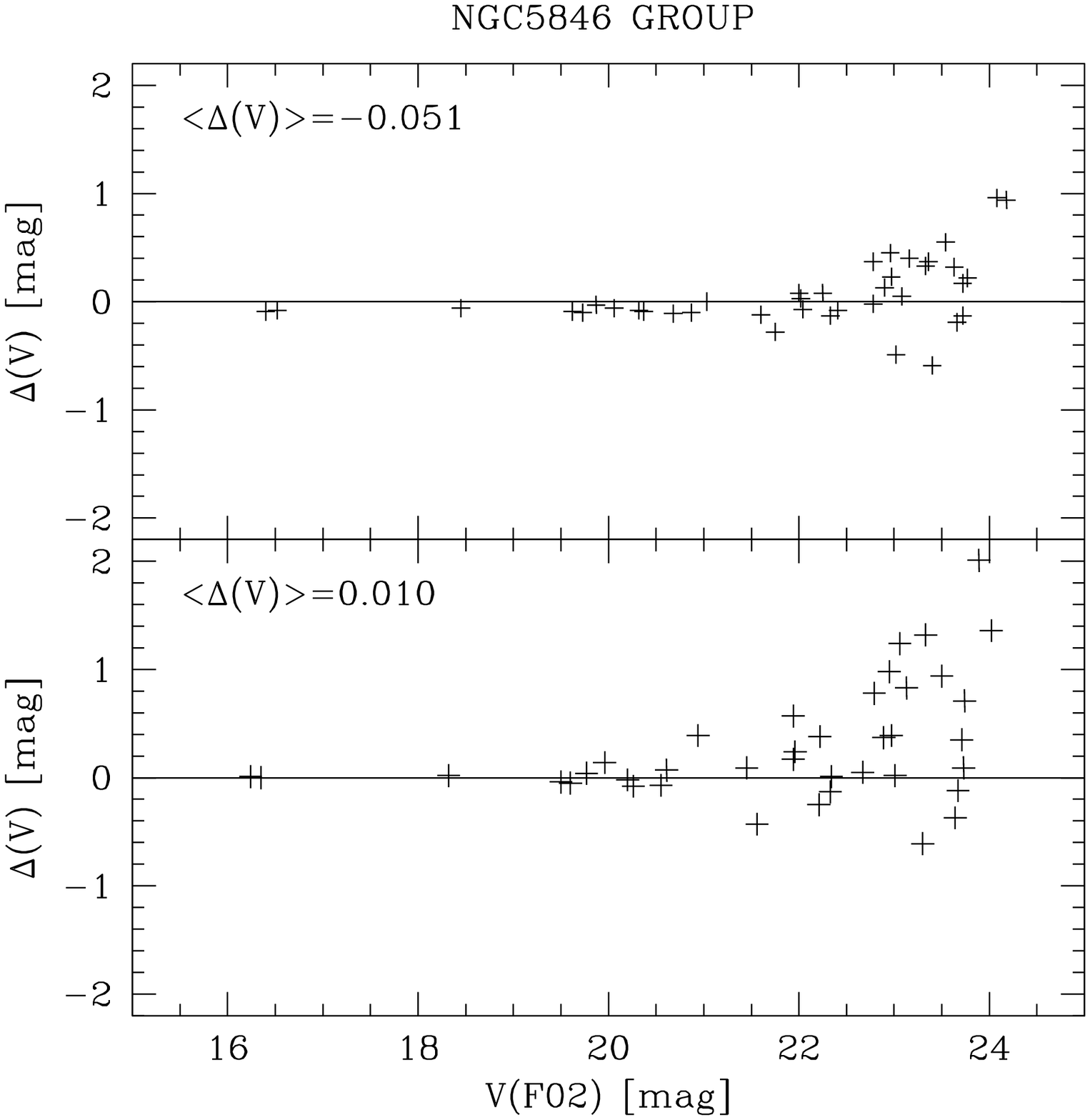}%
\includegraphics[totalheight=4.5cm]{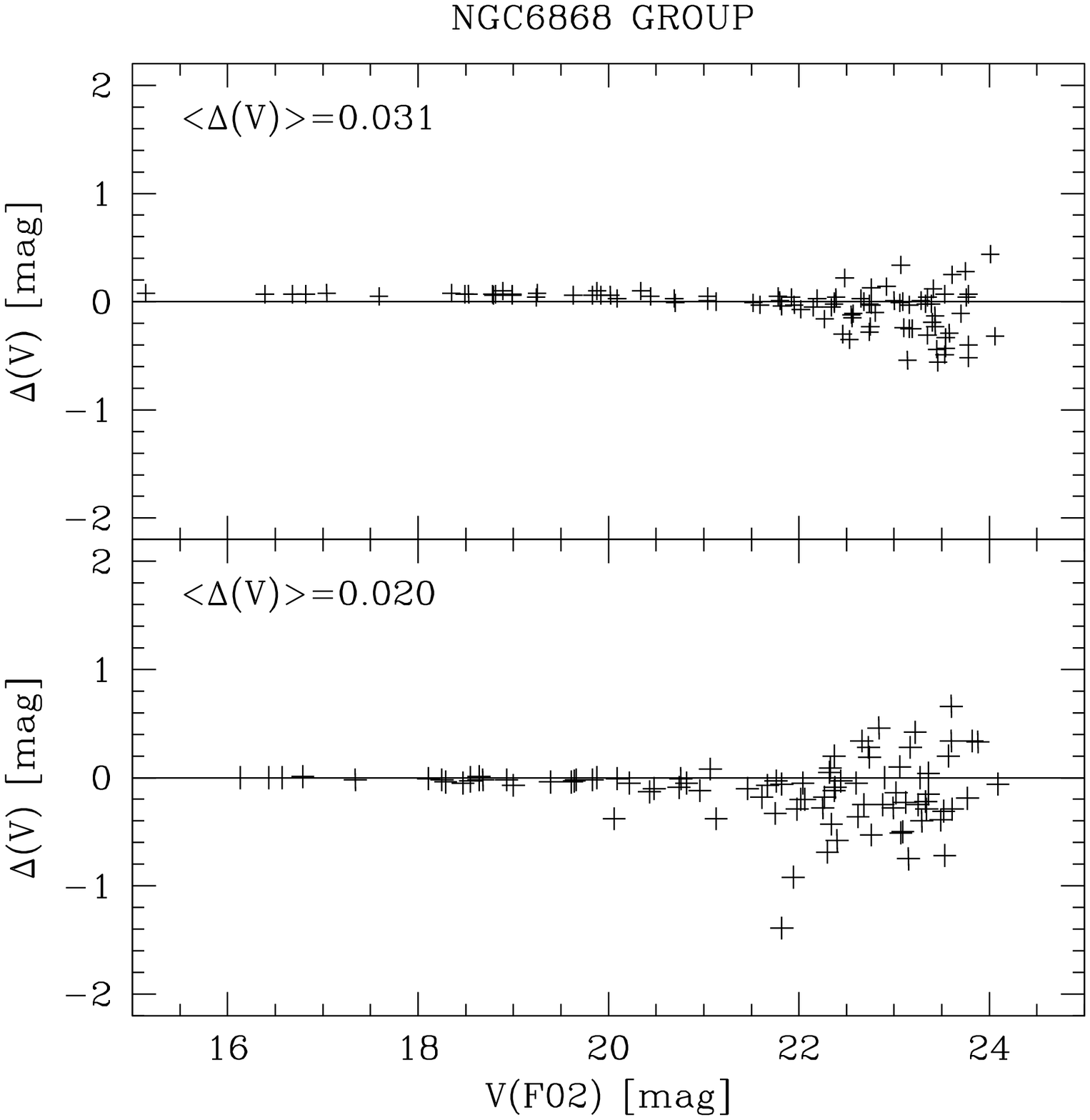} 
\caption{Differences in magnitude  as
a function of the total/aperture magnitudes  for all  objects observed in
common in the overlapping regions of HCG 42  (top, left), NGC 5846 (top, 
right), IC 4765 (bottom, left) and NGC 6868  (bottom, right). On average,
the difference is $\sim0.05$ for objects with $V\le20$ mag.\label{fig2}}
\end{figure}

An accurate determination of the Galaxy Luminosity Function requires
knowledge of any external or systematic errors that could be
affecting the data.  Because the groups were observed following a
mosaic pattern, with a small overlap between consecutive images, we
have used the objects detected in more than one field, to check the
photometry and estimate the errors. We have compared the total
magnitudes and the magnitudes within a fixed aperture, for all
objects in the common regions. These comparisons are shown in Fig.
\ref{fig2}. This figure shows the residual magnitudes as a function
of the magnitude (total/fixed aperture). The residuals for the
galaxies in the four groups are, on average, $\sim 0.05$ mag for 
$V\le20$ mag, with an {\em rms} of the residual of the same order.
For fainter  magnitudes and $V\le21.5$ mag, the scatter is higher
and the average difference  is $\vert\Delta(V)\vert<0.1$ mag. For 
magnitudes fainter than $V=21.5$  the difference is greater than 0.3 
mag. These results are in agreement with the simulations presented in 
section 4.3.

The galaxy catalogs have then been used to select the dwarf galaxies  
with surface brightnesses, luminosities and scale lengths similar to 
the dwarf  spheroidal galaxies present in the Local Group and to select 
the galaxies for spectroscopic follow up. 

\subsection{Spectroscopy}

In addition to the imaging of the groups we have performed a
spectroscopic follow up of galaxies in the group cores.  The spectra
of the galaxies in these central regions were observed with the Wide
Field CCD Camera (WFCCD),  mounted on the 2.5 m Du Pont telescope in
the Las Campanas Observatory in Chile. The spectra were obtained in
two different sessions, during the nights of June  18 -- 22, 1999
and March 9 -12, 2000, in dark conditions, with a good transparency
and under relatively good  seeing (typically between 1.2 and 1.5
arcsec).  The WFCCD camera is a CCD imager and a spectrograph
capable to take images and multi-slit spectra in a 25-arcmin
diameter field of view with a pixel size of 0.774 arcsec. The
galaxies have been observed using a low-resolution grism, giving
a  wavelength coverage of roughly  $3800$\AA $-$ 7400\AA~ with a
dispersion of  $\sim4$\AA/pix and a resolution of $\sim8$\AA~.

All galaxies with apparent magnitudes brighter than V=20 were
selected for spectroscopy and several masks were designed for the
observations of the galaxies in each group.  The spectroscopic
sample included all galaxies detected and classified as low-surface
brightness dwarf galaxies (see  section 4.1) and many of the
brightest galaxies of each of the groups. A total of 669 galaxies
were selected for spectroscopic observations (225 galaxies in
the area of  HCG 42, 184 galaxies in the area of NGC 5846, 169 
galaxies in the area of IC 4765 and 91 galaxies in the area of 
NGC 6868) distributed in 29 masks. To keep the wavelength
coverage between $\sim 3800$\AA~ and 7400\AA, the objects in the
masks were placed within $\pm6$ arcmin from the nominal masks
centers. To maximize the  number of objects per mask, a minimum
separation of 1 arcsec was used between slits. The objects  were
placed with a  minimum distance of 10 arcsec from the slit edges in
order to have enough pixels covered by sky, for a good sky
subtraction.  The slit width was set to 1.5 arcsec (June 1999 run)
and 1.94 arcsec (March 2000).  Unfortunetely, two of the five
nights in the 1999 run and one of the four nights in the March 2000 run
were lost due to bad weather conditions. This affected the 
observational coverage in the area of the IC4765 and NGC 6868 
groups.

Bias, dome and sky flats (for illumination correction) were observed
for each night of observation. Helium-neon comparison lamps were observed 
before and after each exposure, to allow wavelength calibration of the 
spectra. Masks containing bright galaxies were typically observed for 40
minutes to 1 hour. For masks with fainter objects, the total 
exposure times were typically 1.5 to 2 hours. 

\begin{figure}[!htb]
\figurenum{3}
\centering
\includegraphics[totalheight=8.0cm]{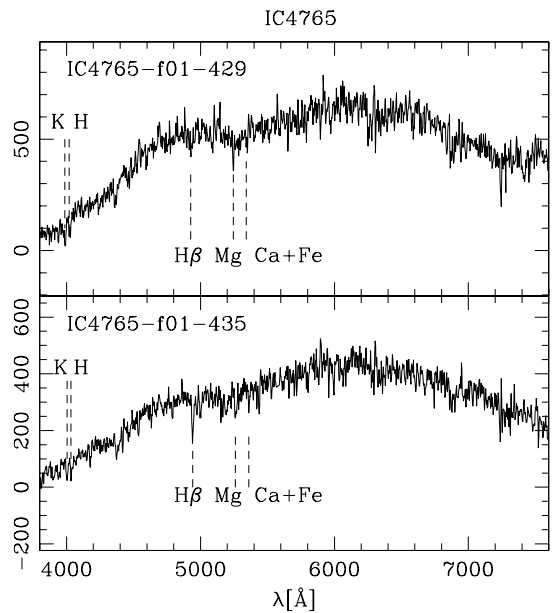}
\caption{Spectra of two newly discovered dwarf elliptical
galaxies in the central region of IC 4765, IC4765-f01-429 (upper panel)
and IC4765-f01-435 (lower panel). The galaxies have apparent magnitude  of
$V=16.8$ and $V=17.1$ ($M_{V}=-17.43$ and M$_{V}=-17.13$) respectively. 
The main absorpions lines detected in the spectra are marked. The y-axis
is in arbitrary units.\label{fig3}}
\end{figure}

A set of scripts, running under IRAF, were written to
reduce the spectra.  All two dimensional images were
bias/overscan subtracted. The science frames and their corresponding
flats and comparison lamps were cut into separate images. Each slit
spectra were then flat-fielded using a normalized flat. The
normalization of the flats was done by fitting a cubic spline of
order 25. After flatfielding, a region was selected to extract the
sky. The sky subtraction was done adjusting a low order chebyshev
polinom (typically of order 1 or 2). The spectra were extracted
using an aperture of two times the full-width half maximum of the
spectral profiles. The comparison lamps were extracted using the
same fitting parameters we used to extract the object  spectra. The
wavelength calibration was performed using between 20 and 30 lines
identified in the He-Ar comparison spectra. All calibrated spectra
have an {\em rms} of the fit less than 0.1\AA. Finally, the residual
of the sky-subtraction (typically, the sky lines at 5577.35\AA,
6300.23\AA~ and 6362.80\AA) were cleaned from the spectra by hand. 
We obtained a final spectral resolution of $\sim 8-10$\AA~ at 6000\AA~
(measured from the  FWHM of the arc and  sky lines). Spectra
observed more than once were then combined by the median.  An
example of the spectra is shown in Fig. \ref{fig3}. The figure
shows  the co-added spectra of two newly discovered dwarf elliptical
galaxies detected in the central region of the poor cluster IC 4765.

\subsubsection{Completeness of the Spectroscopic sample (selection function)}

\begin{figure}[!htb]
\figurenum{4}
\centering
\includegraphics[totalheight=4.5cm]{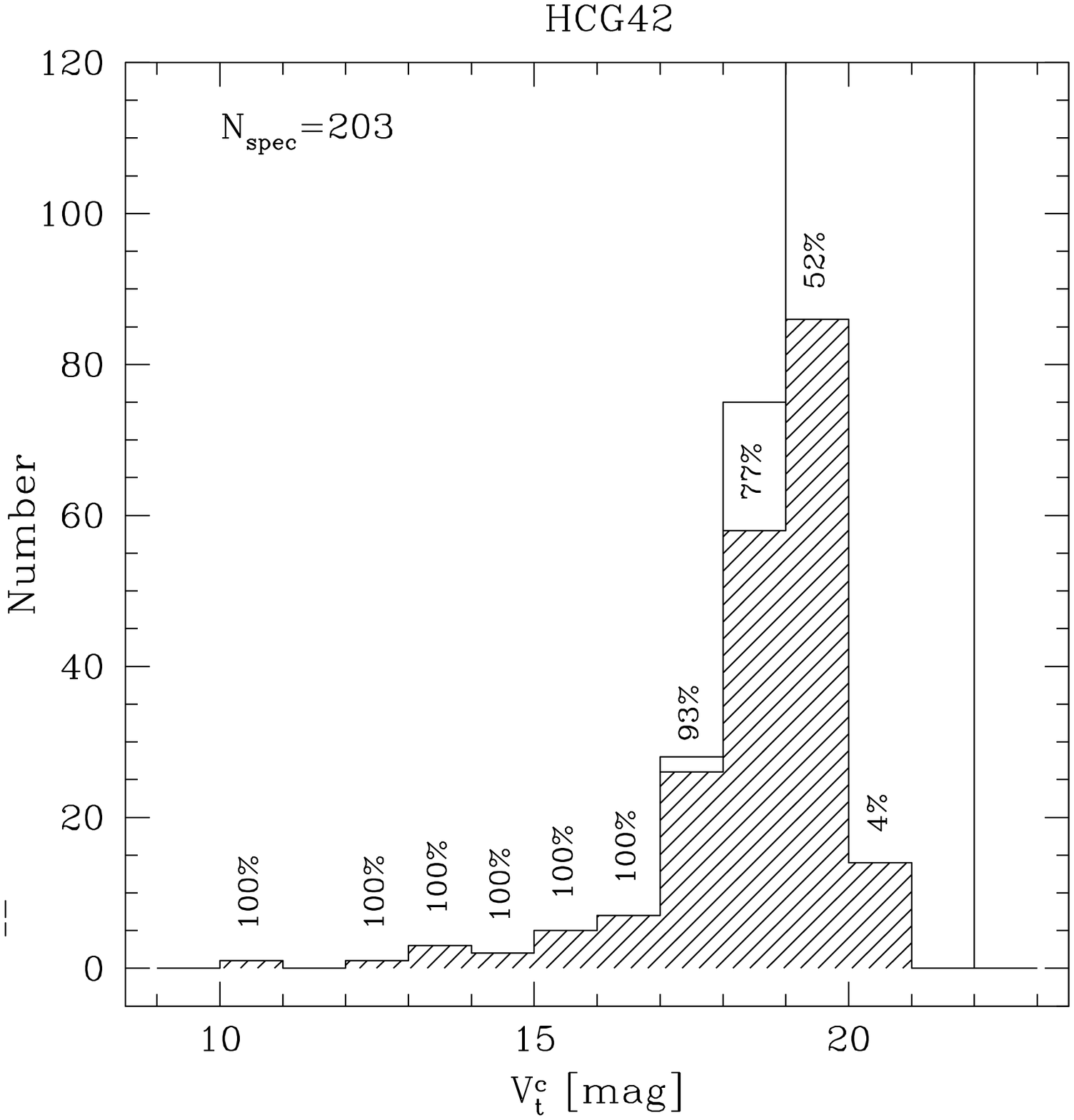}%
\includegraphics[totalheight=4.5cm]{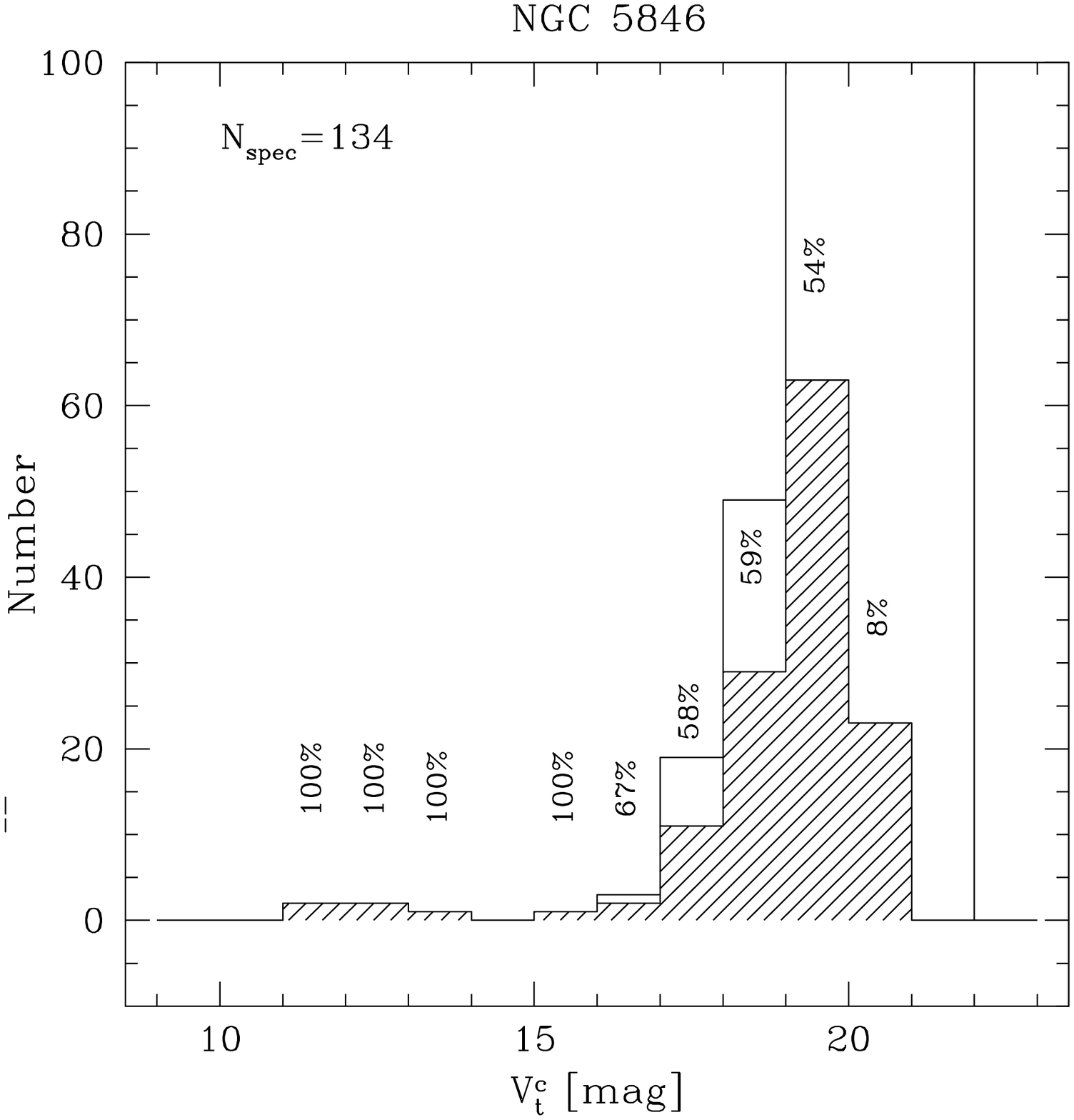}

\includegraphics[totalheight=4.5cm]{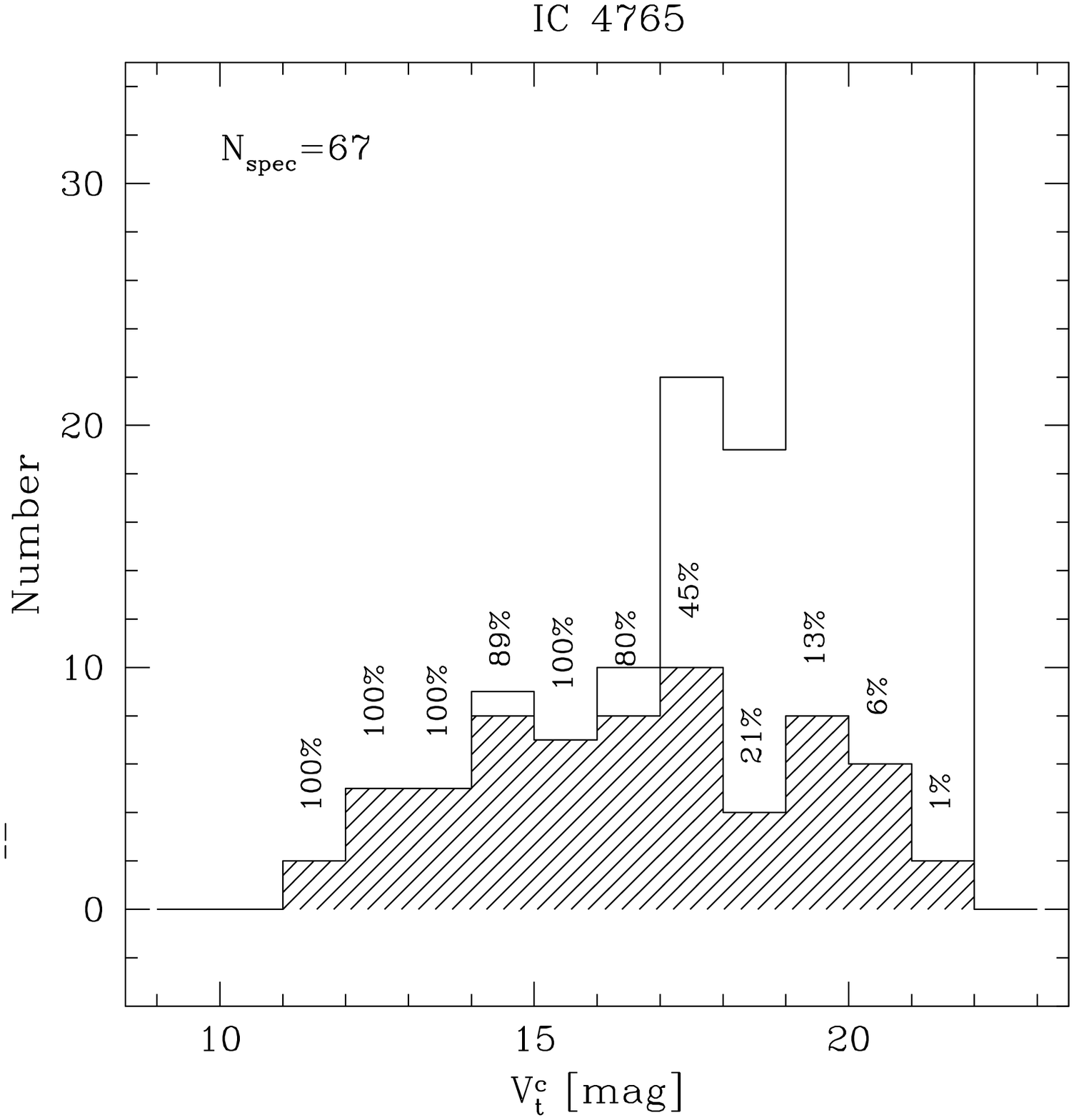}%
\includegraphics[totalheight=4.5cm]{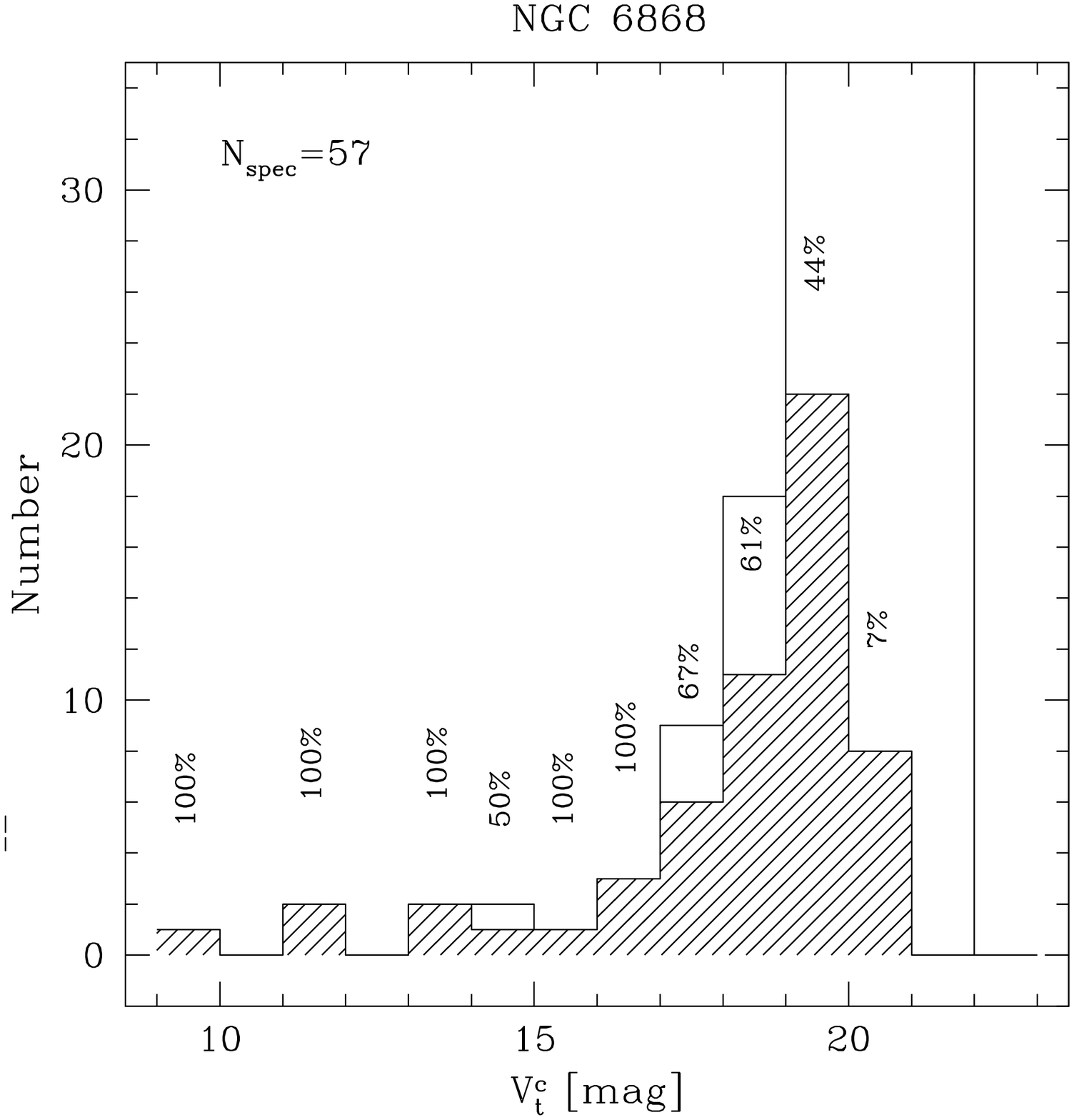}
\caption{Histrogram of the total
magnitude distribution for all galaxies presented in the photometric
catalogs with $V\le21$ mag (open histogram) and for all galaxies with
radial velocities (shaded histogram). For each group the completeness
fraction is given in percentage. The number $N_{spec}$ in each group
includes the galaxies  with radial  velocities from other source and
inside of our observed fields.\label{fig4}}
\end{figure}

With the photometric and spectroscopic catalogs constructed from the
observations it is possible to determine the completeness in magnitude
of the spectroscopic sample. This completeness or selection function is
an important factor that has to be taken into account in the
determination of the Group Galaxy Luminosity Function. Fig. \ref{fig4}
shows the histogram  of the total magnitude distribution in V
(corrected by interestellar extinction)  for all galaxies brighter than
21 mag presented in the photometric catalogs  and for all galaxies with
known radial velocity presented in the observed area (shaded histogram).
For HCG 42 and IC 4765, the completeness fractions are 100\% for $V\le17.5$
mag and drop to $\sim70$\% between $17.5<V\le19.8$ mags. In NGC 5846, 
the completeness fraction is 100\% for $V\le16$ mag and $\sim 65$\% for
$V\le19$ mag. Finally, for NGC 6868 the completeness is 100\% for
$V\le17$ mag and $\sim70$\% for $V\le18.5$ mag. For magnitudes fainter
than $V=18$, the completeness fraction is an upper limit because the
bins were not corrected by incompleteness in the detection of the LSBD
galaxies determined in section 4.3.

\section{Photometric analysis}

\subsection{Selection of the Low surface-brightness dwarf galaxies}

\begin{figure*}[!htb]
\figurenum{5}
\centering
\includegraphics[totalheight=8cm,angle=-90]{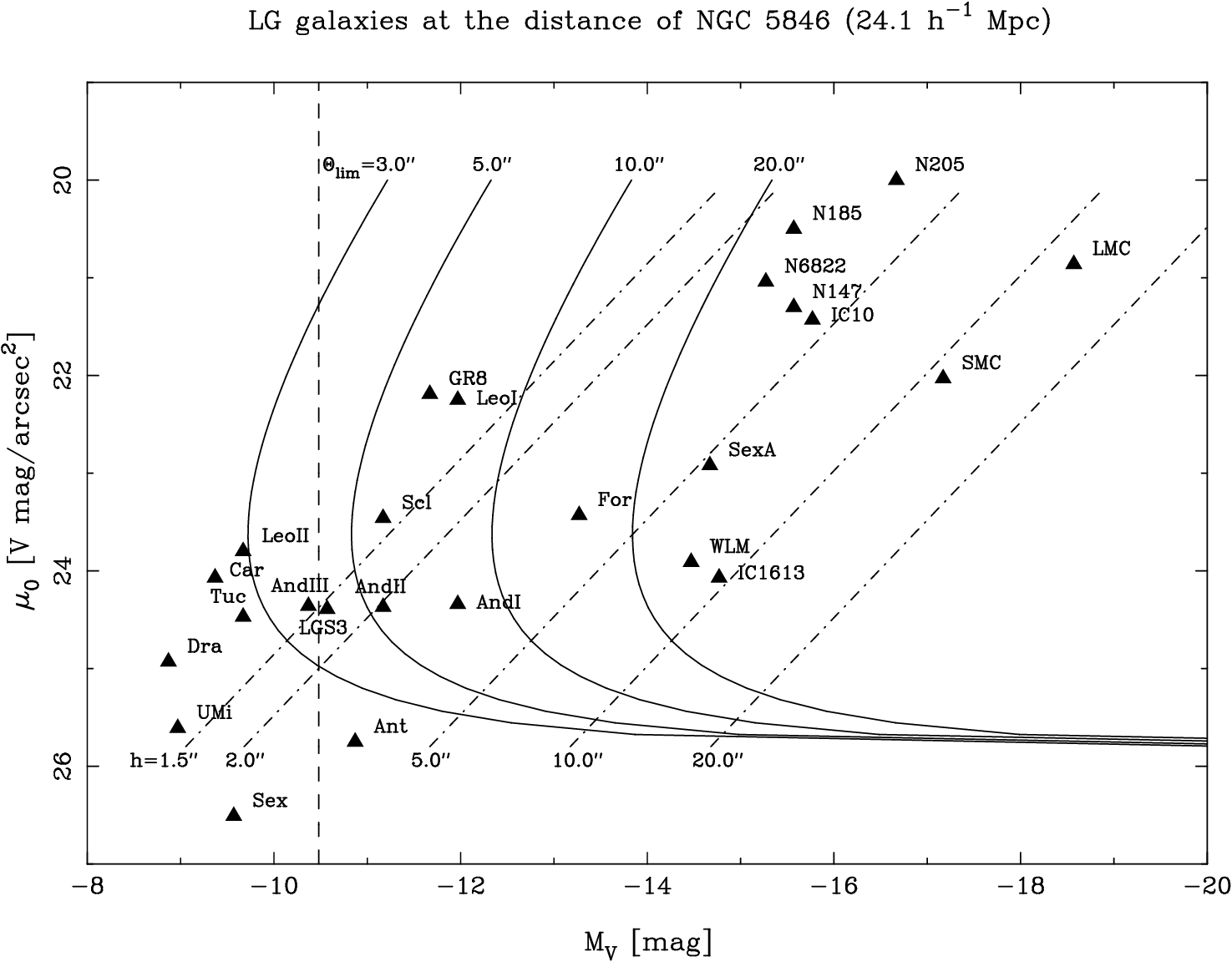}%
\includegraphics[totalheight=8cm,angle=-90]{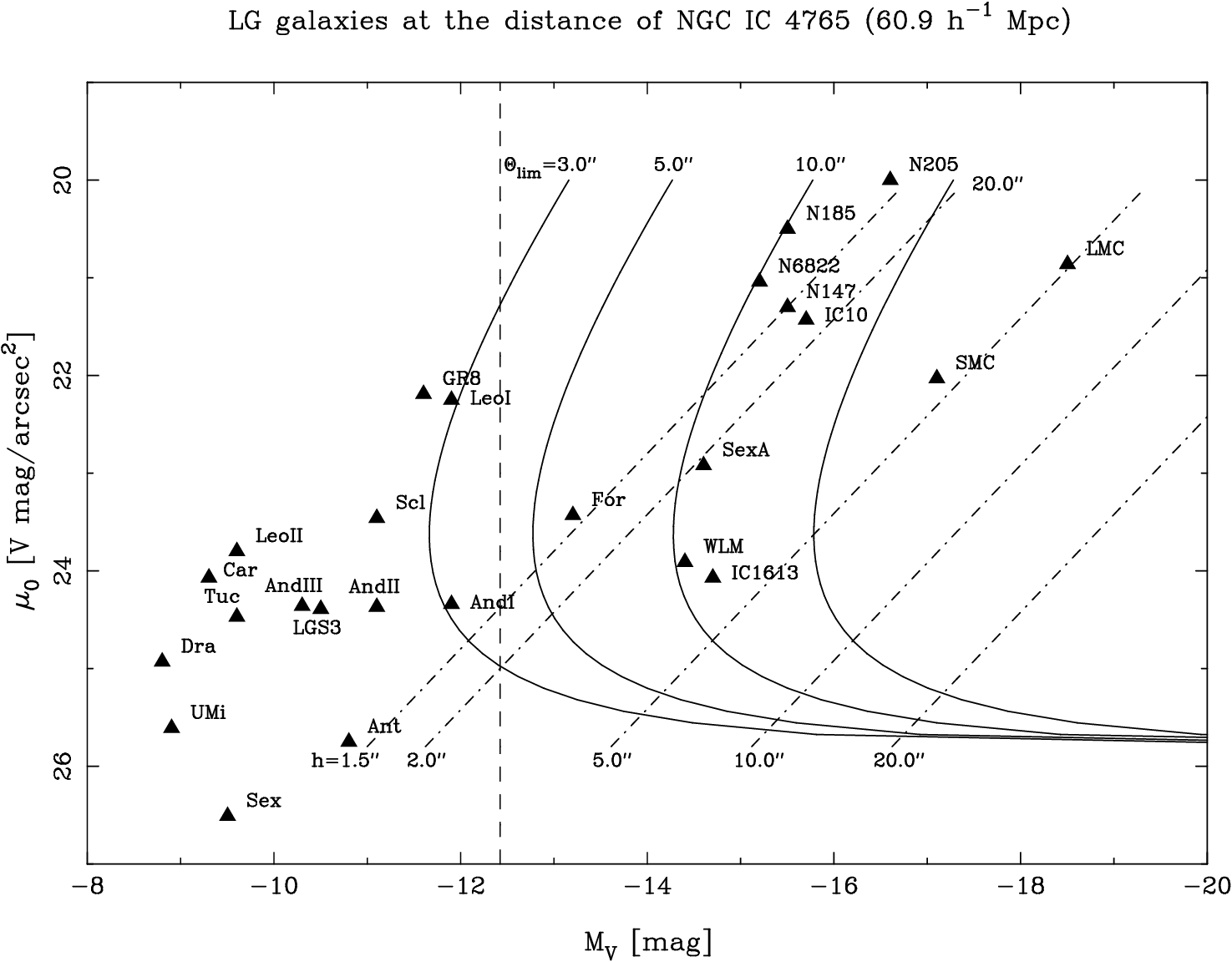}
\caption{Distribution of the Local Group dwarf
galaxies, in the $M_{V} - \mu_{V}$ ,  plane at the distance of NGC 5846
(left panel) and IC 4765 (right panel) groups.  Solid lines are lines of
constant limiting diameters ($\Theta_{lim}$, see text) for a limiting
isophot of 25.8 V mag/arcsec$^2$, assuming an exponential law. The
short-long dashed lines are lines of constant scale lengths. The dashed
lines indicate the limiting absolute magnitude (completeness) of the
sample for each group,  determined by  Monte Carlo simulations, as
described in section 4.3. \label{fig5}}
\end{figure*}

Low surface-brightness dwarf galaxies in the area of the four groups
were searched for by applying a similar criteria to that used in
\citet{car01}, for the Dorado group.  As a first approach,
we are interested in searching for galaxies that have similar
magnitudes, sizes and surface brightnesses to those of the  dwarf
galaxies observed in our own Local Group.  Fig. \ref{fig5} shows the
distribution of the Local Group dwarf galaxies in the $M_{V} - \mu_{V}$
plane as they would be seen at the distance  of NGC 5846 and IC
4765 groups (the closest and most distant groups in our sample). The
magnitudes, scale lengths and central surface brightnesses of Local
Group galaxies were taken from  Mateo (1998) and references therein. 
As we can see in the figure, at the  distance of NGC 5846 (24.1 h$^{-1}$ 
Mpc), we do not expect to detect galaxies like Draco, Carina and Sextans 
A, because these galaxies are too faint and  below the survey detection
limit of $\mu_{lim}=25.8$ V mag/arcsec$^2$ (the dashed line indicate
the limiting absolute magnitude corresponding to an apparent 
magnitude of $V=21.5$ mag at the distance of the groups) . However,
galaxies like Antlia would be possible to identify, following the same 
smoothing method applied for the galaxies in the Dorado
group \citep{car01}. In IC 4765, only the brightest dwarf
galaxies, like Fornax, may be possible  to be detected at the distance
of this group.

The selection criteria applied to detect the LSBD galaxies involve
the parameters given by the exponential profile fit: the central
surface brightness, the scale length and the limiting diameter. 

The galaxy radial profile can be represented by a generalized
exponential profile or Sersic law \citep{ser74} of the form:

\eq
\mu(r) = \mu_{0} + 1.086 (r/h)^{n}, \quad n>0
\label{eq1}
\eeq

\noindent where $\mu_{0}$ is the extrapolated central surface
brightness, $h$ is the scale factor and $n$ is the generalized
exponent. For disk galaxies and dwarf fainter than
$M_{V}\sim-17$, the surface brightness profile is well fit by a
pure exponential law, with $n=1$. 

For each of the eight isophotal areas given by SExtractor, we
determined the   surface brightness and the radius at a given
isophotal area as $\mu_{i}=\mu_{peak} \times i/8 + (1-i/8) \times \mu_{lim}$
and $r_{i}=\sqrt{A_{i}\epsilon/\pi}$, where
$\mu_{peak}$ is the surface brightness at the pixel of maximum
intensity, $\mu_{lim}$ is the surface brightness at the limiting
isophote, $\epsilon=a/b$ is the elongation and $A_{i}$ is the
isophotal area at the {\em i}-label.

We used $\mu_{0}$ and $h$ given by the fit as the main parameters to
allow a primary cut for the LSBD's selection. All galaxies with $\mu_{0}$
fainter than 22.5 V mag/arcsec$^2$ and $h>1.5$ arcsec were
selected (note that the values for the scale factors used here are
slightly larger than the mean seeing value obtained for the images).
The cut in scale length corresponds to a physical size of $\sim 0.4$
$h^{-1}$ kpc, $\sim0.18$  $h^{-1}$ kpc, $\sim0.4$ $h^{-1}$ kpc and
$\sim0.24$ $h^{-1}$ kpc at the distance of HCG 42,  NGC 5846, 
IC 4765 and NGC 6868 groups respectively and is comparable to the 
scale length cut used for the LSBDs selection in the Dorado group.

A second cut was done using the limiting diameter which can be
represented by the parameters given by the exponential fit, in the
following form

\eq
\theta_{lim}=0.735 (\mu_{lim} - \mu_{0}) 10^{0.2(\mu_{0}-m_{tot})}
\label{eq2}
\eeq

\noindent where $\mu_{lim}$ is the surface brightness at the limiting
isophote of 25.8 V mag/arcsec$^2$,  $\mu_{0}$ is the extrapolated
central surface brightness,  $m_{tot}$ is the total magnitude of the
objects, and $\theta_{lim}$ is in arcsec. All galaxies  with a
physical diameter at the limiting isophote larger than  1.2 h$^{-1}$
kpc were selected as LSBD candidates. This is the same  physical
diameter we used to select the LSBD galaxies in the Dorado  group and
corresponds to an apparent size of 4.5\arcsec, 10\arcsec, 
4\arcsec\phn and 7.5\arcsec at the distance HCG 42, NGC 5846, IC
4765  and NGC 6868 respectively. After the selection, all galaxies
were  visually inspected. A few galaxies were discarded based on the
{\em flag}  of the photometric quality given by SExtractor  
\citep[see][for details]{ber96}.  Objects with {\em flag} $>0$ have
a  bright object in their proximity (the additional  contaminating 
light from the nearby bright source can ``create'' false LSBD's)  
and were not included in the final LSBD catalog. 

A total of 71 LSBD galaxy candidates were detected using the 
cut applied above (32 in HCG 42, 14 in NGC 5846, 15 in IC 4765 
and 10 in NGC 6868). Three LSBD galaxy candidates did not have 
reliable color information (two in NGC 6868 and one in IC 4765),
since they were barely detected in the I-band images. 

The color distribution of the 68 LSBD candidates in the groups for which
we have color information range from $0.17 < V - I < 2.31$ with a peak
at $V - I = 1.14$. About 40\% of these galaxies are blue ($V-I<1$) 
and eleven (16\% of the sample) have colors $V-I>1.5$ (four of these are 
very red, with colors $V-I>2$.). In order to minimize the possible 
background contamination due to red, background, galaxies in our 
catalog, we applied an additional cut in color, by selecting 
galaxies with $V-I\le1.5$ mag. Therefore, the final catalog is 
restricted a sample of 60 LSBD galaxies after the color cut applied 
above. Note that most of the LSBD galaxies detected in groups 
\citep[e.g. Dorado, M81 and M101:][1999]{car01,bre98} are blue. 
There is some evidence of the existence of very red LSBD galaxies 
in clusters \citep{oni97} and in groups \citep{car01,flint01}. However,
these are rare. Few red LSDB galaxies were confirmed as members of 
Pegasus and Cancer spiral rich clusters \citep{oni00}.  

\subsection{Searching for large, very low surface-brightness galaxies}

We used the smoothing technique described in \citet{car01} to
search for galaxies with larger sizes and with surface brightnesses near
and/or below  the imposed limit of 25.8 V mag/arcsec$^2$. In summary,
all  objects  above $3\sigma_{sky}$ and additional area around them
were masked and replaced by the appropiate sky noise of the images. The
resulting ``clean'' images were then  convolved with an exponential
profile function similar in form to the  LSBD galaxies that we are
searching for. The SExtractor software was  used to search for
remaining low surface-brightness features in the  convolved images. The
photometry of the detected objects was then  done in the original
frames using these convolved images as templates,  again using the
Sextractor program. At this point, we also did an eye inspection of the
images to check whether the automatic search with SExtractor had lost
any object. The detection of the large, very  low-surface brigthness
galaxies  was performed in the V-band images  (in the I-band images the
detection  was marginal). Using this method, we have detected 20 new
objects that had been previously undetected.

The new LSBD galaxies increases the number of LSBD detected in the
central region of the four groups to eighty galaxies. Of these, 32 are 
in the field of HCG42, 16 in NGC 5846, 20 in IC 4765 and 12 in NGC 6868.
These values correspond to a total of $\sim 115$ gal/degree$^2$, 
$\sim 96$ gal/degree$^2$, $\sim 171$ gal/degree$^2$ and $\sim 108$
gal/degree$^2$ respectively. In addition, a total of 16 LSBD galaxies
were detected in the nine control fields using the same selection
criteria. This corresponds to an average of $\sim 32$  gal/degree$^2$,
for galaxies with $22.5 < \mu_{0} < 25.5$ mag/arcsec$^2$  and with a
scale factor $h>1.5$. \citet{dal97} found that the surface  density of
low surface-brightness galaxies in random fields is  $4.1^{+2.6}_{-2.1}$ 
gal/degree$^2$ for galaxies with $23<\mu_{0}<25$ mag/arcsec$^2$ and with
a scale factor $h>2.5$ arcsec. Similar results were obtained by Sabatini
et al. (2003) in the control fields used for their study of the dwarf
low-surface  brightness galaxy population in the Virgo cluster. They
found a background  galaxy density of $5\pm1$ gal/degree$^2$. Using the
same selection criteria  as in \citet{dal97} and \citet{sab03}, we found
a  total  of 3 LSBD galaxies in our control fields, corresponding to
$\sim6$ gal/degree$^2$.  Our result is, then, in good agreement with both
works mentioned above. 

\begin{figure*}[!htb]
\figurenum{6}
\centering
\includegraphics[totalheight=6cm]{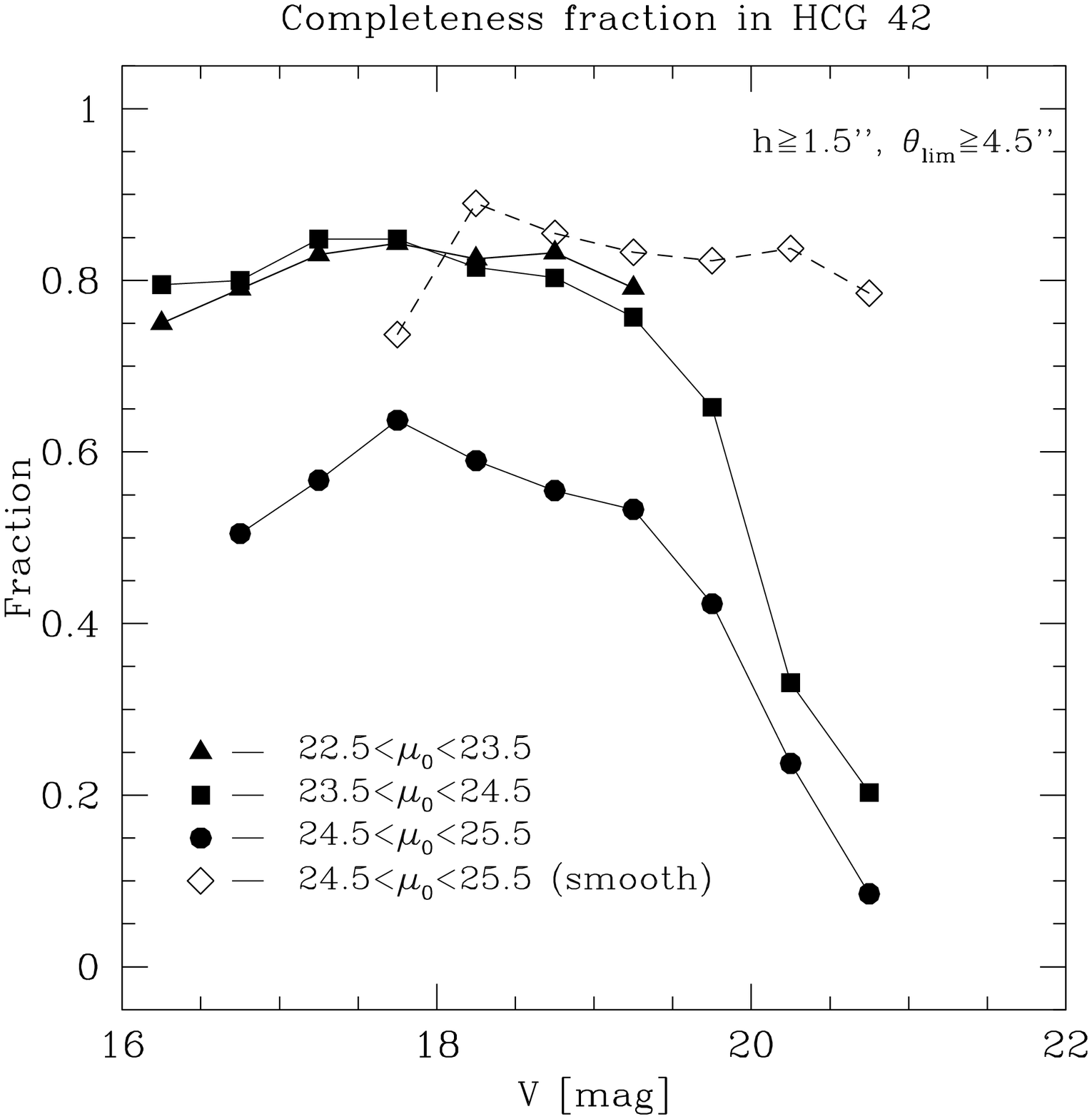}%
\includegraphics[totalheight=6cm]{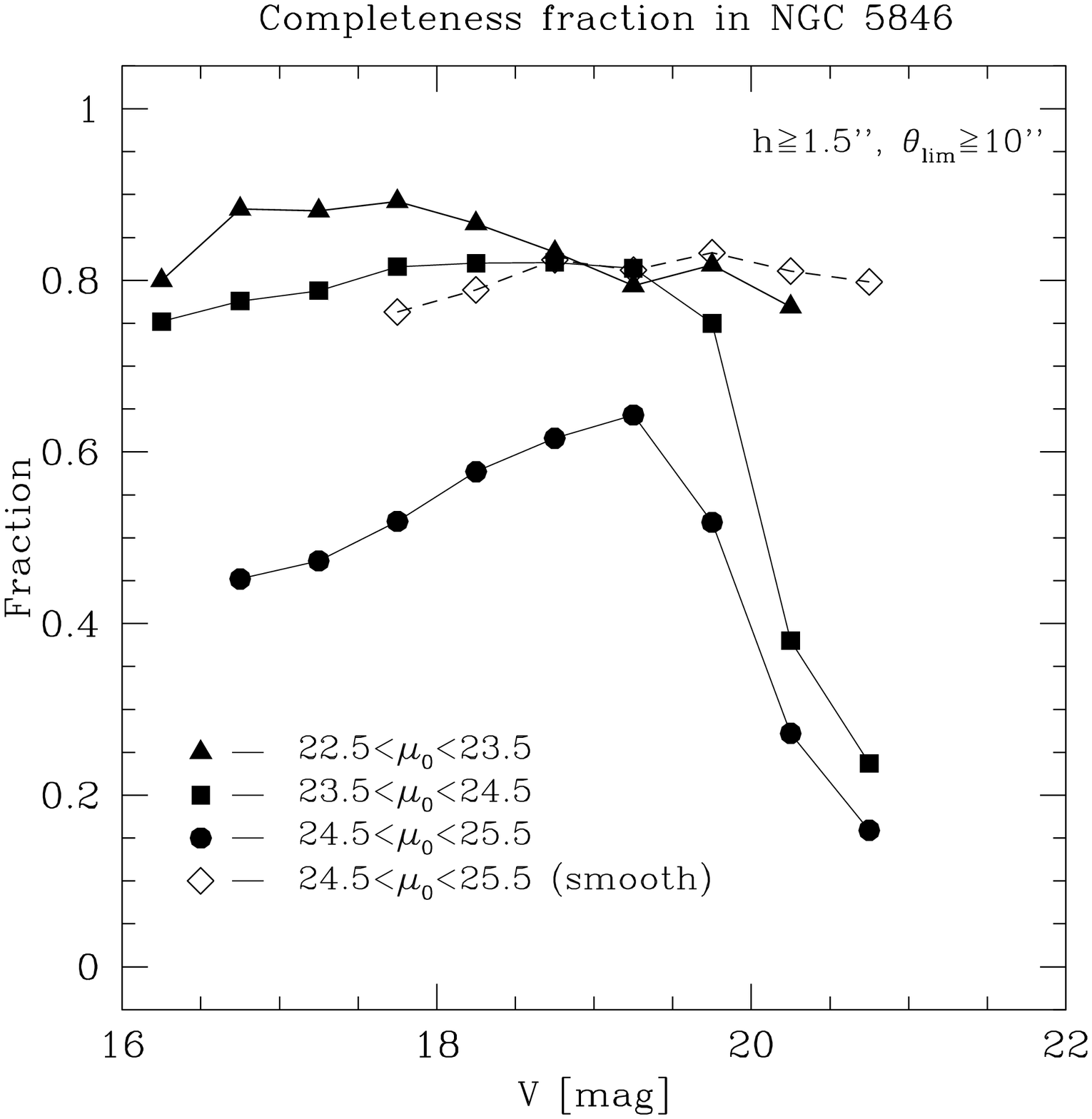}
\caption{Completeness fraction as a function of the 
total magnitude in V for simulated galaxies at the distance of the groups 
HCG 42 (left) and NGC 5846 (right). The different symbols correspond to
different central surface  brightness bins. The open symbols  represent
the completeness fraction in the last bin where we used  the smoothed
technique to search LSBD galaxies.\label{fig6}}
\end{figure*}

We used the LSBD apparent morphology to classify the LSBD galaxies. If
the galaxies showed a smooth extended light  distribution, the objects 
were classified as dE/dSph. If the galaxies were more compact
and bright, then they were classified simply as dE. In some cases,
the galaxies show small halos in the external region, i.e. where the
light distribution fades. These galaxies were classified as dS0. A
reduced group of galaxies shows an extended, smoothed light
distribution but with a bright compact core. These galaxies were
classified as dE,N.  Galaxies with an irregular morphology and with
clear knots that indicate star formation were classified as dIrr.
Finally, those galaxies with morphology similar to the dS0, but with
signs of star formation were classified as dS0/dIrr. The morphological 
classification is recorded in the last column of Table 3.

The photometric data and the profile-fitting parameters  of the LSBDs
detected in the groups are shown in Table 3. The table is  arranged as
follow: (1) galaxy identification, given by the name of the group,
followed by  the field name  and the galaxy number. Those galaxy numbers
starting with ''L'' are the very  low surface-brightness galaxies
detected using the smoothing technique; (2)  and (3) equatorial
coordinates; (3) total V-magnitude; (4) absolute V-magnitude  calculated
assuming the  distances given in Table 1; (5)  $(V-I)$ colour inside an
aperture of 3 arcsec; (7) Isophotal diameter at 25.8 V mag/arcsec$^{2}$;
(8) limiting diameter; (9) extrapolated central
surface  brightness; (10)  the scale factor; (11) and (12) the effective
radius and  mean effective  surface brightness; (13) morphological
classification. 

\subsection{Photometric detection efficiency and errors}

In order to determine the efficiency of our method to detect the
LSBD galaxies, to estimate the photometric errors and the errors in
the calculation of the profile-fit parameters (extrapolated central
surface brightness, scale factor and limiting diameter) a series of
add-star experiments and Monte Carlo simulations were done. 

The simulations were performed on the science frames in order to 
have the same characteristics of a real image (cosmetic defects, 
crowded images, light gradients, noise and seeing). The LSBD
galaxy  detection efficiency was determined as 
$f(V_{t},\mu_{0})=n_{det}/n_{gen}$, where $n_{det}$ is the number
of galaxies detected by SExtractor and $n_{gen}$ is the number of
galaxies generated. We computed the efficiency for galaxies with
magnitudes, central surface brightnesses and scale lengths typical
of low surface-brightness dwarf galaxies at the redshift of the
groups. We generated galaxies with exponential profiles with 
different scale lengths, central surface brightnesses, random
ellipticities and position angles and placed them onto the frames. 
Each image was convolved with a point spread function constructed 
from bright, non-saturated stars in the frame. We generated 400 
galaxies with exponential profiles in bins of 1 mag/arcsec$^{2}$
and  between $22.5 < \mu_{0} < 25.5$ mag/arcsec$^{2}$ in central 
surface brightness. The  galaxies were generated in groups of 20 
(five for the brightest  magnitude bins), randomly distributed over 
the field, and with  magnitudes between 16.0 -- 21.0 mag (in binz of 
0.5 magnitudes).  These galaxies were created using the MKOBJECTS
program in the  NOAO.ARTDATA package in IRAF. 

\begin{figure*}[!htb]
\figurenum{7}
\centering
\includegraphics[totalheight=6cm]{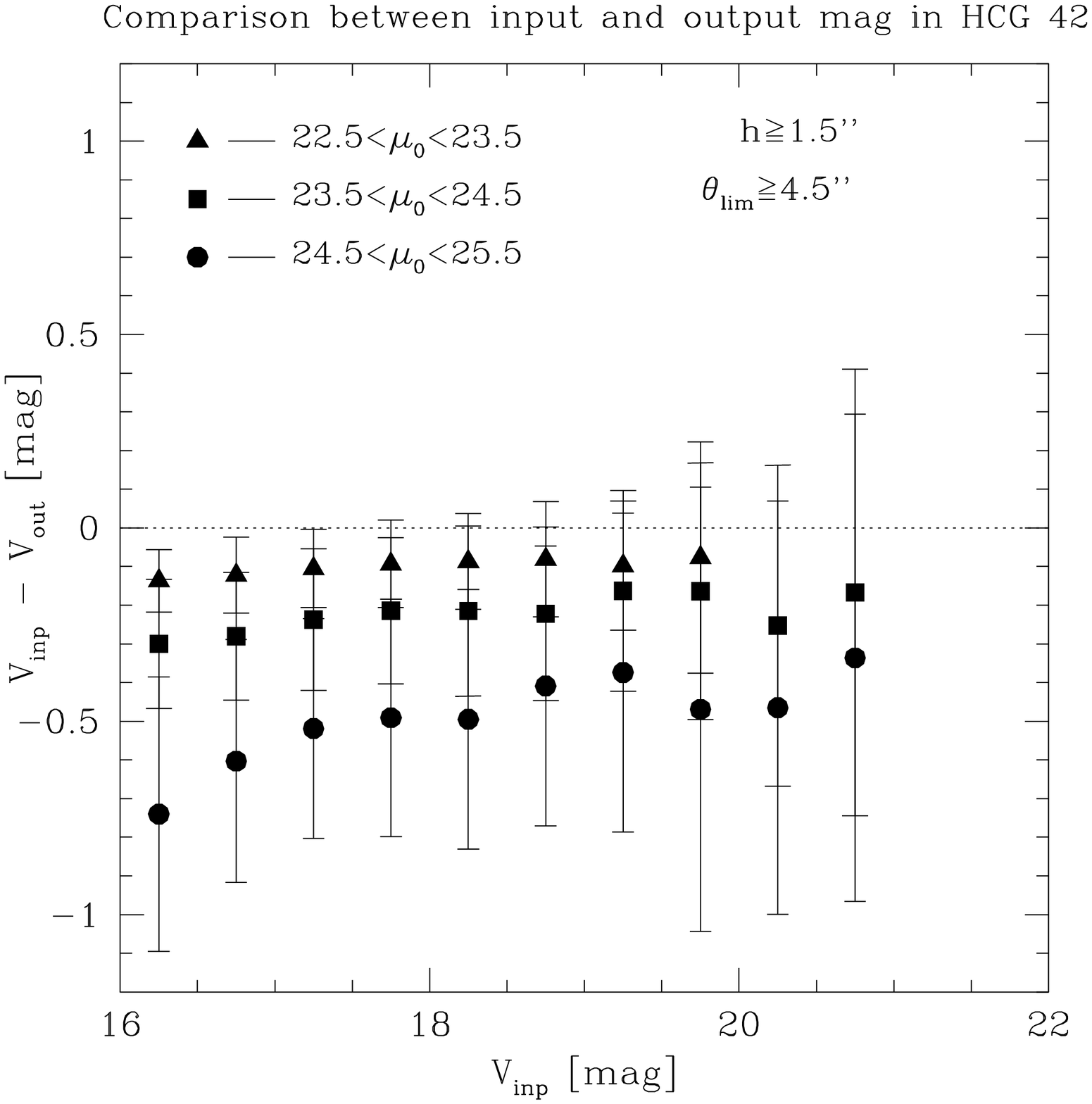}%
\includegraphics[totalheight=6cm]{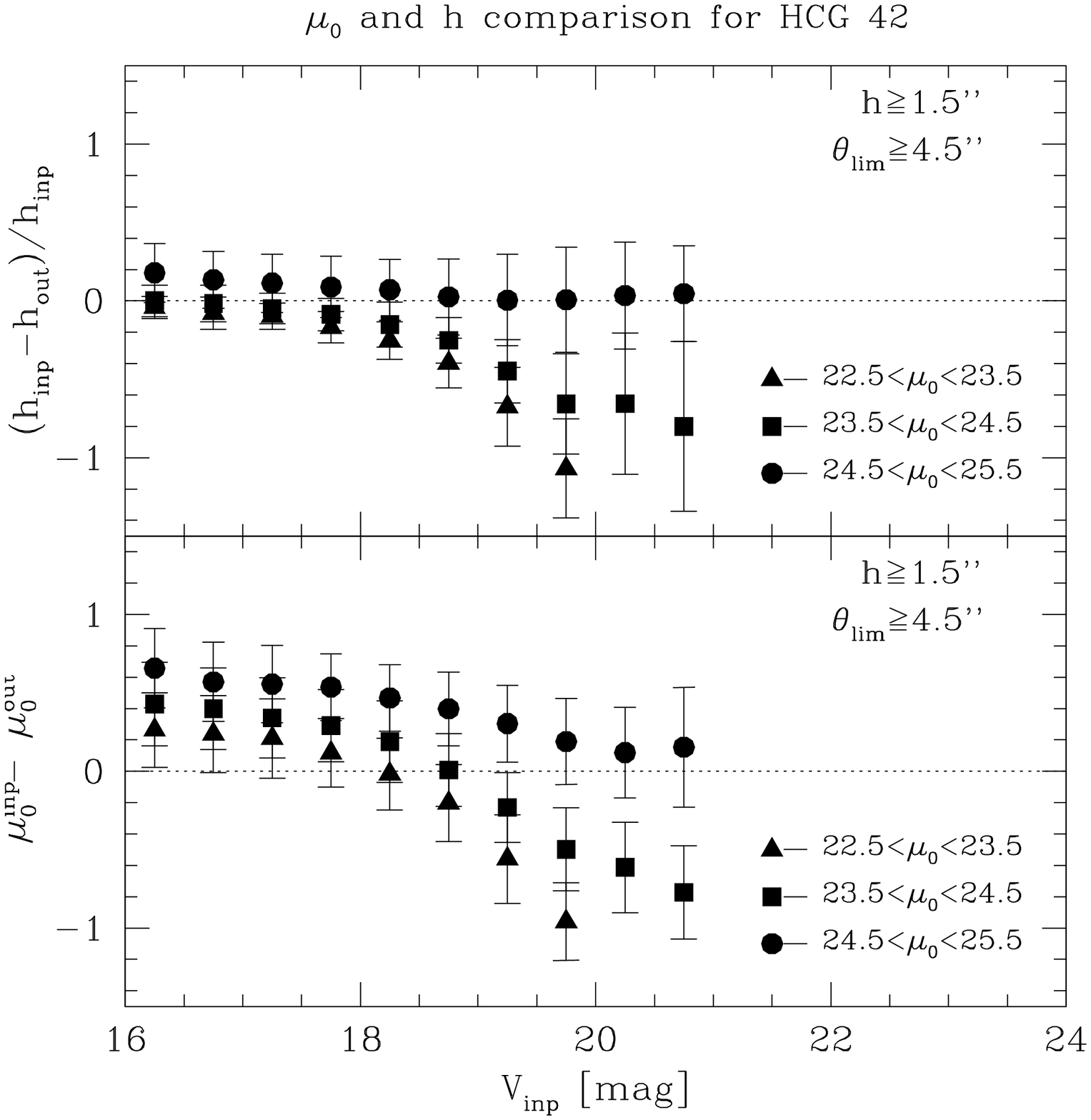}
\caption{Difference in magnitudes (left panel) and in
central surface brightness and scale factor (right panel, bottom and top)
for recovered simulated galaxies. The galaxies presented here were
simulated in one of the observed control field in the area of  HCG 42
group.\label{fig7}}
\end{figure*}

We ran the SExtractor program using the same detection parameters
previously used for the real-galaxy detection (see section
3.1.1). The common objects in the input and output catalogs were
matched. Using the eight isophotal areas, for each matched
galaxy we calculated  the central surface brightness, scale
length and the limiting diameters as in section 4.1. The
resulting values were then compared with the  input values. For
the last bin in surface brightness, we repeated the  simulations,
but using the smoothed technique described in section 4.2. 

The completeness fraction in the detection of the LSBD galaxies
in the interval of 22.5$<\mu_{0}(V)<$25.5 is shown
Fig. \ref{fig6}.  The  completeness is  $\sim 80$\% for $V\le20$
and $\mu_{0}\le24.5$ V  mag/arcsec$^{2}$. For fainter magnitudes
the detection fraction drops  rapidly. For the last interval in
surface brightness, the detection  fraction is below 50\%. 
However,  using the optimized method (smoothing  technique)
described in section 4.2, the  fraction goes up to $\sim 80$\%.

Fig. \ref{fig7} shows the difference between the input and output
magnitudes (left plot) and the difference in scale factor (upper
right panel) and central surface brightness (lower right panel)
as a function of total magnitudes for the recovered simultated
galaxies.  The difference in magnitude for galaxies with
$\mu_{0}\le24.5$   mag/arcsec$^2$ is of the order of 0.2 mag, and
increases to 0.5 mag in  the last central surface brightness bin.
In the case of the central  surface brightness, the difference
is, on  average, 0.5 mag/arcsec$^{2}$. There is a tendency that
for magnitudes  brighter than 19 mag in V, the central surface
brightness values are underestimated, while for fainter
magnitudes the tendency goes in the opposite sense, i.e. the
calculated central surface brightness values are overestimated.
However, the difference in surface brightness is negligible
compared to the errors. In the case of the scale factors, we note
a clear tendency for the  calculated values to be overestimated.
For $V<19$ mag and $\mu_{0}<24.5$, the scale factors are
overestimated in about 30\%. For the last surface brightness bin,
these differences are $\sim60$\%. There is a clear limitation in 
estimating in a correct way the scale length of the real galaxies,
especially at small sizes. This means that the values of scale lengths
listed in Table 3 for $h<2$\arcsec and fainter that $V\sim20$ mag
are only upper limits.

We also estimated the completeness of the sample by  adding point
sources from the observed point spread function   of the frames
using ADDSTAR in the DAOPHOT package, inside IRAF. A total of
1000 stars per 0.5 mag bin and between $15<V<24$ mag were
generated. The recovery fraction was  then compared with the
input and the completeness and the differences in  magnitudes
were computed.  The completeness limit for point sources with
``stellarity index'' $\ge 0.9$ is $\ga 90$\% down to our limit of
$\sim 23$ mag in V. This results is in agreement with the limiting
magnitude derived in section 3.1.1. The difference between the input
and output values magnitudes is $\vert\Delta(V)\vert\sim0.06$ mag. 
This is not surprising, since the Kron's ``first moment'' 
algorithm, used to calculate the total magnitude, measures about 
95\% of the flux for objects with stellar profiles and about 90\% 
for galaxies with exponential profiles \citep{inf92}.

\section{Spectroscopic analysis}

\subsection{Radial velocities}  

The radial velocities were determined using two different methods.
For spectra with emission lines, the routine RVIDLINE in the IRAF RV
package was used employing a line-by-line gaussian fit.  For spectra
dominated by absorption lines, the XCSAO cross-correlation algorithm
in the IRAF RVSAO  package \citep{kur92} was used.  The observed
absorption spectra were correlated with twelve high signal-to-noise
stellar and galaxy  templates. The $R$ values \citep[see][]{ton79}
were used to evaluate the reliability of the measured velocities.
For $R>4$, the template which gave the lowest error was used to
derive the radial velocity.  This method gave more consistent
results than other combinations we tried, such as a mean of the
results of the twelve correlations.  For $R\le 4$, we inspected the
spectra and tried line-by-line gaussian fitting to check which were
the most reliable cross-correlation results.

We were able to measure the radial velocities for 412 of 669
galaxies selected for spectroscopic observations ($\sim62$\% of 
the sample). This correspond to $\sim83$\%, $\sim 68$\%, $\sim 53$\%  
and $\sim62$\% of the selected objects in the area of HCG 42, NGC 5846,
IC 4765 and NGC 6868 respectively. The final radial velocities are 
shown in  Table 4. 
The table lists the  following parameters:  (1) galaxy identification
as in Table 3; (2)--(3) RA and DEC for the group; (4)  V total
magnitudes; (5) (V-I) colour inside an aperture of 3 arcsec; (6) 
galaxy radial velocities corrected to the heliocentric reference 
frame; (7) radial velocity errors; (8) $R$ values (real numbers) 
or the number of emission lines  (integer value) used to calculate 
the velocities; (9) flag for galaxies with multiple measurements:
``0'' $-$ value has been discarded, ``1'' $-$  value has been 
included in the calculation of the final radial velocity; (10)
final radial velocity. For galaxies with multiple measurements,
these values are the mean weighted velocity (see below); (11)
final radial velocity error yielded by the RVSAO/RVIDLINE packages
and by the error of the weighted mean velocity in the case of
multiple measurements. 

In column 6, besides the heliocentric radial velocity, we also
included the radial velocities available in the literature. The
identification and references are shown in the last column of the
table.  For galaxies with multiple radial velocity measurements,
we derived the mean weighted velocity following \citet{qui00}.
These velocities are calculated using the following relation:

\eq 
\overline{V} = \frac{\sum_{i} w_{i}v_{i}}{\sum_{i} w_{i}}
\label{eq3} 
\eeq

\noindent where $v_{i}$ is the galaxy velocity,
$w_{i}=1/\sigma_{i}^{2}$ is the weighting factor, and $\sigma_{i}$
is the uncertainty. The weighting factor $w_{i}$ came  from our 
internal uncertainties (col 7 in Table 4) or from the
published uncertainties. The error of the weighted mean velocity for
galaxies with  multiple measurements is given in column 11 in
Table \ref{tab4} \citep[see][for a full discussion of the criteria used to calculate the errors of the weighted mean velocities]{qui00}. 

\subsection{Estimating the internal and external errors}

\begin{figure}[!htb]
\figurenum{8}
\centering
\includegraphics[totalheight=8cm]{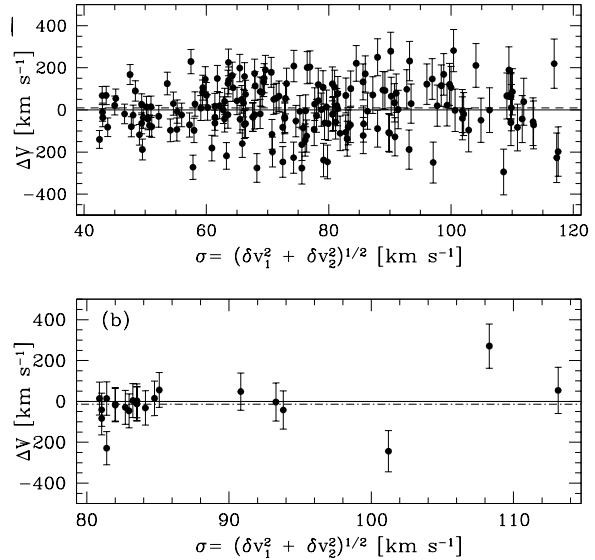}
\caption{(a) Radial velocity differences of 187 galaxies
measured using cross-correlation and line-by-line gaussian fits against
the root square of the sum of the quadratic errors. The average difference
in velocity (dashed line) is 10 km s$^{-1}$  with an rms of the residual
of 100 km s$^{-1}$.  (b) Radial velocity differences for 23 galaxies in
common with Zabludoff \& Mulchaey (1998, 2000). In this case, the average
difference in velocity is $-$14 km s$^{-1}$ with an {\em rms} deviation of
the residual of 100 km s$^{-1}$.\label{fig8}}
\end{figure}

Those absorption spectra for which radial velocities could be measured
both by using a line-by-line gaussian fit and cross-correlation were used
to estimate the internal errors. In Figure \ref{fig8} (top panel) we show
the radial velocity differences of 187 galaxies (cross-correlation minus
line-by-line) as a function of the square root of the sum of the
quadratic errors. The average difference between both data sets is only
10 km s$^{-1}$, with an {\em rms} deviation of the residuals of $\sim100$
km s$^{-1}$. The average difference is negligible compared to the 
{\em rms} suggesting that the  radial velocity errors given by the
cross-correlation and/or by the  line-by-line  gaussian fit are
underestimated and are larger than the values listed in Table 4.
Therefore, we adopted the {\em rms} deviation of the residual of 100 km
s$^{-1}$ as the true radial velocity errors of our typical measurement
for further analysis (determination of the velocity dispersion of the
group). 

We used the radial velocities obtained in common with the literature to
estimate the velocity zero-point corrections (if needed) and any external
error that could be presented in the dataset. We compared our radial
velocities with the values published by Zabludoff \& Mulchaey (1998,
2000) and \citet{deC97} for HCG 42 and NGC 5846 groups, with
\citet{mal92}, \citet{bel89} and \citet{ste96} for IC 4765 group, and
with \citet{gar93} and \citet{ram96} for the NGC 6868 system. Also, for
galaxies that were not presented in the references listed above, we
compared our velocities with velocities from the NASA/IPAC Extragalactic
Database (NED).  The lower plot in Figure \ref{fig8} shows the velocity
differences for 23 galaxies observed in common with \citet{zab98,zab00}
(20 galaxies in HCG 42 and 3 in NGC 5846) as a function of the square
root of the sum of the quadratic errors. We found a mean difference of
$-$14 km s$^{-1}$ (dot-short dashed line) with an {\em rms} deviation of
the residual of $\sim100$ km s$^{-1}$. The mean difference is consistent
with the mean internal difference shown in the top panel of Figure
\ref{fig8}. Similar results were obtained when we compared the data for
the other two groups. Following these results, we did not apply any
zero-point correction to the data.

For a few galaxies we found large discrepancies with respect to the
values given in the literature. An example is the case of the background 
late-type galaxy H42-f04-1528. For this galaxy we obtained a radial
velocity of $44597\pm53$ km s$^{-1}$, while \citet{deC97} and
\citet{zab00} obtained values of $3931\pm33$ km s$^{-1}$ and
$19233\pm80$  km s$^{-1}$ respectively. The spectrum of this galaxy shows
a strong  $H_{\alpha}$ emission line and clear absorption features that
put the galaxy at the redshift calculated by us. In this case, we adopted
our  value as the radial velocity for this galaxy. For other similar 
cases, when the  analysis of the  spectra, the galaxy morphology, or 
other properties showed large discrepancies, we discarded the velocity 
and we did not use it into  the calculation of the final radial velocity.
Column 9 in Table 4 identifies which galaxies were used or not in the 
calculation of the final  radial velocity. 

In addition to the galaxies listed in Table 4, we have
cross-correlated the positions of the galaxies in our photometric
catalogs with available data in the literature to search for
galaxies with radial velocities that were not included in our
spectroscopic survey and are members of the groups.  The result of
this search included 17 galaxies in HCG 42, 8 in NGC 5846, 23 in
IC 4765 and 1 in NGC 6868 groups. In cases where the galaxies
had more than one radial velocity measurement available in the 
literature, we calculated the mean weighted velocity following 
equation \ref{eq3}. 

\begin{figure*}
\figurenum{9}
\centering
\includegraphics[totalheight=6cm]{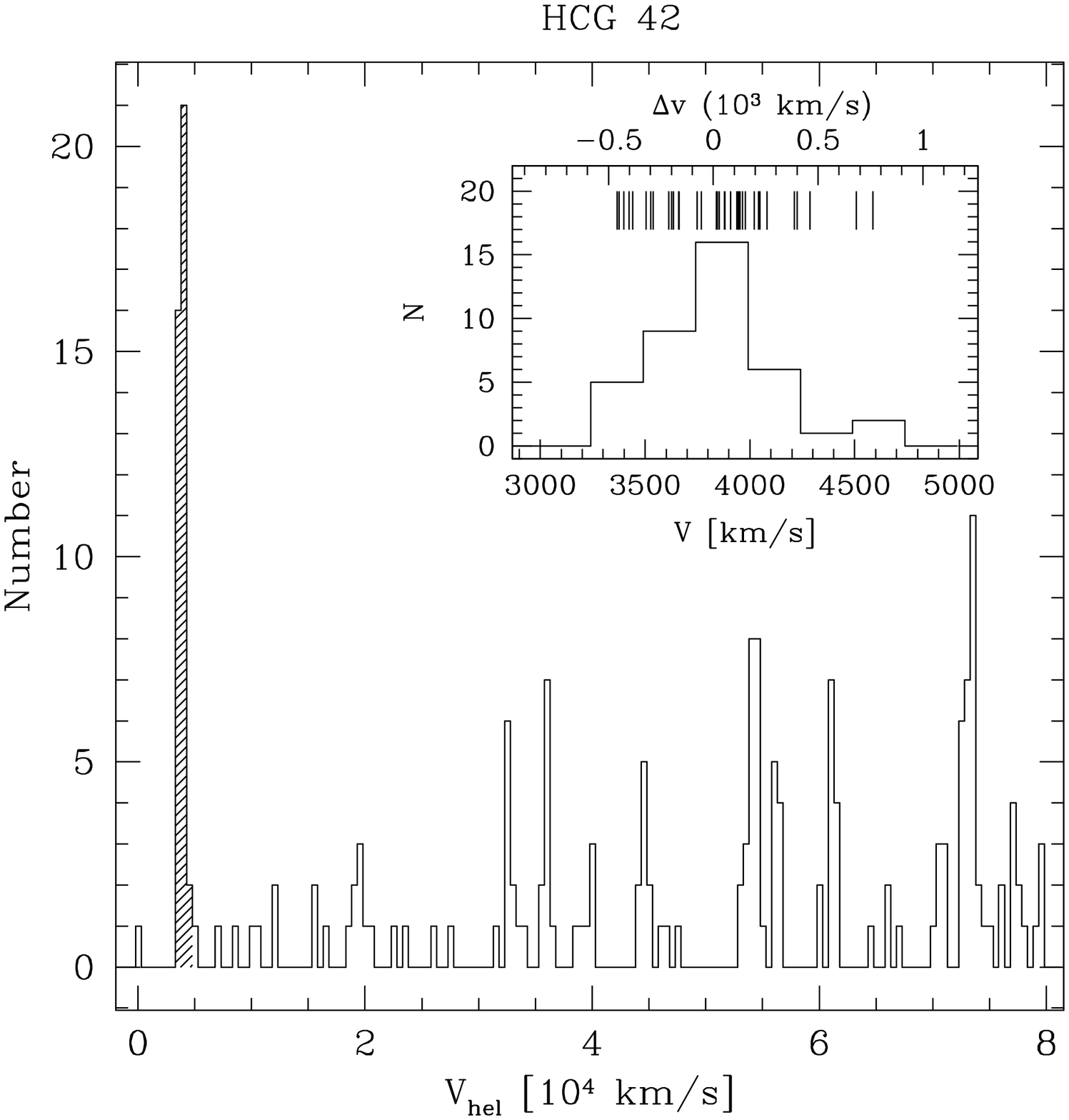}%
\includegraphics[totalheight=6cm]{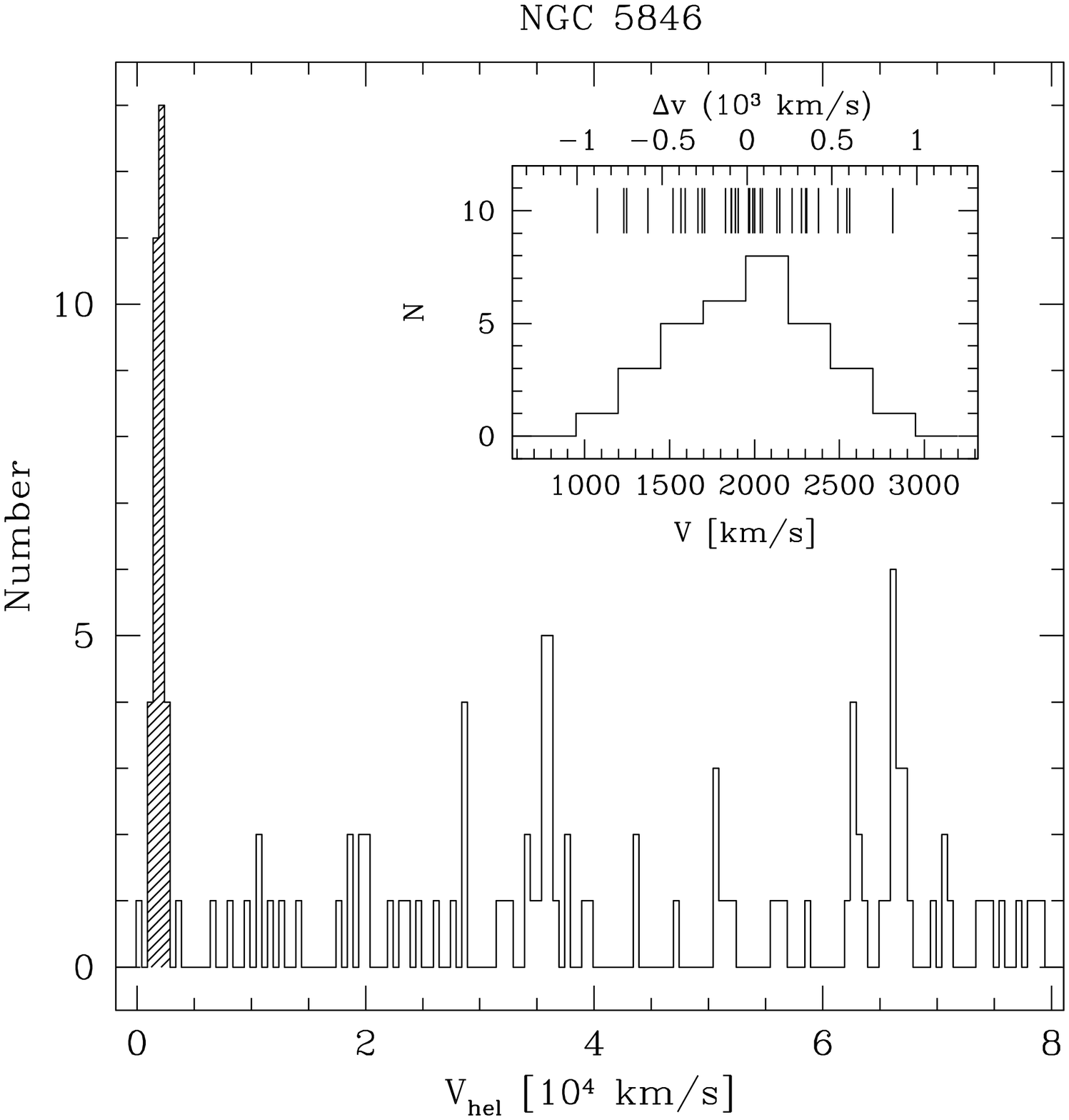}

\includegraphics[totalheight=6cm]{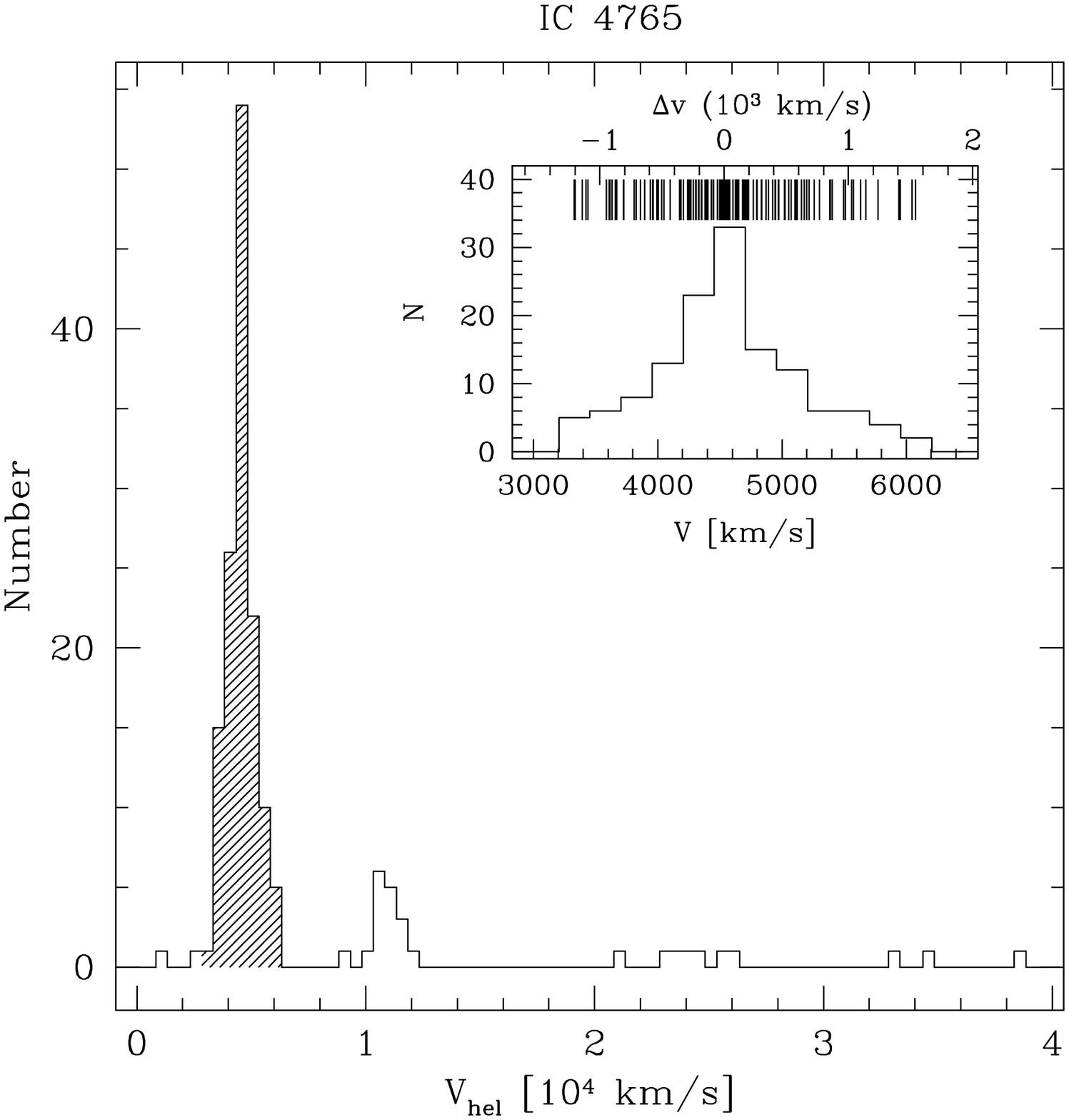}%
\includegraphics[totalheight=6cm]{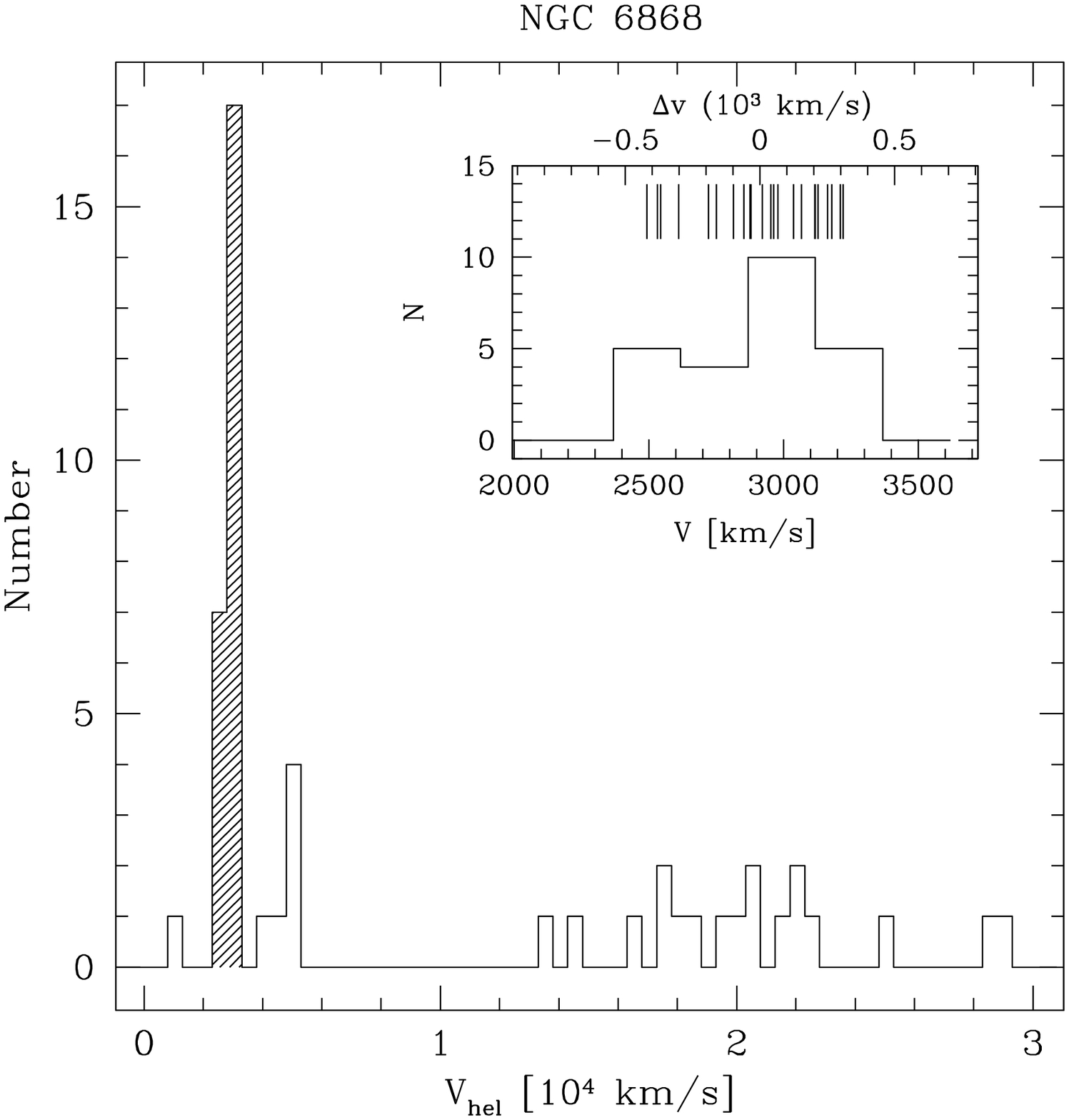}
\caption{Velocity distribution
histograms of all galaxies with measured radial velocities in HCG 42 (top
left), NGC 5846 (top right), IC 4765 (bottom left) and NGC 6868 (bottom
right) in bins of 500 km s$^{-1}$. The histograms also include those
galaxies with radial velocities from the literature that  were not
observed in our survey. The shaded regions indicate all galaxies members
of the groups. {\bf Inset Figures}: velocity distribution of galaxies
members of the groups. The bin size is  250 km s$^{-1}$. The tick marks at
the top of the inset figures represent  the velocities for individual
galaxies.\label{fig9}}
\end{figure*}

\subsection{Background objects among the LSBD galaxies}

We determined the radial velocity for eleven of the 80 LSBD galaxies of
our list, although we placed slits over 60 of them (note that due to
the bad weather we were not able to obtain spectra for galaxies in
four  masks. These masks belong to the groups IC 4765 and NGC 6868 and
contained the non-observed 20 LSBD candidates). Six LSBD's were found
to  have discordant  redshifts and were therefore identified as
background galaxies while 5  were identified as galaxy members of
the groups: 3 in NGC 5846 and 2  in IC 4765. The background  galaxies
correspond to the field of HCG 42  (2),  NGC 5846 (2) and NGC 6868 (2).
Table \ref{tab5} lists the main  structural parameters and the radial
velocities for the six background  galaxies (first six rows) and for
the five LSBD galaxies (last five  rows).

The two background galaxies in the direction of NGC 6868 are dwarf 
members of  a second structure  located at $cz\sim4900$ (i.e. 
2000 km s$^{-1}$ away of  the group - see section  6.4). These two 
galaxies have 
low surface-brightnesses, absolute magnitudes and $(V-I)$ colors typical
for the  LSBD population.  For the remaining four background galaxies,
2 have colors $1.3<(V-I)<1.5$  mag and two are blue galaxies
($(V-I)\sim1$ mag). Further visual inspection showed that these
galaxies have an irregular morphology with knots indicating some star
formation activity. 

Emission lines are present in four of six background galaxies 
(no emission lines are presented in the two dwarf galaxies in the 
background of NGC 6868). Emission lines are also present in the three
LSBD  galaxies members of NGC 5846 group (see Table 5). About 75\% of
our  LSBD sample have morphological types dE/dSph and therefore the
radial velocity  could not be obtained due to the low S/N absorption
spectra they presented.  Only 25\% of the sample are dwarf galaxies
that show an irregular morphology  and clear knots that indicate 
star formation activities (classified as dIrr). The four background
emission line galaxies belong to this sub-sample. They represent
only 30\% of all galaxies we classified as dIrr. If we assume 
that the background galaxies are representative of this sub-sample, 
then we estimate that the fraction contamination of the sample is 
small ($\la 10$\%).

\subsection{Group velocity dispersion}

We have used the 412 galaxies with radial velocities and the galaxies
from the literature to derive the main dynamical parameters of the
groups and to analyse the 3D galaxy distribution. Figure \ref{fig9}
shows the  histograms of the velocity distributions for all  galaxies
with radial  velocities in the area of the four groups. The shaded
areas in the histograms show the location of the group galaxies (the
velocity distributions of the galaxies members of the groups are 
shown in the inserted histogram in Fig. \ref{fig9}). We can clearly see 
several structures along the line-of-sight of the groups. In two groups,  
HCG 42 and IC 4675, we identified background structures associated to at 
least 2 and 1 clusters respectively. In the case of NGC 6868 there are
evidences of a second structure located only 2000 km s$^{-1}$ from the
group. We discuss in more detail this point in section 6.4.  

The average velocities and the one dimensional line-of-sight  velocity
dispersions of the groups were calculated using the bi-weighted  estimators 
of location and scale of \citet{bee90}. We used an iterative procedure by
calculating the location and scale using the ROSTAT program and applying a
3$-\sigma$ clipping algorithm to the results. We repeated this procedure 
until the velocity dispersion converged to a constant value. 

The average group velocities and the corresponding velocity dispersions are
tabulated in Table \ref{tab6}. We have also included in  the table the results
obtained for 3 new background structures detected in  the area of the HCG 42
group and the poor cluster IC 4765.  Table \ref{tab6} contains the following
information: (1) group ID; (2) and (3) equatorial  coordinates of the group
center (in all cases the brightest group galaxy);  (4) total number of
galaxies; (5) number of member galaxies; (6) average  velocities; (7) velocity
dispersion;  (8) radius $r_{200}$. 

Because the groups were observed at different physical areas, we
decided to use the $r_{200}$ radius as the estimator of the virial
radius. In fact, analytical models (e.g. Gunn \& Gott 1972) and
numerical simulations (e.g. Cole \& Lacey 1996) find that the radius
$r_{200}$, defined as the radius in which the density is
$200\rho_{c}$ (critical denstity), contains almost all  the
virialized mass of groups and clusters. Then, the $r_{200}$ is
calculated assuming that $M(r)\propto r$ and using the following
relation:

\eq
r_{200} = \frac{\sqrt{3}\sigma_{r}}{10H_{0}(1+z)(1+\Omega_{0}z)^{1/2}}
\label{eq4}
\eeq

\noindent where $H_{0}=75$ km s$^{-1}$ Mpc$^{-1}$, $\Omega_{0}=0.2$,
$\sigma_r$ is the group velocity dispersion and $z$ is the group redshift.

The average velocities and the velocity dispersions were calculated
assuming an error for the galaxy velocities of 100 km s$^{-1}$ determined
in section 5.2. The results presented in the second line, for each
group, were obtained using all galaxies with available radial
velocities. On average, the velocity dispersions do not
change, if we use only our data or the whole dataset (including
the data from the literature).    

\section{Properties of the galaxies in the studied groups}

\subsection{The magnitude-surface brightness relation}

It is well known that dwarf galaxies follow a tight magnitude-surface
brightness relation (the central surface brightness increases with 
increasing luminosity) in clusters \citep[e.g. Fornax;][]{hil03,inf03} 
and in groups \citep[e.g. NGC 5044;][]{cellone05}. However, the
validation  of this relation has been questioned by a number of authors
that argued against the existence of a magnitude-surface brightness
relation for dwarf galaxies \citep{phil88,irw90,deady02}. It has been
argued that the photometric  sequence seen for dwarf galaxies in the 
$M_{V}-\mu_{0}$ plane could be  produced by selection effects. Very 
compact, high surface brightness  galaxies as well as extended, very 
low surface-brightness galaxies in general are not  present in the 
relation due to observational biases \citep{fer94,imp97}.
 
In the last few years a large number of extended, very-low surface 
brightness objects were discovered in different environments 
\citep[e.g Virgo, Fornax, Leo I and Dorado:][]{imp88,bot91,flint01,car01}.
On the other hand, very
faint, high surface brightness objects were only recently discovered,
in the Fornax \citep[UCD - Ultra Compact Dwarf;][]{drink98,drink99,hil99}
and Virgo \citep{jones06} clusters. Both types of galaxies will fill the 
areas in the parameter space that were completly empty before, suggesting
that there is no observed  correlation between surface brightness and
magnitude.

\begin{figure}[!hbt]
\figurenum{10}
\centering
\includegraphics[totalheight=8cm]{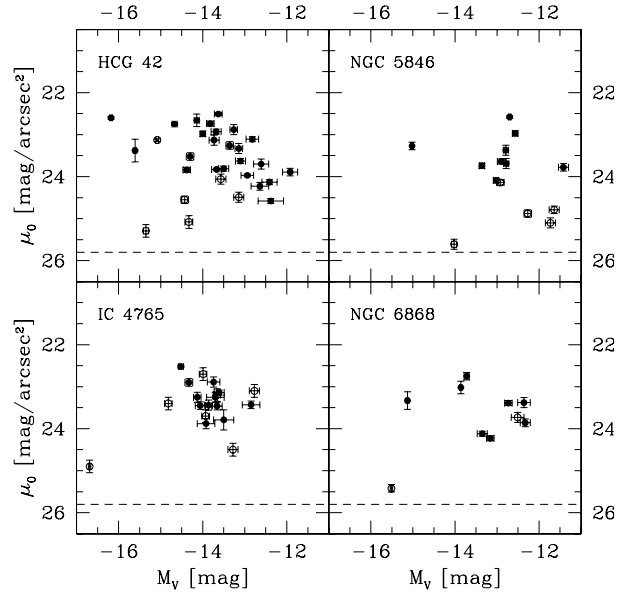}
\caption{Absolute Magnitude - Central surface brightness
relation for LSBD  galaxies in the area of HCG 42, IC 4765, NGC 5846 and
NGC 6868 groups (filled  circles). The extended, very-low surface
brightness galaxies detected using the technique described in section 4.2
are represented by open circles. The limiting surface brightness of 25.8
mag/arcsec$^2$ is represented by a dashed line.\label{fig10}}
\end{figure}

The sky brightness imposes a severe  limitation in the correct
determination of the luminosity distribution  of galaxies. Therefore,
extended, very low surface-brightness galaxies are very difficult to
detect. The use of the smoothing technique  described in section 4.2
gave us the possibility of finding several, previously undetected low
surface-brightness galaxies. Fig. \ref{fig10} shows the $M_{V}-\mu_{0}$ 
relation for all LSBD  galaxies detected in the four groups (we excluded 
the 6 galaxies detected as background objects). We find no clear 
correlation between the central surface brightness and magnitude in
Fig. \ref{fig10}, suggesting that the observed correlation noted by other 
several recent studies could be produced by a selection effect. Similar 
results are found in Dorado \citep{car01} and in Fornax for 
spectroscopically confirmed  dwarf galaxies \citep{deady02}. Note that
the absence of galaxies in the upper right region, where the M32-like 
objects and UCD galaxies are presented, is basically due to the 
limitation imposed by the seeing and the CLASS\_STAR used to select the 
dwarf galaxies.

\subsection{LSBD colors and color-magnitude relation}

The sample of LSBD galaxies has colors between $0<(V-I)<1.5$ with a
peak at $V-I\sim1$ mag. The peak is similar to the value derived for 
LSBD galaxies in other nearby groups and in the field 
\citep[see][for details]{car01}. About 56\% of our sample is formed by 
blue galaxies  ($V-I<1$ mag) and only $\sim10$\% of the sample is
formed by very blue  galaxies ($V-I<0.5$). This is expected since in
low density environments of bright galaxies the blue LSBD galaxies are
more numerous than in dense  environments \citep[e.g Coma][]{secker97}.

\begin{figure}[!htb]
\figurenum{11}
\centering
\includegraphics[totalheight=8cm]{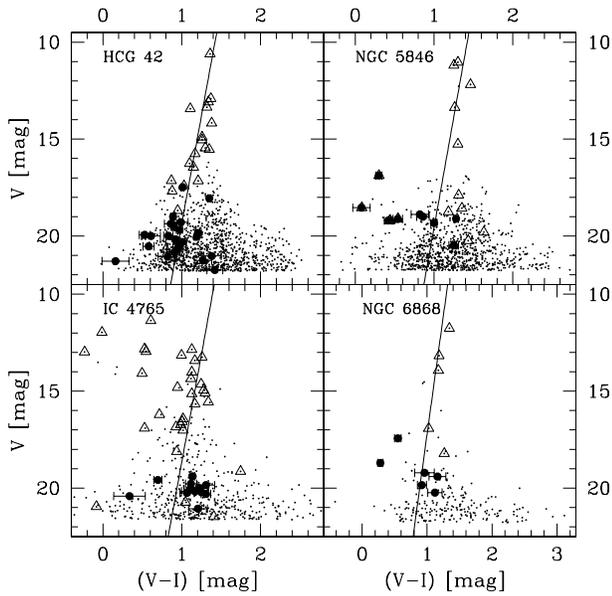}
\caption{Color-magnitude diagrams of all galaxies brighter than
$V=22$ (small dots) identified in the field of HCG 42 (upper left), NGC 5846
(upper right),  IC 4765 (lower left) and NGC 6868 (lower left). The open
triangles are the  galaxies with known redshifts and the filled circles are
LSBD galaxies with  color information. The solid line is the best fit to a
subsample of our group  data (triangles and filled circles) for which
$0<(V-I)<1.5$ mag.\label{fig11}}
\end{figure}

Fig. \ref{fig11} shows the color-magnitude diagrams for all detected
galaxies in the observed areas of the four groups. In the figure,
the filled circles are the LSBD galaxies detected in the groups. The
extended, very low surface brightness galaxies were not included because
no color information is available for them (they were detected in 
the V-band images only). Also in these plots we added all group galaxies 
with known redshifts (triangles).

The early-type galaxies present a well-defined sequence in the CMD 
diagram for more than 10 magnitudes in clusters  
\citep[e.g. Fornax and Coma:][]{hil03,adami06}. In Fig. \ref{fig11} 
we note that this 
relation is also preserved in groups. The color-magnitude relation for
early-type galaxies in groups extends to the region where the LSBD
galaxies (dSph, dE, dIrr) dominate. In all cases, except for some
outliers, the LSBD galaxies follow a well defined color-magnitude
relation (only at fainter  magnitudes, they become bluer). This result
will have implications for  the evolutionary scenarios of dwarf
galaxies. A detailed analysis and discussion of this point will be
presented in a separate paper.  

\subsection{Number density of the LSBD galaxies}

A detailed analysis of the spatial densities and radial distributions of
dwarf galaxies provides important constraints  on the various physical
processes  which are important in understanding galaxy formation. In 
addition, the analysis of the clustering properties of the dwarf 
galaxies is also important from the cosmological point of view. The 
distribution of satellite  galaxies could indicate how luminous and dark
material are distributed at different scales, and therefore can serve as a useful
test for CDM (Cold  Dark Matter) models of galaxy formation. Morever,
it may be used as good  tracer of the dynamic evolution of the dark
halos around bright galaxies.

\begin{figure}[!htb]
\figurenum{12}
\centering
\includegraphics[totalheight=4.5cm]{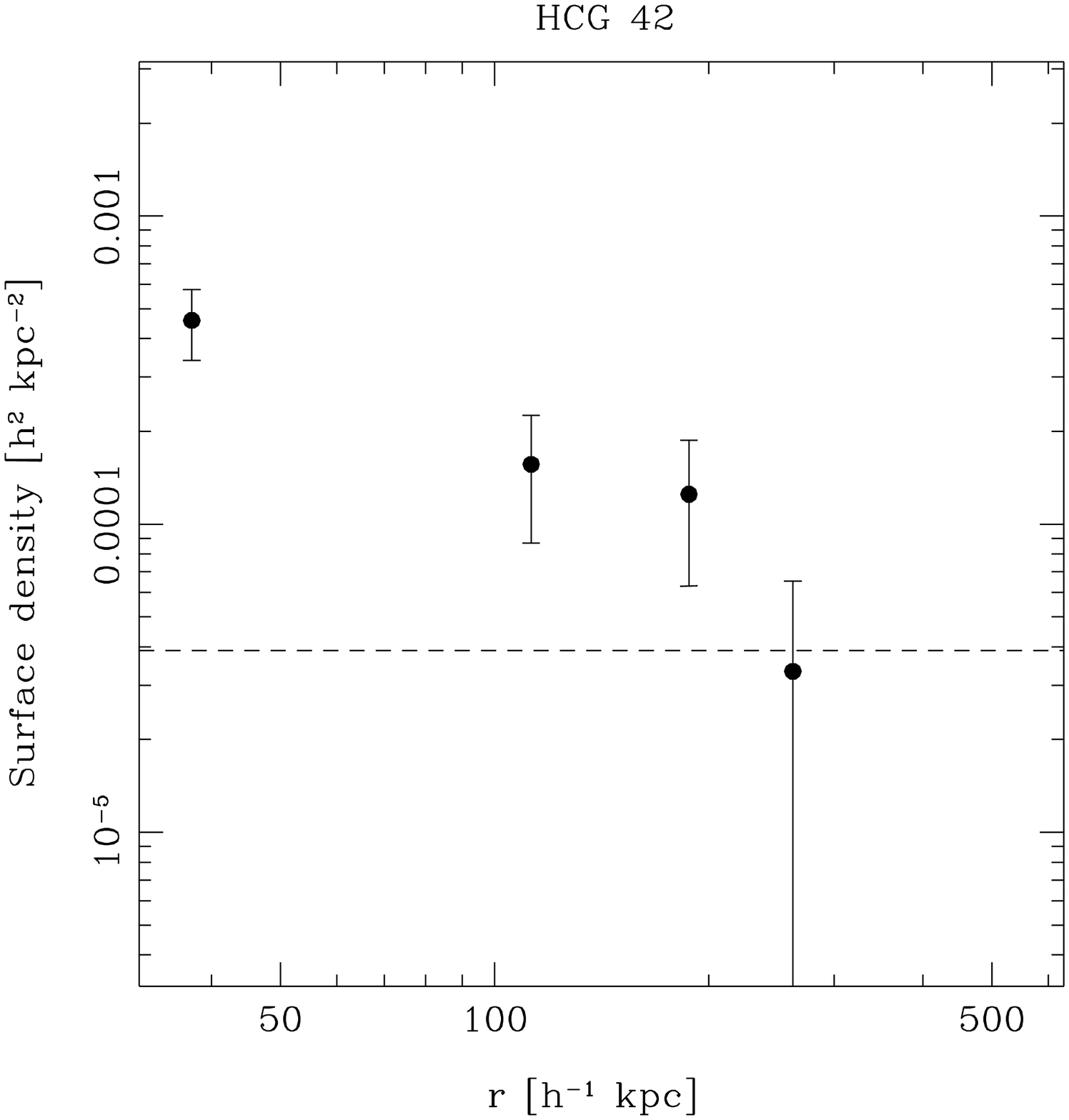}%
\includegraphics[totalheight=4.5cm]{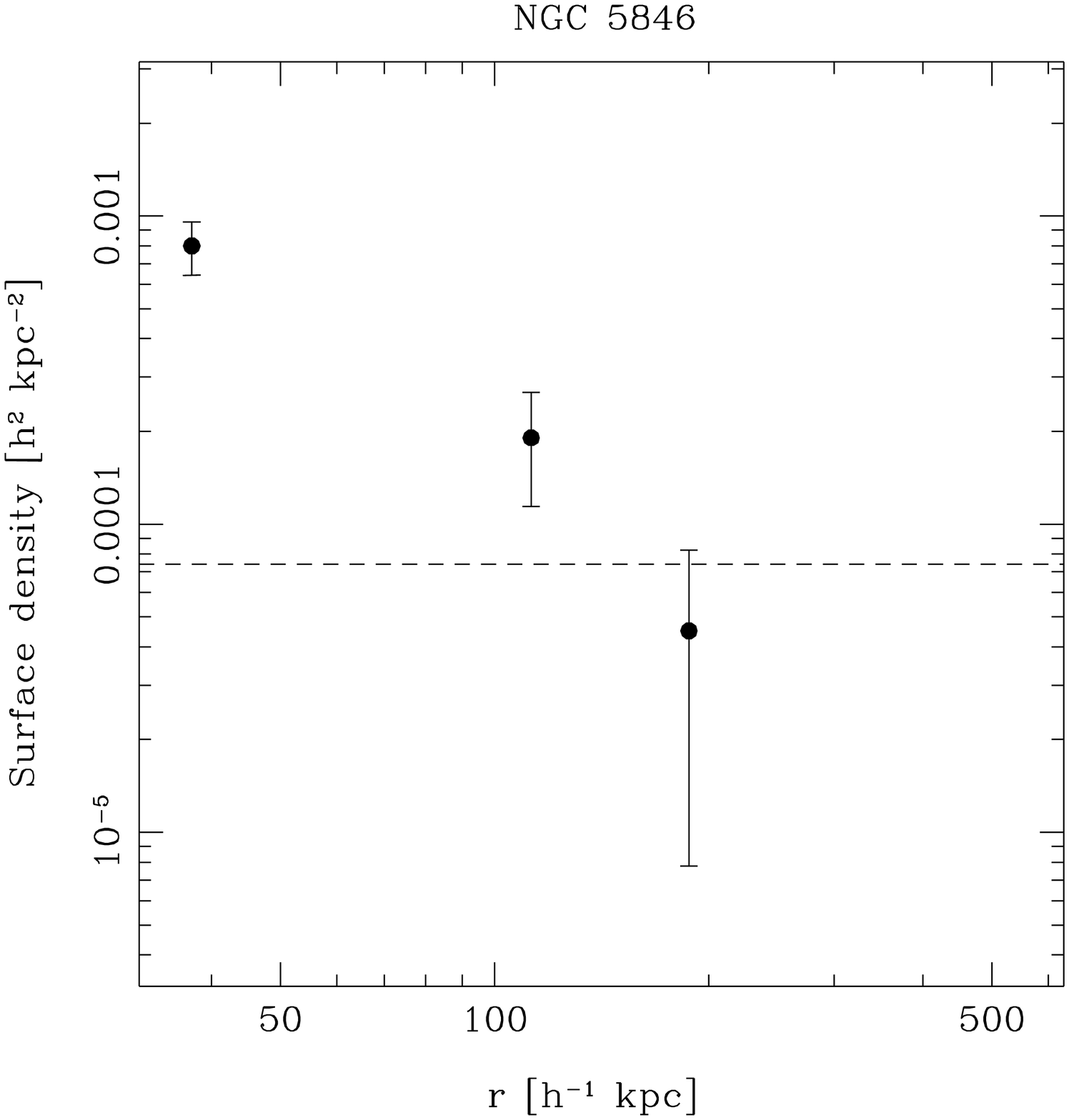}

\includegraphics[totalheight=4.5cm]{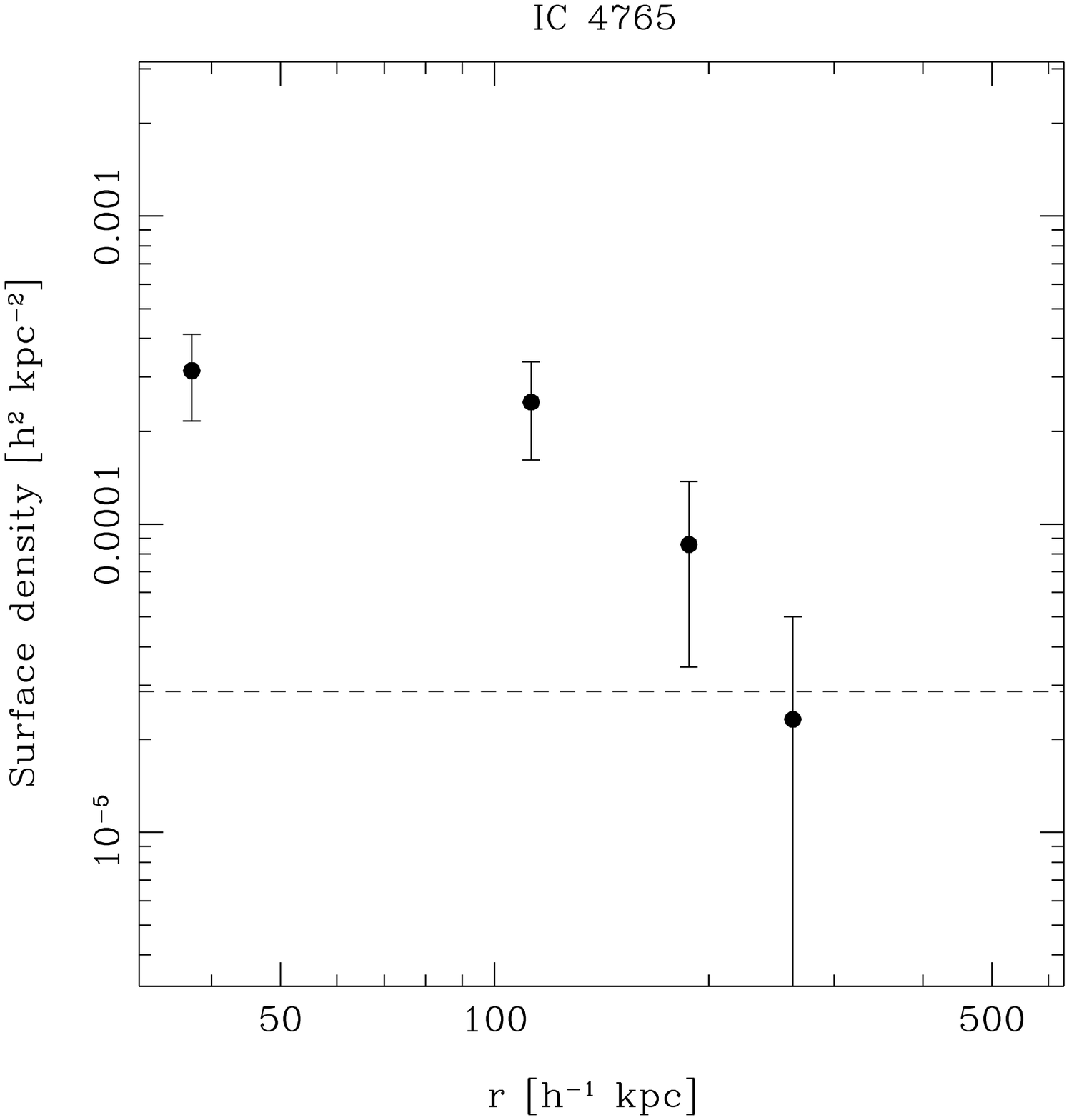}%
\includegraphics[totalheight=4.5cm]{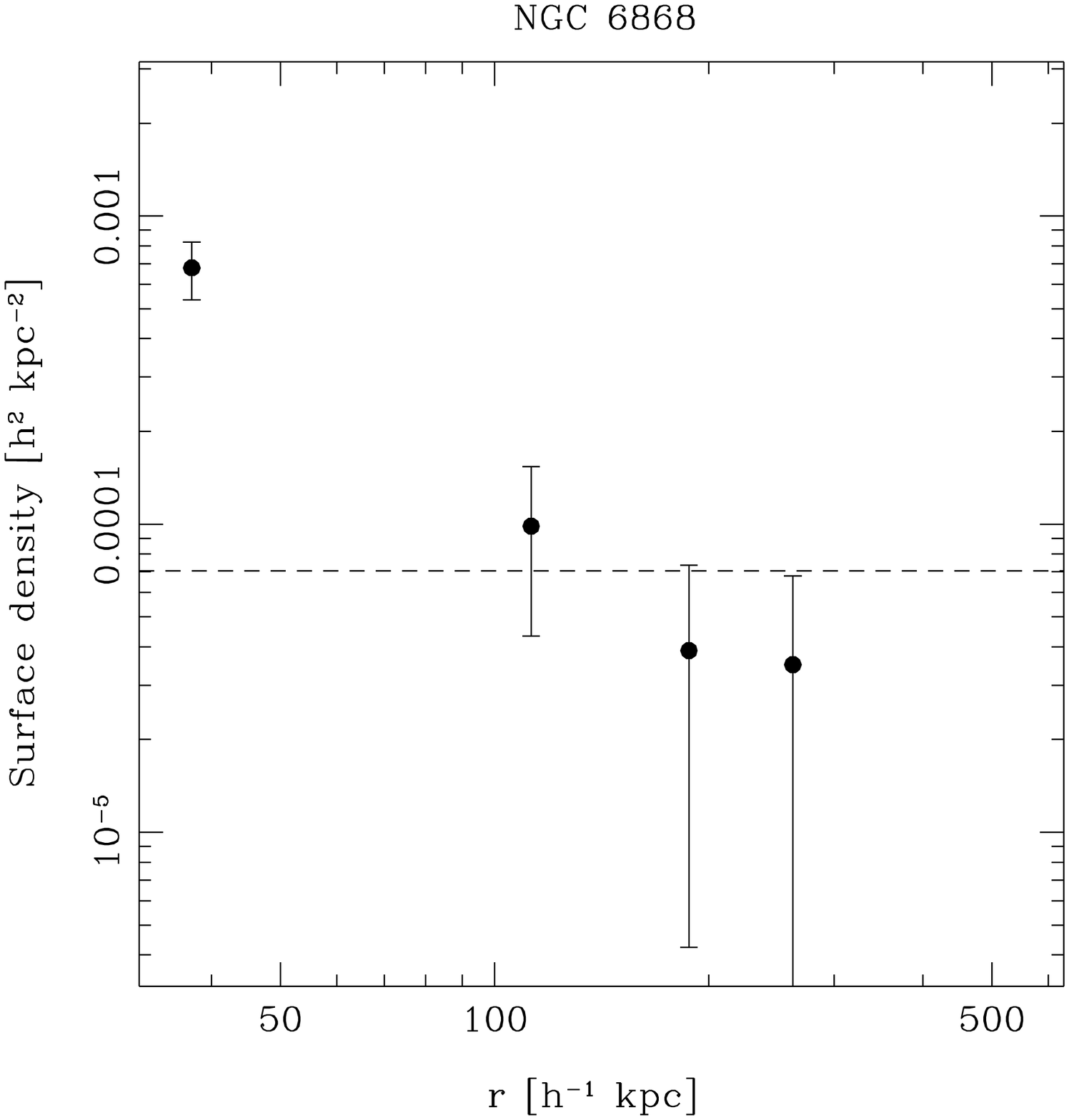}
\caption{Projected number density of
LSBD galaxies as a function radius from  for HCG 42 (top left), NGC 5846 
(top, right), IC 4765 (bottom, left) and  NGC 6868 (bottom, right). The group
centers correspond to the position of the Brightest Group Galaxy of each
group. The data has been divided in bins of 75 h$^{-1}$ kpc. The dashed
lines represent the number density in the control fields.\label{fig12}}
\end{figure}

The dwarf galaxy sample studied here extends from the center of the 
groups to a physical distance of $<$ 0.5 h$^{-1}$ Mpc, covering mainly
the central regions. Fig. \ref{fig12} shows the projected number density
(in  units of h$^{2}$ kpc$^{-2}$) as a function of the distance to the
group centers  for all  LSBD galaxies detected in the four groups. The
data have been divided  in bins of 75 h$^{-1}$ kpc.  

In the figure we clearly see a concentration of LSBD galaxies towards
the center of the four groups in a scale of $<250$ h$^{-1}$   kpc. The
projected number density at the largest measured bin of radius is 
similar to the  value found in our control fields (represented by the 
dahsed lines in the figure). Note that if our sample were dominated  by 
background galaxies, 
then one would expect an almost flat distribution. The results shown  in 
Fig. \ref{fig12} clearly suggests that these galaxies are physically 
associated to the groups. 

We fitted the number density profile of the LSBD sample in Fig.
\ref{fig12}  with a power law of the form $\Sigma(r)=Ar^{\beta}$ whitin
a radius of 250 h$^{-1}$ kpc. The last-square fit to the data gives a
slope, $\beta$, of $-1.27\pm0.21$, $-1.75\pm0.15$, $-1.48\pm0.29$ and
$-1.54\pm0.06$ with a correlation coefficient 0.932, 0.978, 0.900 and
0.995 for HCG 42, NGC 5846, IC 4765 and NGC 6868 groups respectively.
These results are in agreement with those obtained by \citet{vad91} for
satellite of isolated bright elliptical galaxies
($\beta=-1.22\pm0.05$). Also, the results are in a good agreement with
the results presented by \citet{chen05}. These authors  derived a best
fit power-law of $\beta\sim-1.6$, after performing interloper  
contamination corrections, using a volume- and flux-limited Sloan 
Digital  Sky Survey (SDSS) spectroscopic sample of galaxies.  

\begin{figure}[!htb]
\figurenum{13}
\centering
\includegraphics[totalheight=7cm,angle=-90.]{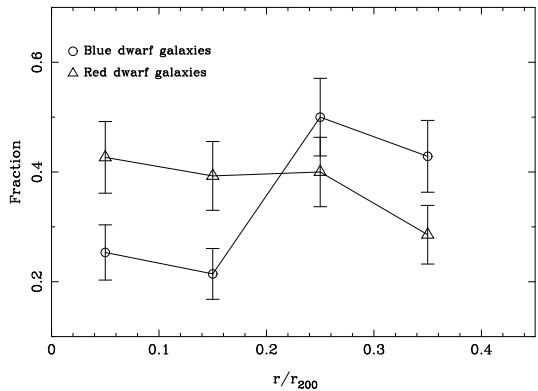}
\caption{Morphological-radius relation for blue ($V-I<0.95$, 
open circles) and  red ($V-I>0.95$, open triangles) dwarf galaxies with
$-17.5<M_{V}<-13$.  The bars denotes Poissonian errors. \label{fig13}}
\end{figure}

We used the velocity catalog of member galaxies and the catalog of LSBD
galaxies  to investigate the environmental dependence of the dwarf
galaxies at scales of  $r<r_{200}$. The dwarf galaxy sample was
constructed by selecting the galaxies  with absolutes magnitudes
between $-17.5<M_{V}<-13$. We analyzed the radial dependence for two
distinct populations of dwarfs:  galaxies which clearly show star
formation activities (blue, with $V-I<0.95$, and  galaxies which do not
have or have little star formation activity (red,  with
$V-I>0.95$). Fig. \ref{fig13} shows the fraction of red and blue dwarf
galaxies as a  function of normalized radius $r/r_{200}$. The 
fraction of red dwarf galaxies (early-type) decreases toward the
outskirts of the groups, while blue, star forming, dwarf galaxies
(late-types) prefer to avoid the central regions of the groups. The 
same relation is seen in denser environments \cite[e.g. Virgo;][]{sab05}.
In their analysis of the properties of galaxy groups in the SDSS,
\citet{weinmann06} found a similar trend for low mass systems. The
fraction of early-type (red) low mass (dwarf) systems  decreases with
the increasing distance from the group's center, while the  fraction of
the late-type (blue) low mass (dwarf) systems increases. The fact that
the dwarf galaxies in groups follow the same ``morphology-radius''
(``mophology-density'') relation as their bright countparts, could be 
an important clue to our understanding of galaxy formation and 
evolution of low-mass systems.

\subsection{Newly discovered background structures}

The histograms in Fig. \ref{fig9} suggest the existence of several
background structures along the group line-of-sights. We used the 
galaxy positions and redshifts to identify the background structures 
in the direction of the groups. 

The velocity distribution of galaxies in the background of HCG 42 group
shows two significant structures at $z\sim0.18$ and at $z\sim0.25$.
The  first structure (HCG42-B1), with RA$\sim10^{h}$  00.2$^{m}$ and
DEC$\sim-19^{\circ}$ 36$^{'}$,  has an elongated  distribution in the
N-S and E-W directions. The second  structure (HCG42-B2), at
RA$\sim10^{h}$ 00.2$^{m}$ and DEC$\sim-19^{\circ}$ 39$^{'}$,  is
elongated in the E-W direction and more concentrated in declination.
The average velocity and the velocity dispersion of these two
structures  are included in Table 6. The most massive structure,
HCG42-B1, has 26 member galaxies. The derived velocity dispersion of
853 km s$^{-1}$ suggest that this is a moderately rich cluster of
galaxies. The second structure, HCG42-B2 with 16 galaxies, has a lower
velocity dispersion  (311 km s$^{-1}$), which is typical of a group of
galaxies. Using the average  clusters redshifts, we derived a distance
of 727 h$^{-1}$ Mpc ($(m-M)=39.3$)   and 972 Mpc h$^{-1}$
($(m-M)=39.3$) for HCG42-B1 and HCG42-B2  respectively. The brightest
galaxy in HCG42-B1  has an absolute magnitude of $M_{V}=-22.5$ 
\citep[the K-correction for an elliptical galaxy at redshift $\sim0.2$
is 0.15 mag;][]{fuk95} similar to the values derived for a central
galaxy in rich clusters, with a morphology similar to a gE/D found in
the center of most rich clusters. For HCG42-B2, the brightest galaxy
has an absolute magnitude of $M_{V}=-21.8$, similar to values found for
the brightest galaxies in groups. 

The galaxy distribution in the direction of IC 4765 shows one well 
defined background structure located at RA$\sim18^{h}$ 47.5$^{m}$, 
DEC$\sim-63^{\circ}$ 18$^{'}$. The average velocity of IC4765-B1 is 
tabulated in Table 6. We derived a value for the velocity dispersion 
of 461 km s$^{-1}$ which is typical for a poor cluster. At the distance 
of the group (146 Mpc,  $(m-M)_{V}=35.8$), the brightest galaxy has an 
absolute magnitude  of $M_{V}=-22.3$, similar to the values obtained 
for the central galaxy  in HCG42-B1 and for central galaxies in other 
poor clusters.

For NGC 5846 and NGC 6868 the picture is less clear. The galaxy
distribution in the direction of NGC 5846 shows several small clumps.
The most important one is located at RA$\sim15^{h}$  06.8$^{m}$,
DEC$\sim+01^{\circ}$ 38$^{'}$ with $cz\sim37000$ km s$^{-1}$.  However,
the existence of this structure as a real cluster is  dubious. In the
direction of NGC 6868 we identified a second structure, $\sim 2000$ km
s$^{-1}$ away from the group. This second  structure is located at
$cz=4980$ km s$^{-1}$  with a velocity  dispersion of 321 km s$^{-1}$.
Two of the low surface-brightness dwarf galaxies identified as
background in section 5.3 belong to this group.

\section{Summary and Conclusions}

Despite the fact that the low surface-brightness dwarf galaxies 
detected in the central region of groups have no membership
determination, we have demonstrated the success of our method in
discovering a population  of dwarf galaxies with low and very low
surface brightnesses that seem to be physically associated to the 
groups (see section 6.3 and below). Using the parameters given by the 
exponential fit profile, we were able to characterize the properties of
the  low surface-brightness dwarf galaxies in groups using an uniform
and well defined criteria. At the 80\% completeness limit, we were able
to detected eighty low surface-brightness dwarf galaxies in the $r<0.5$
h$^{-1}$ Mpc  region of the four groups studied here. These galaxies 
have central surface  brigthnesses $\mu_{0}\la 25$ V mag/arcsec$^{2}$,
scale lengths  $h>1.5$ arcsec, and absolute magnitudes 
$-16.5<M_{V}<-11.5$. The surface brightnesses and scale factors of the
LSBD sample are similar to those seen in Local Group dwarf galaxies.
Eleven of these galaxies have large sizes and  very low 
surface-brightness. Many of the large LSBD galaxies  were not  identified 
in the original images, using the selection criteria  described in section
4.1, but they were detected in the images after we  convolved  them
with an exponential profile function to search for  large and diffuse 
objects. Some of these galaxies are similar in size  and brightness to
the LSBD galaxies discovered in Virgo and Fornax  \citep{imp88,bot91}
and are apparently absent in the Local Group.

A possible limitation in our survey is the criteria used to select the
sample of LSBD galaxies. The criteria used have been optimized to 
search for objects that are similar in magnitudes, sizes and surface 
brightnesses to those dwarf galaxies present in the  Local Group. 
Compact, high surface  brightness galaxies like M32 and very compact, 
high surface brightness, but faint, objects like UCD galaxies
discovered in  Virgo \citep{jones06} and Fornax clusters
\citep{drink98} are virtually  missing from our sample survey. UCD and
M32-like galaxies are very compact and in general they are classified
as stars by SExtractor. We did not detect such objects due  to the
limitation imposed by the seeing and by the CLASS\_STAR limit of  0.35
used to select the LSBD galaxies. Except for the Local group, M32-like
galaxies are very rare and are nearly  absent in other environments.
If these type of galaxies are present in  our sample, their numbers
should be very small. In the case of UCD galaxies the  situation might
be different. These galaxies are not seen in the Local  Group. However,
as mentioned in section 6.1, a large number of these galaxies  were
discovered in nearby clusters. These objects are  apparently absent in
groups, although searches for UCD in poor  environments have just
started \citep{mieske06}. We cannot discard  that  a population of UCD
galaxies could be present in our groups.  This is a pending task that
requires deep, high resolution observations to search for these types
of objects. 

Monte Carlo simulations performed to derive the completeness
fractions of LSBD detection on real images were also used to estimate
the photometric  errors. At $\mu_{0}\sim 25.5$ mag/arcsec$^{2}$ (our
limit in central surface  brightness), the completeness in the
detection is $\sim 80$\%, with an error  in the magnitude and in the
central surface brightness $<0.3$ mag and $<0.5$  V mag/arcsec$^{2}$
respectively. One important result given by the simulation is our
ability to determine correctly the scale factors of the galaxies. The 
results show that the scale factors for galaxies with
$22.5<\mu_{0}<24.5$  mag/arcsec$^{2}$ and  $V<19.0$ mag are
overestimated in $\sim30$\%. At fainter  magnitudes and central surface
brightnesses (the last bins in our simulation),  the overestimation is
$\sim 60$\%. This limitation in estimating correctly the  scale factor
is stronger at fainter magnitudes and could affect the correct 
determination of the faint-end of the Galaxy Luminosity Function by adding
to the sample an undesired population of background galaxies.

In clusters (e.g. Coma and Fornax) there exists
a tight  color-magnitude relation for the early-type galaxies that
extends to the  dwarf regime. We demonstrate, as far as we know for the
first time, that this relation is also present in groups (see Fig.
\ref{fig11}), extending  the relation in more than 10 magnitudes (from
brigth early-type to faint dwarf spheroidal galaxies). We also found 
that the LSBD galaxies do not show a clear  correlation between
magnitude and surface brightness, in contrast  of what is reported in
other  recent studies, perhaps because the apparent  observed
correlation seen  in other environments could be produced by selection
effects. As we discussed in section 6.1, extended, very low-surface 
brightness galaxies could be absent from most of the surveys in groups
and  cluster due to a limitation imposed by the sky brightness. In
fact, we were  able to detect a number of large very low
surface-brightness galaxies only  after we convolved the images with an
exponential profile function that  enhanced the large, low
surface-brightnes  features in the images.

The LSBD galaxies are clearly concentrated around the brightest group
members of the  groups, in a scale of  $\sim 250$ h$^{-1}$ kpc  (see
Fig. \ref{fig12}). At larger radius, the number density is similar  to
the values found in our control fields. A similar correlation is  seen
in other low-density environments \citep{vad91,car01}, where bright
early-type galaxies  are predominant. The central concentration
detected suggests that these galaxies are true dwarf galaxies of the 
groups. The slopes obtained from the fit to the number  density
profile  of LSBD galaxies agree with the values derived for satellites 
dwarfs of isolated bright galaxies \citep{vad91,chen05}. Furthermore, 
the distribution of satellites around the centers of the groups are
similar  in extent to that seen in some fossil groups   
\citep[e.g. NGC 1132][]{mul99}.

Only 78 ($\sim 19$\%) of the 412 galaxies with radial velocities were
confirmed as members of the groups. In addition, several new
background structures were identified along the  line-of-sight of the
groups. In the direction of HCG 42 we identified  one  cluster and one
group of galaxies (at $z=0.18$ and $z=0.25$).  In the direction  of IC
4765 we identified one cluster ($z\sim0.04$) and towards NGC 6868,
only  one poor group ($z\sim0.016$) was identified. 

We have attempted to measure the radial velocities of the majority of
the 80 LSBD galaxies of our list, but only eleven spectra had enough 
S/N for a realiable determination  of their radial velocities. Only 
five LSBD galaxies turned out to be members of the groups. Of the six
background galaxies, two are in the direction of NGC 6868 group and
they are genuine dwarf galaxies which belong to a background structure
located only 2000 km s$^{-1}$ of NGC 6868. For the  remaining four
galaxies, two of them have colors $V-I>1.3$, two are blue and all show
emission lines. It is not surprising that only few galaxies  of the
LSBD catalog, for which we obtained spectroscopy, are members of the 
groups, given that $\sim 75$\% of our sample have morphological types
dE/dSph and therefore could not be observed due to the low S/N
absorption  spectra they present. About 25\% of the LSBD sample are
dwarf galaxies that show an irregular morphology and there is
indication that they are forming stars. The four background
emission-lines galaxies belong to  this sub-sample and they comprise
only the 30\% of the total population  of LSBD galaxies  classified as
dIrr. Hence, we estimated that only a small fraction of our sample
($\sim 10$\% or less) could be contaminated with background galaxies.

The dwarf galaxies ($M_{V}>-17.5)$ in the groups studied follow the same 
morphology-radius relation seen in much denser environments (e.g.
Virgo).  The fraction of red ($V-I>0.95$) dwarfs (dE, dE,N, dSph)
decrease towards the  outskirts of the groups, while the fraction of
blue  ($V-I<0.95$), star  forming dwarfs (dIrr) increase with 
cluster-centric distance. The same  trend is found for low-mass systems
in SDSS galaxy groups analyzed by   \citet{weinmann06}. 
  
In this work we presented a searche for low surface-brightness dwarf 
galaxies in poor environments at distances $>$10 h$^{-1}$ Mpc. We showed 
here that the technique  used
to search for LSBD galaxies is reliable and robust if we can derive 
correctly all errors associated to the detection and photometry. A good
knowledge of the errors and the selection effects may give us the
possibility to construct a complete, well defined sample of
low-luminosity, low surface-brightness galaxies beyond the Local Group,
and characterize their properties and intrinsic nature which is an 
important clue for testing the models of galaxy formation, in particular,
at  low-mass levels.

\acknowledgments We thank the anonymous referee for useful comments
and suggestions. We are grateful to the Director of Las Campanas 
Observatory for generous allocation of telescope time. We are also
grateful to the Las Campanas and Warsaw 1.3 meters telescope personal
for the excelent support and guidence during the observing runs. ERC
acknowledges support for this work provided by FAPESP PhD fellowship
Nr.96/04246-7, and the support of the Gemini Observatory, which is
operated by the Association of Universities for Research in Astronomy,
Inc., on behalf of the international Gemini partnership of Argentina, 
Australia, Brazil, Canada, Chile, the United Kingdom, and the United
States of America. LI wishes to acknowledge the FONDAP ``Center for
Astrophysics'' for support. CMdO thanks the support of FAPESP
through the thematic project 01/07342-7, and CNPq through the PROSUL
project. This work  partially benefitted from the use of the NASA/IPAC 
Extragalactic Database (NED), which is operated by the Jet Propulsion 
Laboratory, California Institute of Technology, under contract with the 
National Aeronautics and Space Administration.

\clearpage

\begin{deluxetable}{lccccccc}
\tabletypesize{\scriptsize}
\tablecaption{Group Parameters \label{tab1}}
\tablenum{1}
\tablecolumns{10}
\tablewidth{0pc}
\tablehead{
\colhead{Group} & 
\colhead{RA} & 
\colhead{DEC} & 
\colhead{D} & 
\colhead{$\sigma_{r}$} & 
\colhead{$\log(L_{X})$} &
\colhead{$A_{V}$\tablenotemark{(a)}} & 
\colhead{$A_{I}$\tablenotemark{(a)}} \\
\colhead{} & 
\colhead{(J2000.0)} & 
\colhead{(J2000.0)} &
\colhead{[h$^{-1}$ Mpc]} &
\colhead{[km s$^{-1}$]} & 
\colhead{[$h^{-2}$ erg s$^{-1}$]} &
\colhead{[mag]} & 
\colhead{[mag]}}
\startdata
HCG 42    & 10 00 11.4 & --19 37 02 & 54.5 & 301 & 41.59$^{(1)}$   &  0.13 & 0.06 \\
NGC 5846  & 15 06 28.5 & $+$01 35 10 & 24.1 & 416 & 41.71$^{(1)}$   &  0.16 & 0.08 \\
IC 4765   & 18 47 32.9 & --63 18 46 & 60.9 & 620 & 43.20$^{(2)}$   & 0.31 & 0.15 \\
NGC 6868  & 20 09 43.0 & --48 17 08 & 32.7 & 238 & $\sim41^{(3)}$ &  0.16 & 0.08 \\
\enddata
\tablecomments{Units of right ascension are hours, minutes and seconds, and units of declination
are degrees, arcminutes and arcseconds.}
\tablenotetext{(a)}{Absorption corrections determined from the reddening maps of 
Schlegel et al. (1998)}
\tablerefs{X-Ray luminosity from: (1) - Mulchaey et al. (2003), (2) - Jones \&
Forman (1999); (3) - Beuing et al. (1999)}
\end{deluxetable}

\clearpage

\begin{deluxetable*}{lccccccccr}
\tabletypesize{\scriptsize}
\tablecaption{Observing log \label{tab2}}
\tablenum{2}
\tablecolumns{10}
\tablewidth{0pc}
\tablehead{
\colhead{Group} &
\colhead{Field Id}&
\colhead{RA}&
\colhead{DEC}&
\colhead{Filters}&
\colhead{Exposure V}&
\colhead{Exposure I}&
\colhead{Seeing V}&
\colhead{Seeing I}&
\colhead{Area\tablenotemark{(a)}}\\
\colhead{}&
\colhead{}&
\colhead{(J2000.0)}&
\colhead{(J2000)}&
\colhead{}&
\colhead{sec}&
\colhead{sec}&
\colhead{\arcsec}&
\colhead{\arcsec}&
\colhead{arcmin$^{2}$}}
\startdata
HCG 42  &  F01  & 10 00 17.5 & --19 38 32& V,I & $6\times600$ & $3\times600$ & 1.2 & 1.2 & 1597.32 \\
        &  F02  & 09 59 41.0 & --19 30 41& V,I &      ''      &      ''      & 1.3 & 1.1 & \nodata \\
        &  F03  & 09 59 41.3 & --19 44 51& V,I &      ''      &      ''      & 1.2 & 1.1 & \nodata \\
        &  F04  & 10 00 41.1 & --19 44 51& V,I &      ''      &      ''      & 1.3 & 1.2 & \nodata \\
        &  F05  & 10 00 39.5 & --19 30 29& V,I &      ''      &      ''      & 1.4 & 1.1 & \nodata \\
        &  F06  & 10 00 40.9 & --18 30 33& V,I &      ''      &      ''      & 1.2 & 1.1 & \nodata \\
        &  F07  & 10 04 39.4 & --19 40 18& V,I &      ''      &      ''      & 1.3 & 1.2 & \nodata \\
        &  F08  & 09 59 56.5 & --21 20 47& V,I &      ''      &      ''      & 1.4 & 1.2 & \nodata \\
NGC 5846&  F01  & 15 05 23.5 &$+$01 34 32& V,I & $4\times900$ & $3\times600$ & 1.3 & 1.2 & 1397.52 \\
        &  F01  & 15 06 16.3 &$+$01 34 55& V,I &      ''      &      ''      & 1.3 & 1.3 & \nodata \\
        &  F03  & 15 07 08.9 &$+$01 34 54& V,I &      ''      &      ''      & 1.3 & 1.2 & \nodata \\
        &  F04  & 15 06 24.3 &$+$02 26 58& V,I &      ''      &      ''      & 1.3 & 1.2 & \nodata \\
        &  F05  & 15 07 29.0 &$+$00 38 33& V,I &      ''      &      ''      & 1.2 & 1.1 & \nodata \\
        &  F06  & 15 01 50.1 &$+$01 20 48& V,I &      ''      &      ''      & 1.2 & 1.1 & \nodata \\
        &  F07  & 15 09 50.6 &$+$01 34 41& V,I &      ''      &      ''      & 1.2 & 1.2 & \nodata \\
IC 4765 &  F01  & 18 47 17.1 & --63 22 27& V,I & $4\times900$ & $3\times600$ & 1.2 & 1.5 & 798.48 \\
        &  F02  & 18 47 34.5 & --63 09 15& V,I &      ''      &      ''      & 1.4 & 1.2 & \nodata \\
        &  F03  & 18 45 57.9 & --62 39 50& V,I &      ''      &      ''      & 1.3 & 1.2 & \nodata \\
        &  F04  & 18 40 02.0 & --63 20 11& V   & $4\times900$ & \nodata      & 1.5 & 1.4 & \nodata \\
NGC 6868&  F01  & 20 09 48.8 & --48 17 25& V,I & $4\times900$ & $3\times600$ & 1.5 & 1.3 & 399.24 \\
        &  F02  & 20 08 31.7 & --48 17 25& V   & $4\times900$ & \nodata      & 1.2 & 1.2 & \nodata \\
\enddata
\tablecomments{Units of right ascension are hours, minutes and seconds, and units of declination
are degrees, arcminutes and arcseconds.}
\tablenotetext{(a)}{The values are the total areas observed}
\end{deluxetable*}

\clearpage

\LongTables
\begin{deluxetable*}{lcccccccccccl}
\tabletypesize{\scriptsize}
\tablenum{3}
\tablecolumns{13}
\tablewidth{0pc}
\tablecaption{Photometric data and profile fit parameters of LSBD galaxies\label{tab3}}
\tablehead{
\colhead{Galaxy id.} & 
\colhead{RA} & 
\colhead{DEC} & 
\colhead{$V_{t}$} & 
\colhead{$M_{V}$} & 
\colhead{(V - I)} & 
\colhead{$D_{iso}$} &
\colhead{$\theta_{lim}$} & 
\colhead{$\mu_{0}$} & 
\colhead{$h$} & 
\colhead{$r_{eff}$} & 
\colhead{$\langle \mu \rangle_{eff}$} & 
\colhead{Type}\\
\colhead{} &
\colhead{[(J2000.0)} &
\colhead{[(J2000.0)]} &
\colhead{[mag]} &
\colhead{[mag]} &
\colhead{[mag]} &
\colhead{[$''$]} &
\colhead{[$''$]} &
\colhead{[mag/$\sq^{''}$]} &
\colhead{[$''$]}&
\colhead{[$''$]}&
\colhead{[mag/$\sq^{''}$]}&
\colhead{}} 
\startdata
H42-f03-2397   & 09 59 19.8 &$-$19 38 11 & 21.03 & -12.64 & 1.38  &  4.5 &  5.5 & 24.2 &  1.9 &  3.2 & 25.3 & dE/dSph\\
H42-f02-2153   & 09 59 24.1 &$-$19 29 05 & 20.04 & -13.63 & 1.19  &  9.4 &  9.9 & 22.5 &  1.6 &  2.7 & 23.6 & dE/dS0?\\
H42-f02-2098   & 09 59 25.2 &$-$19 29 55 & 20.17 & -13.50 & 0.92  &  5.2 &  5.6 & 23.8 &  1.5 &  2.5 & 24.9 & dS0/dIrr\\
H42-f02-738    & 09 59 25.7 &$-$19 32 58 & 20.00 & -13.67 & 0.61  &  5.3 &  5.8 & 23.8 &  1.6 &  2.7 & 24.9 & dS0/dIrr\\
H42-f03-L01    & 09 59 29.2 &$-$19 40 59 & 18.32 & -15.35 &\nodata&  8.5 &  8.5 & 25.3 &  9.0 & 15.0 & 26.4 & dE/dSph\\
H42-f02-2116   & 09 59 29.6 &$-$19 28 16 & 19.94 & -13.73 & 0.53  &  7.8 &  8.8 & 23.1 &  1.8 &  3.0 & 24.2 & dIrr\\
H42-f02-L01    & 09 59 31.0 &$-$19 31 47 & 18.59 & -15.08 &\nodata& 14.6 & 14.6 & 23.1 &  3.0 &  4.9 & 24.2 & dS0,N/dIrr\\
H42-f03-2086\tablenotemark{(a)}   & 09 59 32.2 &$-$19 43 18 & 20.31 & -13.36 & 1.01  &  8.5 &  8.6 & 23.3 &  1.9 &  3.1 & 24.4 & dS0/dIrr\\
H42-f03-1513   & 09 59 33.0 &$-$19 39 54 & 20.73 & -12.94 & 0.94  &  6.8 &  7.7 & 24.0 &  2.3 &  3.8 & 25.1 & dE/dSph\\
H42-f02-408    & 09 59 33.1 &$-$19 35 07 & 19.85 & -13.82 & 1.21  &  8.0 &  8.7 & 22.7 &  1.5 &  2.6 & 23.9 & dS0/dIrr\\
H42-f03-1461   & 09 59 34.7 &$-$19 39 51 & 20.53 & -13.14 & 0.58  &  7.9 &  9.2 & 23.3 &  2.0 &  3.4 & 24.4 & dIrr\\
H42-f03-L02    & 09 59 39.3 &$-$19 38 02 & 19.24 & -14.43 &\nodata&  9.7 &  9.7 & 24.6 &  4.2 &  7.0 & 25.7 & dE,N\\
H42-f02-1939\tablenotemark{(b)}   & 09 59 46.8 &$-$19 27 59 & 18.95 & -14.72 & 0.87  & 13.9 & 15.4 & 22.9 &  2.9 &  4.9 & 24.0 & Sp\\
H42-f03-1819   & 09 59 51.9 &$-$19 41 36 & 19.37 & -14.30 & 0.86  & 11.7 & 12.6 & 23.5 &  3.0 &  5.0 & 24.6 & dE/dSph\\
H42-f02-1972\tablenotemark{(a)}   & 09 59 56.6 &$-$19 28 08 & 20.41 & -13.26 & 0.96  &  8.0 &  8.9 & 22.9 &  1.6 &  2.8 & 24.0 & dS0\\
H42-f02-1774   & 10 00 03.0 &$-$19 26 58 & 19.29 & -14.38 & 0.99  & 11.0 & 12.4 & 23.8 &  3.4 &  5.7 & 25.0 & dE/dSph\\
H42-f01-L01    & 10 00 03.6 &$-$19 42 20 & 20.10 & -13.57 &\nodata&  7.2 &  7.2 & 24.1 &  2.2 &  3.8 & 25.2 & dIrr\\
H42-f01-1158   & 10 00 09.5 &$-$19 38 32 & 20.57 & -13.10 & 0.97  &  7.2 &  8.2 & 23.6 &  2.0 &  3.4 & 24.7 & dE/dSph\\
H42-f01-2072   & 10 00 13.6 &$-$19 36 42 & 17.49 & -16.18 & 1.01  & 23.4 & 25.4 & 22.6 &  4.3 &  7.2 & 23.7 & dE,N\\
H42-f05-2229   & 10 00 17.7 &$-$19 28 42 & 21.06 & -12.61 & 0.82  &  5.5 &  6.1 & 23.7 &  1.6 &  2.6 & 24.8 & dIrr\\
H42-f04-2293\tablenotemark{(a)}	& 10 00 23.7 &$-$19 39 20 & 18.06 & -15.61 & 1.35  & 33.9 & 33.6 & 23.4 &  7.5 & 12.6 & 24.5 & dS0\\
H42-f04-2307\tablenotemark{(b)}	& 10 00 23.7 &$-$19 39 30 & 19.16 & -14.51 & 1.43  & 12.4 & 13.0 & 22.8 &  2.3 &  3.9 & 23.9 & Sp \\
H42-f01-L02    & 10 00 24.5 &$-$19 42 44 & 20.53 & -13.14 &\nodata&  5.5 &  5.5 & 24.5 &  2.3 &  3.8 & 25.6 & dE/dSph\\
H42-f04-1856   & 10 00 41.8 &$-$19 42 21 & 19.67 & -14.00 & 0.97  & 12.7 & 13.6 & 23.0 &  2.6 &  4.4 & 24.1 & dS0\\
H42-f04-1937   & 10 00 45.2 &$-$19 42 58 & 19.53 & -14.14 & 0.90  &  9.0 &  9.6 & 22.7 &  1.7 &  2.8 & 23.8 & dS0/dIrr\\
H42-f04-1618   & 10 00 45.3 &$-$19 40 39 & 19.00 & -14.67 & 0.89  & 12.2 & 13.4 & 22.8 &  2.4 &  4.0 & 23.9 & dE,N\\
H42-f05-L01    & 10 00 45.6 &$-$19 24 27 & 19.34 & -14.33 &\nodata&  6.8 &  6.8 & 25.1 &  5.1 &  8.5 & 26.2 & dE/dSph\\
H42-f04-1064   & 10 00 46.3 &$-$19 45 15 & 21.29 & -12.38 & 0.16  &  4.6 &  5.5 & 24.6 &  2.4 &  4.1 & 25.7 & dIrr\\
H42-f04-1800   & 10 00 53.9 &$-$19 42 02 & 20.85 & -12.82 & 0.91  &  7.0 &  8.1 & 23.1 &  1.6 &  2.7 & 24.2 & dS0/dIrr\\
H42-f04-1063   & 10 00 54.5 &$-$19 44 56 & 19.99 & -13.68 & 0.83  &  7.4 &  8.2 & 22.9 &  1.6 &  2.6 & 24.0 & dS0/dIrr\\
H42-f04-1340\tablenotemark{(a)}	& 10 00 58.0 &$-$19 38 51 & 21.26 & -12.41 & 1.27  &  5.4 &  6.2 & 24.1 &  2.0 &  3.3 & 25.2 & dE/dSph\\
H42-f05-1950\tablenotemark{(a)}	& 10 01 00.3 &$-$19 27 18 & 21.75 & -11.92 & 1.42  &  4.9 &  5.4 & 23.9 &  1.5 &  2.6 & 25.0 & dE/dSph\\
N5846-f01-L01  & 15 04 55.7 &$+$01 33 34 & 17.88 & -14.02 &\nodata&  8.9 &  8.9 & 25.6 & 12.6 & 21.1 & 26.7 & dE/dSph\\
N5846-f01-L02  & 15 04 55.7 &$+$01 41 21 & 19.63 & -12.27 &\nodata& 13.7 & 13.7 & 24.9 &  4.0 &  6.7 & 26.0 & dE/dSph\\
N5846-f01-L03  & 15 04 55.7 &$+$01 32 47 & 20.17 & -11.73 &\nodata&  9.0 &  9.0 & 25.1 &  3.5 &  5.8 & 26.2 & dE/dSph\\
N5846-f01-1819 & 15 05 01.0 &$+$01 36 35 & 19.33 & -12.57 & 1.00  & 11.9 & 12.7 & 23.0 &  2.4 &  4.1 & 24.1 & dS0,N/dIrr\\
N5846-f01-1443\tablenotemark{(b)} & 15 05 21.8 &$+$01 39 04 & 19.11 & -12.79 & 0.99  & 11.8 & 12.9 & 22.6 &  2.2 &  3.7 & 23.8 & Sp\\
N5846-f01-1996\tablenotemark{(d)} & 15 05 27.9 &$+$01 35 40 & 20.48 & -11.42 & 1.25  & 11.9 & 12.8 & 23.8 &  2.4 &  3.9 & 24.9 & dS0,N/dIrr\\
N5846-f01-343\tablenotemark{(b)}  & 15 05 30.6 &$+$01 30 31 & 18.48 & -13.42 & 1.22  & 15.6 & 17.1 & 23.7 &  4.4 &  7.4 & 24.8 & S0/Sp\\
N5846-f02-L01  & 15 05 48.5 &$+$01 34 50 & 18.98 & -12.92 &\nodata& 11.9 & 11.9 & 24.1 &  3.9 &  6.5 & 25.3 & dE/dSph\\
N5846-f02-L02  & 15 05 48.5 &$+$01 28 20 & 20.26 & -11.64 &\nodata& 10.8 & 10.8 & 24.8 &  2.9 &  4.8 & 25.9 & dE/dSph\\
N5846-f02-1812 & 15 06 33.5 &$+$01 39 19 & 19.00 & -12.90 & 0.86  & 11.4 & 12.7 & 23.6 &  3.2 &  5.3 & 24.8 & dS0,N/dIrr\\
N5846-f03-669  & 15 06 59.4 &$+$01 33 16 & 19.12 & -12.78 & 0.54  & 17.1 & 18.4 & 23.7 &  4.7 &  7.9 & 24.8 & dIrr/HIIreg?\\
N5846-f03-1821 & 15 07 00.9 &$+$01 36 19 & 19.11 & -12.79 & 1.28  & 19.2 & 19.6 & 23.4 &  4.4 &  7.3 & 24.5 & dE\\
N5846-f03-1341 & 15 07 01.1 &$+$01 39 38 & 18.88 & -13.02 & 0.82  & 16.1 & 18.9 & 24.1 &  6.0 & 10.0 & 25.2 & dE/dSph\\
N5846-f03-1\tablenotemark{(d)}    & 15 07 02.9 &$+$01 30 49 & 16.88 & -15.02 & 0.30  & 60.3 & 65.6 & 23.3 & 14.1 & 23.5 & 24.4 & dIrr/HIIreg?\\
N5846-f03-147\tablenotemark{(d)}  & 15 07 05.1 &$+$01 30 41 & 18.54 & -13.36 & 0.08  & 26.4 & 29.3 & 23.7 &  7.7 & 12.9 & 24.9 & dIrr/HIIreg?\\
N5846-f03-268  & 15 07 08.3 &$+$01 30 28 & 19.20 & -12.70 & 0.42  & 10.8 & 11.8 & 22.6 &  2.0 &  3.3 & 23.7 & dE/HIIreg?\\
IC4765-f01-1220& 18 46 25.1 &$-$63 23 34 & 20.30 & -13.61 & 1.32  &  7.7 &  5.7 & 23.1 &  1.8 &  3.0 & 24.3 & dE/dS0/dIrr?\\
IC4765-f01-L01 & 18 46 28.3 &$-$63 16 31 & 19.92 & -13.99 &\nodata&  6.6 &  6.6 & 22.7 &  1.2 &  2.0 & 23.8 & dE/dSPh\\
IC4765-f01-1894& 18 46 30.2 &$-$63 16 37 & 19.39 & -14.52 & 1.15  &  7.9 &  8.2 & 22.5 &  1.5 &  2.5 & 23.6 & dE,N\\
IC4765-f01-L03 & 18 46 46.1 &$-$63 16 46 & 19.98 & -13.93 &\nodata&  6.6 &  6.6 & 23.7 &  1.9 &  3.2 & 24.8 & dE/dSph\\
IC4765-f02-518 & 18 46 46.2 &$-$63 13 41 & 19.99 & -13.92 & 1.20  &  6.5 &  6.3 & 23.9 &  2.2 &  3.7 & 25.0 & dE/dSph\\
IC4765-f02-L01 & 18 46 48.7 &$-$63 11 12 & 17.22 & -16.69 &\nodata& 16.8 & 16.8 & 24.9 & 11.7 & 19.5 & 26.0 & dE/dSph\\
IC4765-f01-1457& 18 46 57.4 &$-$63 22 31 & 20.17 & -13.74 &\nodata&  7.5 &  5.9 & 22.9 &  1.6 &  2.7 & 24.0 & dIrr\\
IC4765-f01-1707& 18 47 00.3 &$-$63 20 39 & 19.85 & -14.06 & 1.32  &  8.5 &  7.1 & 23.4 &  2.3 &  3.9 & 24.6 & dE/dSph\\
IC4765-f01-L04 & 18 47 01.8 &$-$63 25 03 & 19.09 & -14.82 &\nodata& 10.1 & 10.1 & 23.4 &  2.5 &  4.2 & 24.5 & dE/dSph\\
IC4765-f01-L02 & 18 47 05.7 &$-$63 17 07 & 20.25 & -13.66 &\nodata&  5.6 &  5.6 & 23.2 &  1.3 &  2.2 & 24.3 & dE,N/dSph\\
IC4765-f02-1620& 18 47 09.1 &$-$63 03 49 & 19.78 & -14.13 & 1.10  & 10.4 &  7.1 & 23.2 &  2.9 &  4.8 & 24.4 & dIrr\\
IC4765-f02-L02 & 18 47 16.5 &$-$63 15 21 & 20.63 & -13.28 &\nodata&  4.4 &  4.4 & 24.5 &  2.0 &  3.3 & 25.6 & dE/dSph\\
IC4765-f01-560\tablenotemark{(d)} & 18 47 21.8 &$-$63 26 44 & 20.05 & -13.86 & 1.12  &  7.6 &  6.5 & 23.4 &  2.2 &  3.7 & 24.6 & dE,N/dSph\\
IC4765-f01-1921& 18 47 25.1 &$-$63 17 09 & 20.25 & -13.66 & 1.08  &  6.7 &  5.9 & 23.5 &  1.9 &  3.2 & 24.6 & dE,N/dSph\\
IC4765-f02-529 & 18 47 25.9 &$-$63 13 42 & 20.41 & -13.50 & 0.32  &  5.5 &  5.3 & 23.8 &  2.0 &  3.3 & 24.9 & dIrr\\
IC4765-f01-1843& 18 47 51.7 &$-$63 21 45 & 21.06 & -12.85 & 1.22  &  5.7 &  4.1 & 23.4 &  1.5 &  2.6 & 24.5 & dS0/dIrr?\\
IC4765-f02-1627& 18 47 54.0 &$-$63 03 48 & 20.21 & -13.70 & 1.23  &  7.8 &  5.9 & 23.2 &  1.9 &  3.2 & 24.4 & dE,N?/dSph\\
IC4765-f02-L03 & 18 48 00.9 &$-$63 15 21 & 21.14 & -12.77 &\nodata&  4.0 &  4.0 & 23.1 &  1.8 &  2.4 & 24.2 & dE/dSph\\
IC4765-f01-435 & 18 48 08.2 &$-$63 26 22 & 16.73 & -17.19 & 0.99  & 11.1 & 11.6 & 22.6 &  3.3 &  4.2 & 23.8 & dE/dSph\\
IC4765-f02-2297& 18 48 17.2 &$-$63 07 21 & 19.58 & -14.33 & 0.68  &  9.5 &  7.6 & 22.9 &  2.1 &  3.6 & 24.0 & dS0,N?/dIrr\\
N6868-f02-L01  & 20 08 03.6 &$-$48 21 18 & 20.05 & -12.51 &\nodata&  7.2 &  7.2 & 23.7 &  2.0 &  3.3 & 24.8 & dE/dSph\\
N6868-f02-1826 & 20 08 48.5 &$-$48 12 13 & 18.84 & -13.72 &\nodata& 14.4 & 14.5 & 22.8 &  2.9 &  4.9 & 23.9 & dIrr\\
N6868-f02-2427 & 20 09 06.9 &$-$48 15 59 & 20.20 & -12.36 &\nodata&  8.3 &  8.7 & 23.4 &  2.3 &  3.8 & 24.5 & dE/dSph\\
N6868-f01-309  & 20 09 11.4 &$-$48 22 56 & 19.83 & -12.73 & 0.92  &  9.2 & 10.4 & 23.4 &  2.4 &  4.1 & 24.5 & dIrr\\
N6868-f01-457\tablenotemark{(c)}  & 20 09 12.4 &$-$48 20 41 & 18.07 & -14.49 & 1.00  & 23.8 & 25.0 & 23.4 &  5.9 &  9.8 & 24.5 & dS0/dIrr\\
N6868-f01-1688 & 20 09 35.7 &$-$48 14 22 & 17.43 & -15.13 & 0.55  & 43.2 & 42.5 & 23.3 &  9.7 & 16.2 & 24.4 & dIrr\\
N6868-f01-1971 & 20 09 42.6 &$-$48 15 05 & 19.21 & -13.35 & 0.96  &  9.0 &  9.9 & 24.1 &  3.4 &  5.7 & 25.2 & dE/dSph\\
N6868-f01-2402\tablenotemark{(c)} & 20 09 54.1 &$-$48 11 48 & 18.66 & -13.90 & 0.67  & 21.6 & 24.7 & 23.2 &  5.4 &  9.0 & 24.3 & dIrr\\
N6868-f01-L01  & 20 09 55.7 &$-$48 17 22 & 17.05 & -15.51 &\nodata& 17.6 & 17.6 & 25.4 & 17.0 & 28.4 & 26.5 & dE/dSph,cloud?\\
N6868-f01-2361 & 20 10 00.1 &$-$48 16 23 & 20.23 & -12.33 & 1.12  &  7.6 &  8.6 & 23.9 &  2.5 &  4.2 & 25.0 & dE/dSph\\
N6868-f01-1135 & 20 10 11.9 &$-$48 18 33 & 18.70 & -13.86 & 0.28  & 17.8 & 17.6 & 23.0 &  3.5 &  5.9 & 24.1 & dS0?/dIrr\\
N6868-f01-2303 & 20 10 13.1 &$-$48 16 11 & 19.40 & -13.16 & 1.16  &  9.2 & 10.6 & 24.2 &  3.9 &  6.5 & 25.3 & dE/dSph\\
\enddata
\tablecomments{Units of right ascension are hours, minutes and seconds, and units of declination
are degrees, arcminutes and arcseconds.}
\tablenotetext{(a)}{Photometry contaminated by a close object (insede 2 isophotal radius)}
\tablenotetext{(b)}{Background galaxy (confirmed by velocity measurement): Sp - spiral, S0 - S0}
\tablenotetext{(c)}{LSBD galaxy in the background of the group NGC 6868}
\tablenotetext{(d)}{LSBD galaxy member of the group (confirmed by velocity measurement)}
\end{deluxetable*}

\clearpage

\begin{deluxetable*}{lccccrrrcrrl}
\tabletypesize{\scriptsize}
\tablenum{4}
\tablecolumns{12}
\tablewidth{0pc}
\tablecaption{Galaxy radial velocities\label{tab4}}
\tablehead{
\colhead{Galaxy} &
\colhead{RA}&
\colhead{DEC}&
\colhead{V} &
\colhead{(V - I)} &
\colhead{$v_{r}$} &
\colhead{$\delta v$} &
\colhead{R/$\#$ll} &
\colhead{F} &
\colhead{$v_{r}^{f}$} &
\colhead{$\delta v$} &
\colhead{References} \\
\colhead{} &
\colhead{[(J2000.0)} &
\colhead{[(J2000.0)} &
\colhead{[mag]} &
\colhead{[mag]} &
\colhead{[km s$^{-1}$]} &
\colhead{[km s$^{-1}$]} &
\colhead{} &
\colhead{} &
\colhead{[km s$^{-1}$]} &
\colhead{[km s$^{-1}$]} &
\colhead{}}
\startdata
H42-f02-616    & 09 59 12.9 &$-$19 33 48 &  19.47 &  0.89 &  61532 &  92 &   11 & 1 &  61532 &  92 & \\
H42-f02-956    & 09 59 13.7 &$-$19 31 45 &  19.60 &  1.73 &  92801 &  53 &  6.6 & 1 &  92801 &  53 & \\
H42-f03-2406   & 09 59 14.9 &$-$19 38 11 &  19.50 &  1.11 &  76014 &  73 &  5.4 & 1 &  76014 &  73 & \\
H42-f03-1928   & 09 59 15.3 &$-$19 42 06 &  19.24 &  1.01 &  36122 &  38 &   10 & 1 &  36122 &  38 & \\
H42-f03-990    & 09 59 17.0 &$-$19 45 59 &  18.88 &  1.00 &  52923 &  45 &  7.2 & 1 &  52923 &  45 & \\
H42-f03-805    & 09 59 17.6 &$-$19 46 12 &  16.21 &  1.34 &  19200 &  18 & 28.5 & 1 &  19200 &  30 & \\
               &	    &		 &	  &	  &  19218 &  80 &	& 1 &	     &     & ZM00, H42-68 \\
H42-f03-964    & 09 59 17.8 &$-$19 45 16 &  17.85 &  1.10 &  18972 &  44 & 12.6 & 1 &  18972 &  44 & \\
H42-f02-1345   & 09 59 18.6 &$-$19 24 38 &  19.81 &  1.60 &  70304 &  94 &  4.4 & 1 &  70304 &  94 & \\
H42-f02-4007   & 09 59 18.7 &$-$19 28 22 &  14.17 &  1.38 &   3635 &  15 & 28.4 & 1 &	3634 &  61 & \\
               &	    &		 &	  &	  &   3621 &  80 &	& 1 &	     &     & ZM00, H42-28 \\
H42-f03-2088   & 09 59 19.1 &$-$19 43 10 &  19.26 &  1.44 &  55954 &  36 & 10.7 & 1 &  55954 &  36 & \\
H42-f02-2234   & 09 59 22.7 &$-$19 30 29 &  20.25 &  2.19 & 134686 &  66 &   11 & 1 & 134686 &  66 & \\
H42-f03-2101   & 09 59 22.7 &$-$19 43 34 &  19.53 &  1.38 & 114587 & 102 &  3.1 & 1 & 114587 & 102 & \\
H42-f02-1898   & 09 59 23.2 &$-$19 27 45 &  19.67 &  1.32 &  56366 &  43 &  9.5 & 1 &  56366 &  43 & \\
H42-f02-1806   & 09 59 25.8 &$-$19 27 07 &  18.36 &  1.19 &  61411 &  47 &  7.4 & 1 &  61411 &  47 & \\
H42-f03-1208   & 09 59 25.9 &$-$19 44 23 &  19.17 &  1.54 &  70919 &  32 & 11.6 & 1 &  70919 &  32 & \\
H42-f03-1495   & 09 59 26.6 &$-$19 38 57 &  17.15 &  0.87 &   3399 &  48 &  8.9 & 1 &	3399 &  73 & \\
               &	    &		 &	  &	  &   3402 &  80 &	& 1 &	     &     & ZM00, H42-85 \\
H42-f03-1089   & 09 59 27.1 &$-$19 45 16 &  18.69 &  1.63 &  70553 &  34 & 14.0 & 1 &  70553 &  34 & \\
H42-f03-2190   & 09 59 28.5 &$-$19 43 53 &  18.56 &  1.28 &  70664 &  45 &  6.5 & 1 &  70664 &  45 & \\
H42-f02-4008   & 09 59 29.0 &$-$19 29 30 &  12.92 &  1.37 &   3934 &  22 & 28.1 & 1 &	3937 &  44 & \\
               &	    &		 &	  &	  &   3980 &  80 &	& 1 &	     &     & ZM00, H42-5 \\
H42-f03-403    & 09 59 29.6 &$-$19 49 12 &  18.89 &  1.07 &  44052 &  54 &  5.9 & 1 &  44052 &  54 & \\
H42-f03-897    & 09 59 29.9 &$-$19 46 24 &  18.29 &  1.15 &  39928 &  58 &  6.5 & 1 &  39928 &  58 & \\
H42-f02-40     & 09 59 30.2 &$-$19 37 17 &  19.68 &  1.18 &  81018 &  62 &  5.3 & 1 &  81018 &  62 & \\
H42-f03-663    & 09 59 30.7 &$-$19 47 46 &  19.03 &  1.62 &  73883 &  40 & 12.1 & 1 &  73883 &  40 & \\
H42-f02-2030   & 09 59 31.7 &$-$19 28 31 &  19.16 &  1.40 &  73390 &  26 &   15 & 1 &  73390 &  30 & \\
H42-f03-783    & 09 59 31.7 &$-$19 47 17 &  19.27 &  1.47 &  73742 &  89 &  4.2 & 1 &  73742 &  89 & \\
\enddata
\tablecomments{Units of right ascension are hours, minutes and seconds, and units of declination
are degrees, arcminutes and arcseconds. Table 4 is published in its entirety in the electronic 
edition of the {\em Astronomical Journal}. A portion is shown here for guidance regarding its
format and content.}
\tablenotetext{(a)}{Galaxy observed in two different masks}
\tablenotetext{(b)}{Background galaxy selected as LSBD}
\tablenotetext{(c)}{LSBD galaxy in the background of the group NGC 6868}
\tablenotetext{(d)}{LSBD galaxy member of the group}
\tablerefs{R96 - Ramella et al. (1996); MKV92 - Malamuth et al. (1992); S96 - Stein (1996); G93 - Garcia (1993); 
ZM98 - Zabludoff \& Mulchaey (1998); ZM00 - Zabludoff \& Mulchaey (2000), dC97 - de Carvalho et al. (1997), 
NED - NASA/IPAC Extragalactic database  (\url{http://nedwww.ipac.caltech.edu/)}}
\end{deluxetable*}

\clearpage

\begin{deluxetable*}{lcccccrrrl}
\tabletypesize{\scriptsize}
\tablenum{5}
\tablecolumns{9}
\tablewidth{0pc}
\tablecaption{LSBD select galaxies with radial velocities\label{tab5}}
\tablehead{
\colhead{Galaxy} &
\colhead{RA}&
\colhead{DEC}&
\colhead{$V_{t}$} &
\colhead{(V - I)} &
\colhead{$\mu_{0}$} &
\colhead{$h$} &
\colhead{$D_{iso}$} &
\colhead{$v_{r}\pm\delta v$} \\
\colhead{} &
\colhead{[(J2000.0)} &
\colhead{[(J2000.0)} &
\colhead{[mag]} &
\colhead{[mag]} &
\colhead{[mag/$\sq^{''}$]} &
\colhead{$''$} &
\colhead{$''$} &
\colhead{[km s$^{-1}$]}}
\startdata
H42-f02-1939  & 09 59 46.8 &$-$19 27 59 & 18.95 & 0.87 & 22.9 & 2.9 & 14.0 & 79534$\pm$169 \\
H42-f04-2307  & 10 00 23.7 &$-$19 39 30 & 19.16 & 1.43 & 22.8 & 2.3 & 12.4 & 82455$\pm$94 \\
N5846-f01-343 & 15 05 30.6 &$+$01 30 31 & 18.48 & 1.22 & 23.7 & 4.4 & 15.6 & 43856$\pm$127 \\
N5846-f01-1443& 15 05 21.8 &$+$01 39 04 & 19.11 & 0.99 & 22.7 & 2.2 & 11.8 & 18697$\pm$30 \\
N6868-f01-457 & 20 09 12.4 &$-$48 20 41 & 18.07 & 1.00 & 23.4 & 5.9 & 23.8 &  4166$\pm$74 \\
N6868-f01-2402& 20 09 54.1 &$-$48 11 48 & 18.66 & 0.67 & 23.2 & 5.4 & 21.6 &  4891$\pm$30 \\ 
              &            &            &      &      &      &      &     &              \\ \hline     
              &            &            &      &      &      &      &     &              \\     
N5846-f01-1996& 15 05 27.9 &$+$01 35 40 & 20.48 & 1.25 & 23.8 & 2.4 & 11.9 & 1707$\pm$73  \\
N5846-f03-1   & 15 07 02.9 &$+$01 30 49 & 16.88 & 0.30 & 23.3 &14.1 & 60.3 & 2815$\pm$71  \\
N5846-f03-147 & 15 07 05.1 &$+$01 30 40 & 18.54 & 0.08 & 23.7 & 7.7 & 26.4 & 2047$\pm$92  \\
IC4765-f01-560& 18 47 21.8 &$-$63 26 44 & 20.05 & 1.12 & 23.4 & 2.2 &  7.6 & 3392$\pm$99  \\ 
IC4765-f01-435& 18 48 08.2 &$-$63 26 22 & 16.73 & 0.99 & 22.6 & 3.3 & 11.1 & 5198$\pm$31  \\
\enddata
\end{deluxetable*}

\clearpage

\begin{deluxetable*}{lccrrccc}
\tabletypesize{\scriptsize}
\tablenum{6}
\tablecolumns{8}
\tablewidth{0pc}
\tablecaption{Group dynamical parameters\label{tab6}}
\tablehead{
\colhead{Group} & 
\colhead{RA} & 
\colhead{DEC} &
\colhead{N$_{tot}$} &
\colhead{N$_{mem}$} &
\colhead{$langle v \rangle$} &
\colhead{$\delta v$} &
\colhead{$r_{200}$} \\
\colhead{} &
\colhead{(J2000.0)} &
\colhead{(J2000.0)} &
\colhead{} &
\colhead{} &
\colhead{[km s$^{-1}$]} &
\colhead{[km s$^{-1}$]} &
\colhead{h$^{-1}$ Mpc}}
\startdata
HCG42  & 10 00 11.4 & $-$19 37 02 &  203  & 19  & 3845$\pm$69  & 296$_{-64}^{+39}$  & 0.67  \\
       &            &            &       & 36\tablenotemark{(a)}  & 3851$\pm$51\tablenotemark{(a)}  & 301$_{-43}^{+30}$\tablenotemark{(a)}  &       \\
HCG42-B1 & 10 00 09.0 & $-$19 35 57 &       & 26  &54541$\pm$202 & 853$_{-159}^{+103}$&       \\
HCG42-B2 & 10 00 11.8 & $-$19 39 10 &       & 16  &72981$\pm$101 & 311$_{-91}^{+42}$  &       \\ 
       &            &            &       &     &              &                    &       \\
NGC5846  & 15 06 28.5 & $+$01 35 10 &  134  & 16  & 1983$\pm$135 & 513$^{+69}_{-121}$ & 1.18  \\
       &            &            &       & 32\tablenotemark{(a)}  & 1950$\pm$75\tablenotemark{(a)}  & 416$^{+43}_{-63}$\tablenotemark{(a)}  &       \\
       &            &            &       &     &              &                    &       \\    
IC4765 & 18 47 32.9 & $-$63 18 46 &   67  & 33  & 4331$\pm$131 & 733$^{+76}_{-111}$ & 1.66  \\
       &            &            &      & 94\tablenotemark{(a)}  & 4488$\pm$65\tablenotemark{(a)}  & 620$^{+40}_{-50}$\tablenotemark{(a)}  &       \\
IC4765-B1 & 18 47 29.1 & $-$63 16 42 &    & 16 & 10975$\pm$125 & 461$^{+66}_{-115}$ &     \\
       &            &             &       &     &              &                    &       \\ 
NGC6868  & 20 09 22.5 & $-$48 18 59 &   57  &  9  & 2807$\pm$80  & 217$^{+37}_{-80}$  & 0.50  \\
       &            &             &       & 19\tablenotemark{(a)}  & 2844$\pm$57\tablenotemark{(a)}  & 238$^{+29}_{-49}$\tablenotemark{(a)}  &       \\
       &            &             &       &     &              &                    &       \\ 
\enddata
\tablecomments{Units of right ascension are hours, minutes and seconds, and units of declination
are degrees, arcminutes and arcseconds. The errors are at the 68\% confidence level.}
\tablenotetext{(a)}{Results obtained using all available velocities (our$+$literature)}
\end{deluxetable*}

\end{document}